\documentclass[aps,prb,twocolumn,superscriptaddress,longbibliography]{revtex4-2}
\usepackage{graphicx}
\usepackage{amsmath}
\usepackage{bbold}
\usepackage[colorlinks=true, citecolor=blue, urlcolor=blue, linkcolor=blue,bookmarks=false,hypertexnames=true]{hyperref} 
\usepackage[normalem]{ulem}

\def \DSME {Department of Semiconductor Materials Engineering, Wroc\l{}aw University of Science and Technology, Wybrze\.{z}e Wyspia\'{n}skiego 27, 50-370 Wroc\l{}aw, Poland}

\def \DTP {Department of Theoretical Physics, Wroc\l{}aw University of Science and Technology, Wybrze\.{z}e Wyspia\'{n}skiego 27, 50-370 Wroc\l{}aw, Poland}

\def \UOTT {Department of Physics, University of Ottawa, Ottawa, Ontario, Canada K1N 6N5}

\begin{document}
	
\title{Valley two-qubit system in a MoS$_2$-monolayer gated double quantum dot}

\author{J. Paw\l{}owski}
\email[]{jaroslaw.pawlowski@pwr.edu.pl}
\affiliation{\DTP}
\author{M. Bieniek}\affiliation{\DTP}\affiliation{\UOTT}
\author{T. Wo\'zniak}\affiliation{\DSME}

\begin{abstract}
We explore a two-qubit system defined on valley isospins of two electrons confined in a gate-defined double quantum dot created within a MoS$_2$ monolayer flake. We show how to initialize, control, interact and read out such valley qubits only by electrical means using voltages applied to the local planar gates, which are layered on the top of the flake. By demonstrating the two-qubit exchange or readout via the Pauli blockade, we prove that valley qubits in transition-metal-dichalcogenide semiconductors family fulfill the universality criteria and represent a scalable quantum computing platform. Our numerical experiments are based on the tight-binding model for a MoS$_2$ monolayer, which gives single-electron eigenstates that are then used to construct a basis of Slater-determinants for the two-electron configuration space. We express screened electron-electron interactions in this basis by calculating the Coulomb matrix elements using localized Slater-type orbitals. Then we solve the time-dependent Schr\"odinger equation and obtain an exact time-evolution of the two-electron system. During the evolution we simultaneously solve the Poison equation, finding the confinement potential controlled via voltages applied to the gates.
\end{abstract}

\maketitle

\section{Introduction}
In recent years quantum computing has experienced a return of great interest mainly due to advances in scaling multi-qubit registers into devices composed of dozens of qubits, such as the Google superconducting quantum computer \cite{scq,scq1}, in case of which they claim achieving quantum supremacy \cite{google}. However, the road to realizing systems that scale to hundreds of logical qubits (each of them storing dozens of noisy qubits with applied quantum error-correction codes \cite{qec}) is still a long way off and intensive work is currently put into other approaches to implement qubits in the solid-state quantum computer \cite{Zwanenburg_Eriksson_2013, Kim_Eriksson_2014, Zajac_Petta_2018, Watson_Vandersypen_2018, Mi_Petta_2018, Landig_Ihn_2018, Zheng_Vandersypen_2019, Petit_Veldhorst_2020}.

Two-dimensional (2D) crystals consisting of single layers of atoms are modern materials that can be used to implement quantum computations. 2D monolayers of transition-metal-dichalcogenides (TMDCs), e.g. MoS$_2$, seem to be better candidates than graphene because they have wide direct band gaps and strong spin-orbit coupling \cite{prx,kos}. The spin-orbit coupling allows to perform quantum operations on a qubit defined on spin of a confined electron. However, TMDCs monolayers have no inversion centers that allow to access an extra degree of freedom of charge carriers, the so-called $\mathcal{K}$-valley index, which opens up an intriguing prospect to define a valley-based qubit \cite{valq0,valq,mymos2}, or create a spin-valley two-qubit system \cite{rohling,mynjp,kotekar}. 

The valley-based information carrier is also postulated in many other materials and arrangements. It was explored in carbon nanotubes \cite{tubes,nt2}, TMDCs \cite{bil,bil2} or graphene bilayers \cite{bil1}, and very recently a fast valley qubit in silicon has been put forward \cite{svq1,svq2}. Deformations of the structure of a graphene flake induces pseudo-magnetic fields that couple the $K$ and $K'$ valleys with opposite signs \cite{valq1} resulting in the valley splitting, or generating valley polarization, thus working as a \emph{valleytronic} filtering device \cite{valq2,filter}.
Also, the electrically controlled valley degree of freedom was reported in twisted WS$_2$ bilayers \cite{twist}.
It is worth noting, that TMDCs monolayers are attractive also from the point of view of optical manipulation of valleys \cite{Xiao_Yao_2012, Cao_Feng_2012, Mak_Heinz_2012, Zeng_Cui_2012, Wang_Urbaszek_2018, baimuratov}, which has been widely used to study initialization and coherence of this degree of freedom \cite{Jones_Xu_2013, Wang_Urbaszek_2016, Hao_Li_2016, Srivastava_Imamoglu_2015b, Aivazian_Xu_2015}, also including interesting many-body effects related to valley polarization \cite{Scrace_Hawrylak_2015, Braz_Castro_2018, Miserev_Loss_2019}. The excess electron gas in these systems \cite{Mak_Shan_2013, Jadczak_Bryja_2017, Back_Imamoglu_2017, Roch_Warburton_2019, Jadczak_Hawrylak_2019, Jadczak_Bryja_2020} potentially provides additional means of control as has recently been examined experimentally \cite{Wang_Kim_2018, BrotonsGisbert_Gerardot_2019}. In this work, however, we focus on all-electrical manipulation protocols due to their potentially improved scalability in large multi-qubit systems.

It is known that for carriers that are spatially localized the valley degree of freedom is still well defined. There are many theoretical \cite{klinovaja, Kormanyos_Burkard_2014, Liu_Yao_2014, Pavlovic_Peeters_2015, Wu_Yao_2016, Dias_Qu_2016, Brooks_Burkard_2017, Qu_Azevedo_2017, mymos2, Szechenyi_Palyi_2018, David_Kormanyos_2018, Chen_Peeters_2018, Chen_Wu_2020, Brooks_Burkard_2020} and experimental \cite{Song_Guo_2015, Zhang_Guo_2017, Pisoni_Ensslin_2018, Wang_Kim_2018, Lau_Goh_2019, Davari_Churchill_2020} studies of quantum dots (QDs) based on MoS$_{2}$ and related TMDCs pointing towards promising routes to various spin-valley massive-Dirac-fermion-based qubit realizations, as summarized recently in Ref. [\onlinecite{Goh_Yee_2020}]. We also note that rich physics of single- \cite{Chirolli_SanJose_2019, MB2} and many-body \cite{MB3} properties of such dots, combined with tunability via heterostructure details or proper substrate engineering, may lead to even greater amount of interest in these systems. Also, experimental demonstrations of TMDCs-based QDs are still in an early phase and significant progress in this direction is expected in coming years.
	
In our recent work [\onlinecite{mymos2}] we proposed a valley-qubit implementation and proved that it is possible to perform single valley-qubit operations in monolayer TMDCs. Here we extend this idea to two-electron systems. We achieve this by using a pair of qubits and coupling them in a controlled manner to perform two-qubit operations. For this we propose a nanodevice based on a gate-defined \cite{gated_haw} double QD \cite{dgd_haw1,dqd_haw2,dqd_haw3,dqd_haw4} within a MoS$_2$ monolayer in which we confine two electrons. 
Using quantum computing language, we identify two-qubit states of two-electron system as described by left-dot ($L$) and right-dot ($R$) valley index (showing which valley, i.e. $K$ or $K'$ is occupied) for a chosen spin orientation (e.g. spin-up, $\uparrow$).
This means that single qubit state is pinned to the specific dot, not to the electron, which is delocalized between the dots. In our proposal the two-qubit $|00\rangle$ state is associated with the $|K_{\uparrow, L}^{'}K_{\uparrow, R}^{'}\rangle$, where $L$ and $R$ indicate the left and right dot, respectively. The remaining two-qubit basis states are defined as: $|01\rangle = |K_{\uparrow, L}^{'}K_{\uparrow, R}^{}\rangle$, $|10\rangle = |K_{\uparrow, L}^{}K_{\uparrow, R}^{'}\rangle$, and $|11\rangle = |K_{\uparrow, L}^{}K_{\uparrow, R}^{}\rangle$. For brevity, in the next Sections we will drop L and R indices. By applying control voltages to the gates we modulate the confinement potential forcing this way intervalley transitions \cite{valtr} of each qubit associated with each dot, and adjust the potential barrier between them. Accurate modeling of the Coulomb interaction enables us to correctly describe the valley-swap operation between qubits, as well as the qubit initialization and readout via the valley Pauli blockade \cite{nt2,nt3}.

The paper is organized as follows: In Section II we propose and describe a realistic nanodevice structure. In Section III we discuss theoretical tools used, i.e. single- and two-electron theory, details of calculating the Coulomb integrals, the electrostatic potential model and the time-dependent simulations. Readers interested in results may omit Section III and proceed to Section IV, in which states of the double QD are discussed. Then, in Sections V and VI results of the electrically-driven valley exchange and the valley-Pauli-blockade readout mechanism are discussed. We conclude with a summary in Section VII.

\section{Device structure}
\begin{figure}[tb]
	\center
	\includegraphics[width=8.7cm]{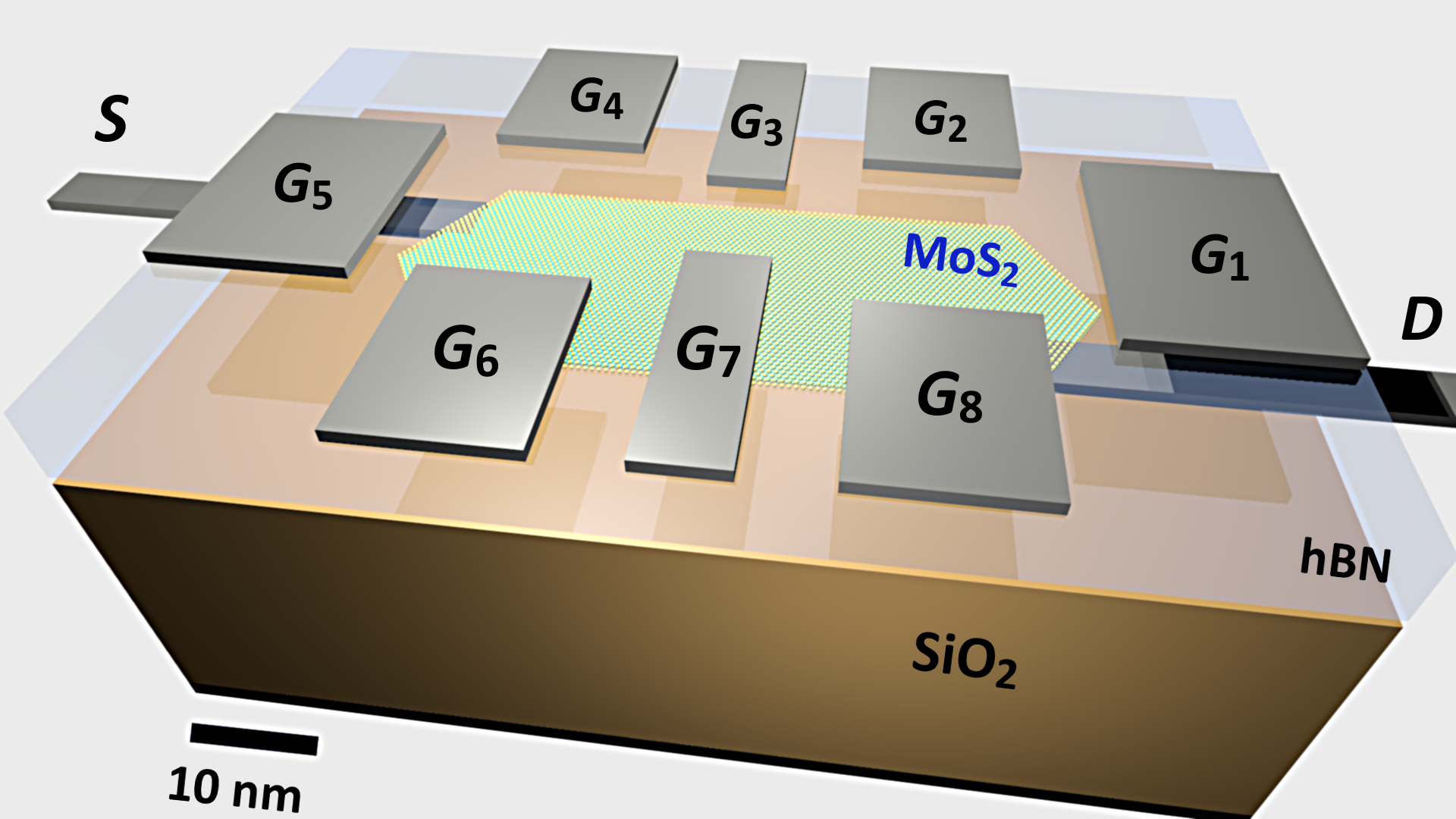}
	\caption{\label{fig:1} Structure of the proposed device consisting of a MoS$_2$ nanoflake with nearby source (S) and drain (D) electrodes deposited on a SiO$_2$ layer placed on highly doped substrate serving as the back gate, together with the layout of eight gates ($G_{1...8}$) separated from the flake by a hBN barrier layer.}
\end{figure}
The structure of the proposed nanodevice is presented in Fig.~\ref{fig:1}. On a strongly doped silicon substrate (Si$^{n++}$) we place a 25-nm-thick layer of SiO$_2$. Then we place two electrodes that serve as a source ($S$) and a drain ($D$) and deposit a MoS$_2$ monolayer ribbon of the shape and dimensions presented in Fig.~\ref{fig:2}. The monolayer is then covered with a $5$-nm-thick insulating layer of hexagonal boron nitride (hBN), which has a large band gap \cite{hbn}, forming a tunnel barrier. Finally, on top of this layered structure we lay down eight control gates $G_{1...8}$, as presented in Fig.~\ref{fig:1}. Three of them: 15-nm-wide $G_1$, and 10-nm-wide pair $G_{2,8}$ are placed around the right-dot region and form its confinement. Similarly, 15-nm-wide $G_5$, and 10-nm-wide pair $G_{4,6}$ form the left dot. Left and right dots are separated by a controllable barrier generated by a pair of elongated gates $G_{3,7}$. The proposed device structure is very similar to the one described in Ref. [\onlinecite{Davari2020}], albeit with a reversed ordering of layers, i.e. in our case the role of the top gate is taken over by the strongly doped substrate.
	
The presented gate layout enables us to	create a confinement potential forming a double QD structure within the flake. With the tunable barrier height between the dots (controlled via $G_{3,7}$ gates), variable locations of the dot-potential minima (via $G_{1,2,8}$ or $G_{4,5,6}$ gates) and confinement depth (by tuning the negative bias voltage applied to all gates), we can efficiently control each dot confinement, as well as the potential offset between the dots (via $G_1$ and $G_5$ gates). Potential $\phi(\mathbf{r})$ in the entire nanodevice, controlled by the gate voltages, is calculated by solving the generalized Poisson's equation \cite{mydrut}, while the electron states in the flake are described with the tight-binding formalism.

\section{Model and methods}

\subsection{Single-electron tight-binding theory}

The flake used in our device is a monolayer made out of molybdenum disulfide. This semiconductor is successfully described by several tight-binding (TB) models with various numbers of orbitals used \cite{MB1, azgari, ridolfi}. Although, at least six \cite{MB1} Mo and S orbitals with next-nearest-neighbor hoppings are necessary to construct a minimal TB model that reproduces low-energy physics around the Fermi level in the entire Brillouin zone. It has been shown that a simple three-Mo-orbital TB model \cite{xiao} on a triangular lattice can correctly represent the dispersion relation and the orbital composition close to the K point in the BZ near the band edges, where the Bloch states mainly consist of Mo $d$ orbitals \cite{haw}. Because in our calculations we are concerned solely with states derived from the minimum in the conduction band (CB) at $\pm K$ points, we can safely ignore the multi-valley structure of the conduction band \cite{MB1, MB2} and focus on simple, effective description of the confined electron states.  
\begin{figure}[tb]
	\center
	\includegraphics[width=8cm]{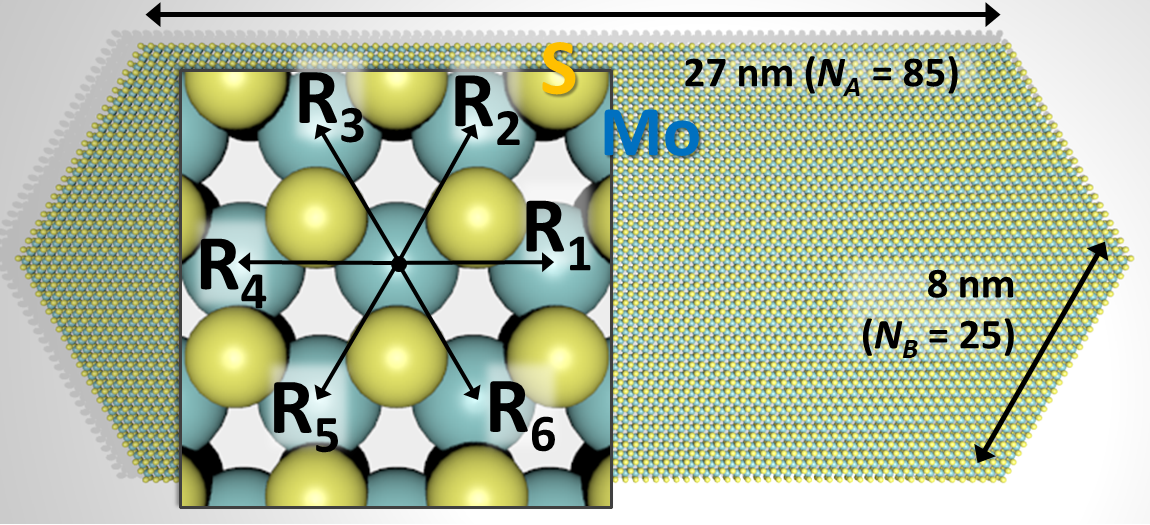}
	\caption{\label{fig:2} The MoS$_2$ monolayer flake used in the device has elongated hexagonal shape with sides made of $N_A=85$ or $N_B=25$ Mo atoms, giving the bottom side $27$-nm-long, and lateral sides $8$-nm-long (the distance between Mo nodes is $0.319$~nm); (inset) The MoS$_2$ crystal lattice structure formed of hexagonally packed Mo and S atoms arranged in triangular lattices rotated relative to each other by $\pi$. The Mo lattice vectors $R_k$ determine the hopping directions in our nearest-neighbors TB model.}
\end{figure}
Consequently, we have described the monolayer structure using three Mo orbitals:  $d_{z^2}$, $d_{xy}$, $d_{x^2-y^2}$, with the nearest-neighbors hoppings \cite{xiao}:
\begin{align}\label{ham1e}
&H_{1e}=\nonumber\\
&\sum_{m\,\sigma \sigma'\alpha \beta }\left(\delta_{\sigma \sigma'}\delta_{\alpha \beta}(\epsilon_{\alpha }+{\varphi }_m)\,{\hat{n}}_{m\alpha \sigma }+{s}^z_{\sigma \sigma'}{\lambda }_{\alpha \beta }\,{\hat{c}}^{\dagger }_{m\alpha \sigma }{\hat{c}}_{m\beta \sigma '}\right)\nonumber\\
+&\sum_{\langle mn\rangle\,\sigma \sigma'\alpha \beta }{\delta_{\sigma \sigma'}t_{\alpha \beta }^{\langle mn\rangle}\,{\hat{c}}^{\dagger }_{m\alpha \sigma }{\hat{c}}_{n\beta \sigma }}
+H_Z.
\end{align}
Indices $\{m,n\}$, $\{\sigma, \sigma'\}$ and $\{\alpha,\beta\}$ enumerate lattice sites, spins, and orbitals, e.g. operator ${\hat{c}}^{\dagger }_{m\alpha \sigma }$ (${\hat{c}}_{m\alpha \sigma }$) creates (annihilates) electron with orbital $\alpha$ and spin $\sigma$ at $m$-th lattice site. Also, we abbreviate $\hat{n}=\hat{c}^{\dagger}\hat{c}$.  On-site energies for orbitals  $\alpha$ are parametrized by $\epsilon_{\alpha}$. The potential energy of the electrostatic confinement at the $m$-th lattice site $\varphi_m=-|e|\phi(x_m,y_m)$ together with the on-site energies $\epsilon_\alpha$ enter the diagonal matrix elements. Parameters $\lambda_{\alpha\beta}$ express the intrinsic spin-orbit coupling \cite{kos}, $s_z$ stands for the $z$-Pauli-matrix, and $H_Z$---the Zeeman Hamiltonian which is added whenever response to magnetic field is studied.

The off-diagonal electron hopping element from the $\beta$ Mo orbital localized in the $n$-th lattice site to the $\alpha$ orbital localized in the $m$-th site is denoted by $t_{\alpha \beta }^{\langle mn\rangle}\equiv t_{\alpha \beta }\!\left(R_{p\left(m,n\right)}\right)$. We note that it does not flip spin, as explicitly written using Kronecker delta's $\delta_{\sigma\sigma'}$. It depends on the hopping direction between $\langle m\,n\rangle$ nearest neighbor pair, that is described by the $R_p$ ($p=1...6$) vectors for the molybdenum (Mo) lattice, see Fig.~\ref{fig:2}.  They form two non-equivalent families: $R_{1,3,5}$ and $R_{2,4,6}$, which differ by the nearest sulphur (S) neighbor position, either on the left or the right side. This symmetry constraint reflects on the reciprocal lattice where the K points in the corners of the hexagonal Brillouin zone form two non-equivalent families: $K$ and $K'$. We note  that opposite hoppings are mutually transposed: $t_{\alpha \beta}(R_p)=t_{\beta \alpha }(-R_p)$. Their explicit forms, together with the on-site energies $\epsilon_\alpha$ and spin-orbit coupling parameters $\lambda_{\alpha\beta}$, can be found in [\onlinecite{mymos2,xiao}]. 

Knowing the tight-binding representation of the flake lattice we solve the eigenproblem for the single-electron Hamiltonian (\ref{ham1e}): $H_{1e}\boldsymbol{\psi}_i = \mathcal{E}_i\boldsymbol{\psi}_i$, with eigenenergies $\mathcal{E}_i$. Calculation results will be discussed in Section \ref{sec:dqd}.

\subsection{Two-electron theory}
We use the found single-electron eigenstates $\boldsymbol{\psi}_i$ to construct the two-electron spinor 
$\langle \mathbf{r}_1\mathbf{r}_2\tilde{|ij\rangle}\equiv \boldsymbol{\Psi}_{ij}(\mathbf{r}_1,\mathbf{r}_2)$ of an antisymmetric form:
\begin{equation}\label{slater}
 \boldsymbol{\Psi}_{ij}(\mathbf{r}_1,\mathbf{r}_2)=
 \frac{1}{\sqrt{2}}
 \left(
 \boldsymbol{\psi}_i(\mathbf{r}_1)\otimes\boldsymbol{\psi}_j(\mathbf{r}_2)
 -\boldsymbol{\psi}_j(\mathbf{r}_1)\otimes\boldsymbol{\psi}_i(\mathbf{r}_2)
 \right),
\end{equation}
with spin-orbitals notation: $\boldsymbol{\psi}_i(\mathbf{r}_a)\equiv
\left(\psi^{\sigma\alpha}_i(\mathbf{r}_a)\right)$, where $a=1,2$ (two electrons), $\sigma=1,2$ (1/2-spin-vector elements), and $\alpha=1,2,3$ (orbital number). We abbreviate: $\boldsymbol{\psi}_i(\mathbf{r}_1)\otimes\boldsymbol{\psi}_j(\mathbf{r}_2)\equiv \langle \mathbf{r}_1\mathbf{r}_2|ij\rangle$, and then $\tilde{|ij\rangle}=\frac{1}{\sqrt{2}}\left(|ij\rangle-|ji\rangle\right)$.
Taking $m$ single-electron low-energy spin-orbitals near the CB minimum we combine them into $n=\binom{m}{2}$ two-electron spinors $\tilde{|ij\rangle}$. Next we expand the full two-electron spinor in this basis:
\begin{equation}\label{ci}
\boldsymbol{\Psi}(\mathbf{r}_1,\mathbf{r}_2)=\sum_{k=1}^{n}d_k\, 
\langle \mathbf{r}_1\mathbf{r}_2\tilde{|ij\rangle},
\end{equation}
where $i,j\in\{1,2,\dots,m\}$, and $n=m(m-1)/2$.

In order to include electron-electron interactions the two-electron Hamiltonian is written as
\begin{equation}\label{ham2e}
H_{2e}(\mathbf{r}_1,\mathbf{r}_2)=H_{1e}(\mathbf{r}_1)+H_{1e}(\mathbf{r}_2)+\bar{V}_C(\mathbf{r}_1,\mathbf{r}_2),
\end{equation}
where $\bar{V}_C(\mathbf{r}_1,\mathbf{r}_2)=\bar{V}_C(|\mathbf{r}_1-\mathbf{r}_2|)$ is the Coulomb interaction screened by the dielectric environment of the nearby layers, described in details further in the text.

\subsection{Configuration-interaction method}
To describe a two-electron state exactly,
we utilize the configuration-interaction method \cite{ci1,ci2,ci3,ci4} where the Hamiltonian given in Eq. (\ref{ham2e}) is represented in the two-electron basis [defined in Eq. (\ref{ci})] of two-particle antisymmetric Slater determinants [given by Eq. (\ref{slater})], constructed from the single-electron states. In such a basis, with
$\langle ij| \bar{V}_C|kl\rangle$ abbreviated as $V_{ijkl}$, one has \cite{ci5,ci61,ci6}:
\begin{equation}\label{h2e}
H_{2e} = \sum_{i<j} d^\dagger_i d^\dagger_j d_i d_j\,\mathcal{E}_{ij} +\!\sum_{i<j,k<l}\!d^\dagger_i d^\dagger_j d_k d_l \left(V_{ijkl}\!-\!V_{ijlk}\right).
\end{equation}
Here $d^\dagger_i d^\dagger_j$ creates an electron pair in the $\tilde{|ij\rangle}$ state, and the two-electron energy is defined as $\mathcal{E}_{ij}=\mathcal{E}_i+\mathcal{E}_j$, with $\mathcal{E}_i$ being the energy of a single-electron state $|i\rangle$. An explicit derivation of the Hamiltonian in Eq.~(\ref{h2e}) can be found in Appendix~\ref{apx:a}. We also note that due to the symmetry constraints, Coulomb-matrix elements $V_{ijkl}$ obey the following relation:
\begin{equation}
\sum_{i<j,k<l} d^\dagger_i d^\dagger_j d_k d_l \left(V_{ijkl}-V_{ijlk}\right)= \frac{1}{2}\sum_{i,j,k,l} d^\dagger_i d^\dagger_j d_k d_l V_{ijkl}.
\end{equation}

\subsection{Coulomb integrals} 
Two-electron scattering-matrix elements $\langle ij| \bar{V}_C|kl\rangle$ are calculated from the two-body localized on-site Coulomb-matrix elements 
$\langle sp|\bar{V}_C|df\rangle\equiv \mathcal{V}_{spdf}$. To get the latter we expand states $|i\rangle$ in a basis of atomic orbitals $\eta^\alpha_s$ centered at the lattice nodes $\mathbf{r}_s=(x_s,y_s,0)$. For every two nodes $s$ and $p$, located in $\mathbf{r}_s$ and $\mathbf{r}_p$, we have $\eta^\alpha_s(\mathbf{r}-\mathbf{r}_s)=\eta^\alpha_p(\mathbf{r}-\mathbf{r}_p)$.
On-site states will be indexed by $spdf$ indices, e.g. for $s$:
\begin{equation}
\langle\mathbf{r}|i\rangle = \boldsymbol{\psi}_i(\mathbf{r}) = \sum_{s,\sigma_s,\alpha_s} \psi^{\sigma_s\alpha_s}_{i,s} \eta^{\alpha_s}_s(\mathbf{r}).
\end{equation}
In some places we will abbreviate on-site states as $|s^{\alpha_s}\rangle$, i.e. $\eta^{\alpha_s}_s(\mathbf{r})=\langle\mathbf{r}|s^{\alpha_s}\rangle$. Let us now expand the scattering-matrix elements $V_{ijkl}$ in this new on-site basis of atomic-orbitals. We abbreviate $\langle \eta^{\alpha_s}_s\eta^{\alpha_p}_p|\bar{V}_C| \eta^{\alpha_d}_d\eta^{\alpha_f}_f\rangle \equiv
\langle s^{\alpha_s}p^{\alpha_p}|\bar{V}_C| d^{\alpha_d}f^{\alpha_f}\rangle \equiv \mathcal{V}^{\alpha_s\alpha_p\alpha_d\alpha_f}_{spdf}$, or simply $\mathcal{V}_{spdf}$. In general we have
\begin{equation}\label{vijkl}
V_{ijkl}=\sum_{\substack{s,\sigma_s,\alpha_s\\ p,\sigma_p,\alpha_p\\ d,\sigma_d,\alpha_d\\ f,\sigma_f,\alpha_f}} 
\left(\psi^{\sigma_s\alpha_s}_{i,s}
\psi^{\sigma_p\alpha_p}_{j,p}\right)^\ast
\psi^{\sigma_d\alpha_d}_{k,d}
\psi^{\sigma_f\alpha_f}_{l,f}
\mathcal{V}_{spdf}.
\end{equation}
When calculating $V_{ijkl}$ the main contribution is from one- and two-center integrals, i.e. $s=d$ and $p=f$. In addition, the Coulomb interaction does not change spin, thus $\sigma_s=\sigma_d$ and $\sigma_p=\sigma_f$. Therefore, we can simplify calculations taking leading order-of-magnitude elements 
\begin{equation}\label{vijkl2}
V_{ijkl}\approx\sum_{\substack{s,\sigma_s,p,\sigma_p\\ \alpha_s\alpha_p,\alpha_d,\alpha_f}} 
\left(\psi^{\sigma_s\alpha_s}_{i,s}
\psi^{\sigma_p\alpha_p}_{j,p}\right)^\ast
\psi^{\sigma_s\alpha_d}_{k,s}
\psi^{\sigma_p\alpha_f}_{l,p}
\mathcal{V}^{\alpha_s\alpha_p\alpha_d\alpha_f}_{spsp},
\end{equation}
with atomic elements defined as
\begin{equation}\label{vspdf}
\begin{split}
\mathcal{V}^{\alpha_s\alpha_p\alpha_d\alpha_f}_{spsp}=
\int\!\!\int d^3&r_1d^3r_2\left(\eta^{\alpha_s}_s(\mathbf{r}_1) \eta^{\alpha_p}_p(\mathbf{r}_2)\right)^\ast\times\\  
&\times\eta^{\alpha_d}_s(\mathbf{r}_1) \eta^{\alpha_f}_p(\mathbf{r}_2) 
\bar{V}_C(\mathbf{r}_1,\mathbf{r}_2).
\end{split}
\end{equation}
 
Atomic Coulomb-matrix elements $\mathcal{V}_{spdf}$ were calculated using a Monte-Carlo approach with adaptive sampling, via the VEGAS algorithm \cite{lepage}. We assume that localized orbitals $\eta^{\alpha_s}_s(\mathbf{r})$ have hydrogen-like Slater form \cite{sc3} with appropriate atomic shielding parameters \cite{ci5,sc2}.
The molybdenum atomic-shielding constant $\zeta$ is in the range between $2.85$ and $3.11$~(a.u.) \cite{sc4}. However, due to screening by the sulfur dimers, a smaller $\zeta$ value is taken since the actual orbital is slightly widened. We estimate this effect through density functional theory (DFT) calculations using the projector augmented waves (PAW) method \cite{PAW} for atoms, and the Perdew-Burke-Ernzerhof (PBE) parametrization \cite{PBE} of generalized gradients approximation (GGA) for exchange-correlation functional, as implemented in VASP \cite{VASP}. We have used a plane-waves-basis cutoff of 400 eV and a $12 \times 12 \times 1$ $k$-points grid. The unit cell contains 15\AA~of vacuum in the direction perpendicular to the monolayer. The spin-orbit interaction was taken into account during all calculation steps. 
This setup ensures consistency with the employed tight-binding model parametrization from Ref. [\onlinecite{xiao}]. 

\begin{figure}[tb]
	\center
	\includegraphics[width=7.7cm]{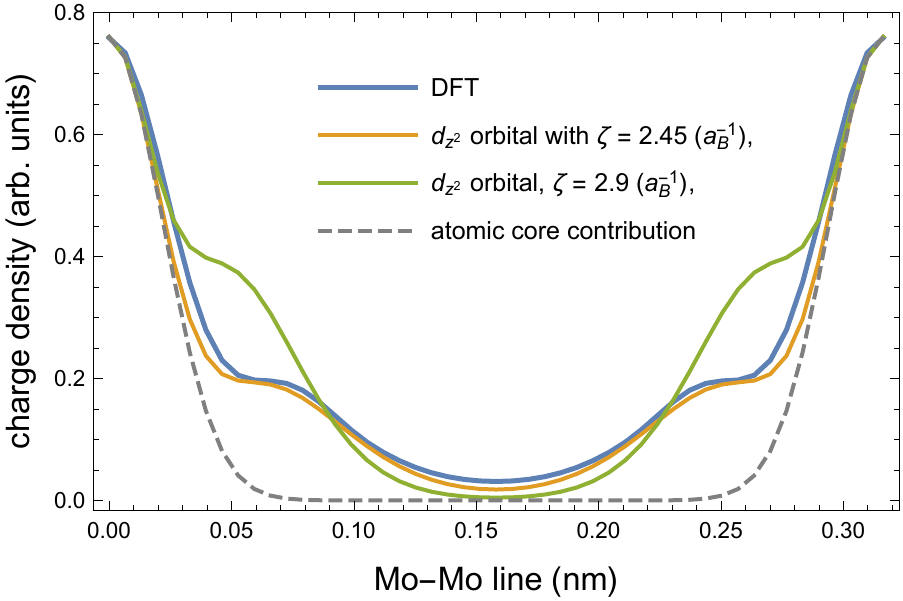}
	\caption{\label{fig:3} Fitting of the molybdenum atomic shielding constant $\zeta$. On-site electron density along Mo-Mo interatomic line in MoS$_2$ monolayer -- calculated within DFT (blue curve) versus density obtained for the Slater-type orbitals with $\zeta=2.45$~$a^{-1}_B$ (orange), or $\zeta=2.9$~$a^{-1}_B$ (green) -- together with the assumed atomic core density (dashed grey).}
\end{figure}
In Fig.~\ref{fig:3} we plot an
electronic charge density for the MoS$_2$ monolayer along a Mo-Mo line at the CB minimum at $K$ point obtained from the DFT calculations (blue curve), and compare it with a charge density calculated 
using Slater-type orbitals $Nr^{n-1}e^{-\zeta r}\,Y_{lm}$ for molybdenum, which for $4d_{z^2}$ has the form:
\begin{equation}
\eta_{0}^{4d_{z^{2}}}(r)=\frac{\zeta^{9/2}}{2\sqrt{63\pi}} re^{-\zeta r}(3z^2-r^2),
\end{equation}
and contributes to 88\% of the orbital composition in the CB at $K$. To include the contribution from the Mo atomic core, we add a charge density approximated by a Gaussian fit and shown as a dashed gray curve in Fig.~\ref{fig:3}. It turns out that $\zeta=2.9$~(a.u.), taken from the literature, gives a less satisfactory fit to the DFT electron density (green curve), than the tuned value $\zeta=2.45$ (orange), which is assumed when calculating $\mathcal{V}_{spdf}$, as defined in Eq. (\ref{vspdf}).

During $\mathcal{V}_{spdf}$ calculations, we found that the most significant contributions come from one-center \textit{direct} integrals, i.e. involving orbitals centered at the same node: $s=p=d=f$ ($3^4=81$ different integrals; in practise it is less due to orbital symmetries), and two-center integrals: $s=p$ and $d=f$ between the nearest sites, i.e. $\langle12|\bar{V}_C|12\rangle$ (also 81 elements). To simplify the calculations of $V_{ijkl}$ we take the explicit values only for the 75 integrals which are greater than $0.3$~eV. They are grouped in Table~\ref{tab:1} in Appendix~\ref{apx:b}, with the largest value reported for each group. The small spread of values within each group of integrals comes from the probabilistic nature of the method used in the calculations. Taking more integrals explicitly, i.e. smaller than the arbitrarily set $0.3$~eV threshold, would have no practical impact on the calculations. However, what is important, remaining long-range elements were also taken into account in $V_{ijkl}$, albeit modelled as a classical point-like density-density terms of screened Coulomb interaction $\bar{V}_C(r)$, defined in Appendix~\ref{apx:c}.

The $V_{ijkl}$ elements were calculated once, by summing up $\mathcal{V}_{spdf}$ integrals, as described in Eq.~(\ref{vspdf}), and stored on hard disk ($80\times80\times80\times79/2$ elements in total) for further use.

\subsection{Electrostatic potential}
 \begin{figure}[tb]
	\center
	\includegraphics[width=8.5cm]{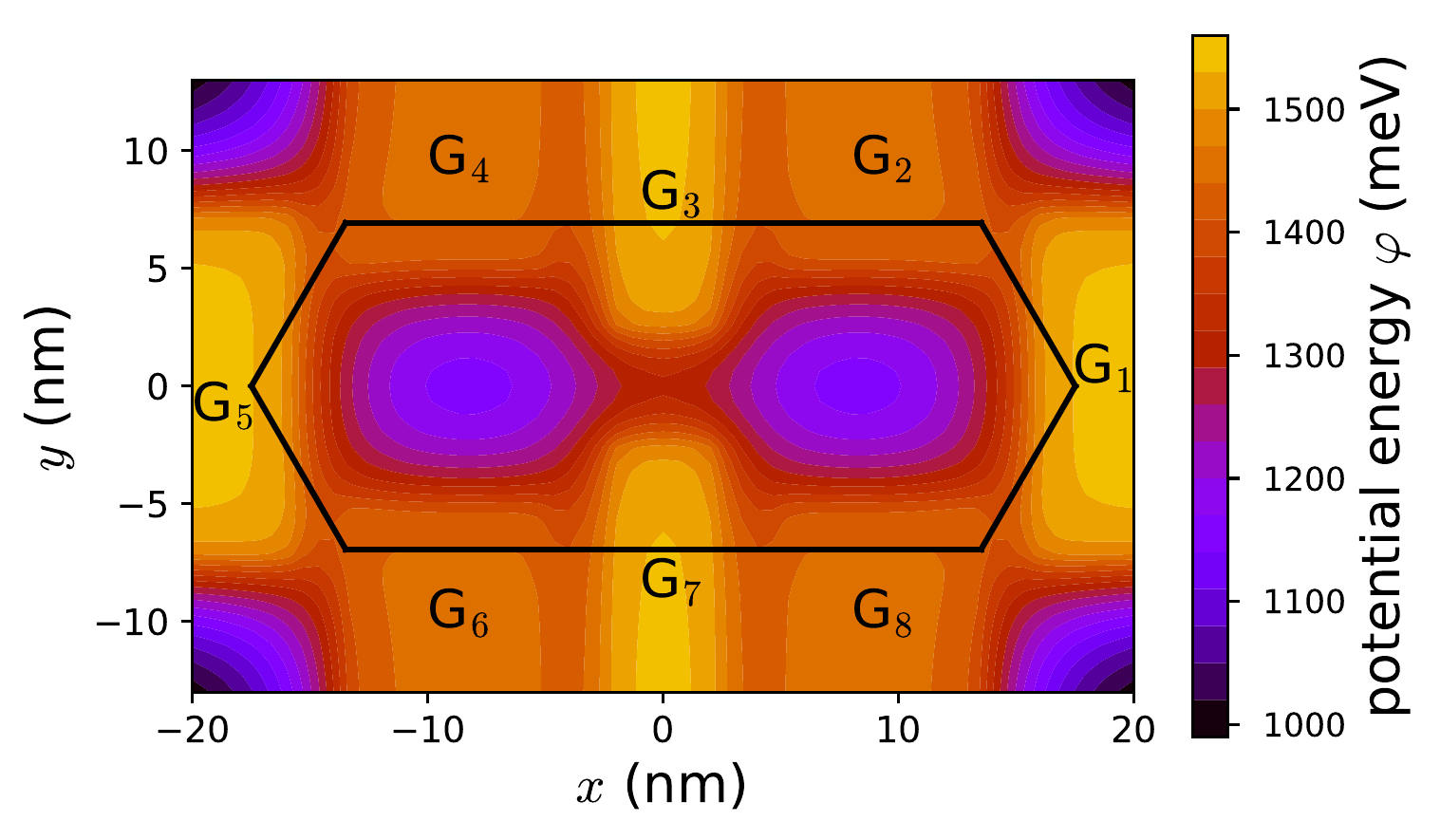}
	\caption{\label{fig:4} Confinement potential in the area of the nanoflake (black outline) calculated via the Poisson equation for voltages: $V_{1,3,5,7}=-1600$~mV, $V_{2,4,6,8}=-1500$~mV applied to the nanodevice local gates G$_{1...8}$ (see Fig.~\ref{fig:1}). Such gating results in a double quantum dot structure created within the nanoflake area.}
\end{figure}
As discussed previously, the layout of gates is presented in Fig.~\ref{fig:1}. Voltages applied to these gates (relative to the substrate) are used to create confinement in the flake. To calculate realistic electrostatic potential $\phi(\mathbf{r})$ we solve the generalized Poisson equation \cite{mydrut} taking into account voltages $V_{1..8}$ applied to the control gates $G_{1..8}$ and to the highly doped substrate (kept at the referential potential $V_0=0$), together with space-dependent permittivity of different materials in the device. \textbf{At the lateral and top sides of the computational box we apply Neumann boundary conditions with zeroing normal component of the electric field}. Further details of the used method can be found in Ref. [\onlinecite{mymos2}]. Resulting potential in the area between SiO$_2$ and hBN layers, where the flake is sandwiched, is presented in Fig.~\ref{fig:4}, and calculated for voltages $V_{1,3,5,7}=-1600$~mV and $V_{2,4,6,8}=-1500$~mV applied to the local gates. At the same time, voltage  $V_B=-1600$~mV applied to gates $V_3$ and $V_7$ controls the height of the interdot barrier.

\subsection{Time-dependent simulations}
We have learnt how to calculate eigenstates of a two-electron system confined within a double quantum dot with an included Coulomb interaction. Such dressed in interaction states, each described by a set of $m(m-1)/2$ ($m=80$) $d_{ij}$ amplitudes, will serve as initial states for the time-dependent simulations which we will now introduce.

To control qubits we will apply time-varying voltages to the device gates. Evolution of the system wavefunction induced in this way will be described within our configuration-interaction base, albeit now with time-dependent amplitudes $d_{ij}(t)$, constituting the time-dependent configuration-interaction method. Insertion of $\boldsymbol{\Psi}(\mathbf{r}_1,\mathbf{r}_2,t)=\sum_{i<j}d_{ij}(t)\, 
\boldsymbol{\Psi}_{ij}(\mathbf{r}_1,\mathbf{r}_2)e^{-\frac{\imath}{\hbar}\mathcal{E}_{ij}t}$ into the Schr\"{o}dinger equation with the time-dependent Hamiltonian
\begin{equation}
H_{2e}(\mathbf{r}_1,\mathbf{r}_2,t)=H_{2e}(\mathbf{r}_1,\mathbf{r}_2)+\delta\varphi(\mathbf{r}_1,t)+\delta\varphi(\mathbf{r}_2,t),
\end{equation}
and the variable potential energy $\varphi(\mathbf{r},t)=\varphi(\mathbf{r})+\delta\varphi(\mathbf{r},t)$, together with the Coulomb matrix elements, gives a recipe time-evolution of the system (the dot denotes the time derivative):
\begin{equation}
\label{timeq}
\dot{d}_{ij}(t)=-\frac{\imath}{\hbar}\sum_{k<l} d_{kl}(t)
\!\left\{V_{ijkl}\!-\!V_{ijlk}\!+\!\delta_{ijkl}(t)\right\}
\!e^{\frac{\imath}{\hbar}(\mathcal{E}_{ij}\!-\!\mathcal{E}_{kl})t},
\end{equation}
with energy $\mathcal{E}_{ij}=\mathcal{E}_i+\mathcal{E}_j$ in the $\tilde{| ij\rangle}$ basis-state. Matrix elements related to the potential energy are given by $\delta_{ijkl}(t)=\tilde{\langle ij|} \delta\varphi(\mathbf{r}_1,t)+\delta\varphi(\mathbf{r}_2,t)\tilde{|kl\rangle}= \delta\varphi_{ijkl}(t)-\delta\varphi_{ijlk}(t)$, where
\begin{equation}
\begin{split}
\delta\varphi_{ijkl}(t)&=\langle ij| \delta\varphi(\mathbf{r}_1,t)+\delta\varphi(\mathbf{r}_2,t)|kl \rangle=\\
&= \langle i | \delta\varphi(\mathbf{r},t) | k \rangle \delta_{jl}+
\langle j | \delta\varphi(\mathbf{r},t) | l \rangle \delta_{ik}.
\end{split}
\end{equation}

The full time-dependent potential energy $\varphi(\mathbf{r},t)=\varphi(\mathbf{r})+\delta\varphi(\mathbf{r},t)$ contains a variable component
$\delta\varphi(\mathbf{r},t)$, generated by modulation of the gate voltages. The whole is calculated as $\varphi(\mathbf{r},t)=-|e|\phi(\mathbf{r},t)$, with the electrostatic potential $\phi(\mathbf{r},t)$ obtained by solving the Poisson equation for the variable density $\rho(\mathbf{r},t)$ at every time step. Note that the charge density originates from the actual wavefunction, thus the Schr\"{o}dinger and Poisson equations are solved in a self-consistent way.

\subsection{Evaluation of valley indices}
In order to follow the valley index corresponding to each dot, as well as the total valley index, we have to calculate the Fourier transform of the time-dependent two-electron wave function:
\begin{equation}
\begin{split}
\tilde{\boldsymbol{\Psi}}(\mathbf{k}_1,\mathbf{k}_2,t)&= 
\int_{F}\!d^2r_1 d^2r_2\,\boldsymbol{\Psi}(\mathbf{r}_1,\mathbf{r}_2,t) e^{-\imath(\mathbf{k}_1\mathbf{r}_1+\mathbf{k}_2\mathbf{r}_2)}\\
&=\sum_{i<j}d_{ij}(t)\tilde{\boldsymbol{\Psi}}_{ij}(\mathbf{k}_1,\mathbf{k}_2) e^{-\frac{\imath}{\hbar}\mathcal{E}_{ij}t},
\end{split}
\end{equation}
with $\tilde{\boldsymbol{\Psi}}_{ij}(\mathbf{k}_1,\mathbf{k}_2)=
[\tilde{\boldsymbol{\psi}}_i(\mathbf{k}_1) \tilde{\boldsymbol{\psi}}_j(\mathbf{k}_2)-
\tilde{\boldsymbol{\psi}}_j(\mathbf{k}_1) \tilde{\boldsymbol{\psi}}_i(\mathbf{k}_2)]/\sqrt{2}$.
For single-electron states $\boldsymbol{\psi}_i(\mathbf{r})$ the Fourier transform is defined as:
\begin{equation}\label{sfour}
\tilde{\boldsymbol{\psi}}_i(\mathbf{k})=\int_{F}\!d^2r\,\boldsymbol{\psi}_i(\mathbf{r}) e^{-\imath\mathbf{k}\mathbf{r}},
\end{equation} 
integrated over the flake surface $F$, with the 2D-wave vector $\mathbf{k}\equiv(k_x,k_y)$. The Fourier transform naturally exhibits periodicity in the reciprocal space, so we can restrict the $k$-area to $\tilde{F}$: $k_{x,y}\in\left[-2\pi/a,2\pi/a\right]$, which encompasses the first Brillouin Zone (BZ). 

Knowing $\tilde{\boldsymbol{\psi}}_i(\mathbf{k})$ we can calculate the probability density in the reciprocal space as: 
\begin{equation}
\begin{split}
&\tilde{\rho}(\mathbf{k},t)=2\int_{\tilde{F}}\!d^2k'\,|\tilde{\boldsymbol{\Psi}}(\mathbf{k},\mathbf{k}',t)|^2=\\
&=\!\sum_{i<j,k<l} d^\ast_{ij}(t)d_{kl}(t)
\left\{\tilde{\boldsymbol{\psi}}_i^\dag(\mathbf{k})\tilde{\boldsymbol{\psi}}_k(\mathbf{k})\delta_{jl} -\tilde{\boldsymbol{\psi}}_i^\dag(\mathbf{k}) \tilde{\boldsymbol{\psi}}_l(\mathbf{k})\delta_{jk}\right.\\
&\qquad\left.-\tilde{\boldsymbol{\psi}}_j^\dag(\mathbf{k}) \tilde{\boldsymbol{\psi}}_k(\mathbf{k})\delta_{il}
+\tilde{\boldsymbol{\psi}}_j^\dag(\mathbf{k}) \tilde{\boldsymbol{\psi}}_l(\mathbf{k})\delta_{ik}\right\}
e^{\frac{\imath}{\hbar}(\mathcal{E}_{ij}-\mathcal{E}_{kl})t}.
\end{split}
\end{equation}
Finally, the total valley index of the two-electron system is calculated as
\begin{equation}\label{totk}
\mathcal{K}(t)=\frac{3a}{4\pi}\int_{\tilde{F}_{1/3}}\!d^2k\,\tilde{\rho}(\mathbf{k},t)k_x,
\end{equation}
where the integration of $k_x$ component is performed over the reciprocal space area $\tilde{F}_{1/3}$ defined as two opposite $\pi/3$ sectors within the $\tilde{F}$ area encompassing exactly one $K$ point and one $K'$ point (note that they have opposite $k_x$-components). Point $K$($K'$) in $\tilde{F}_{1/3}$ has coordinates $1$($-1$)$\times(4\pi/(3a),0)$, thus the valley index for one electron would be in the interval $\mathcal{K}_{1e}\in\left[-1,1\right]$, with $\mathcal{K}_{1e}=1$ representing the $K$ valley, whereas $\mathcal{K}_{1e}=-1$ the $K'$ valley. The total valley index spans the interval $\mathcal{K}\in\left[-2,2\right]$, with e.g $\mathcal{K}=2$ for $|K_\downarrow K_\downarrow\rangle$ or $|K_\uparrow K_\uparrow\rangle$ state. 

To get the expectation value of the valley isospin in each dot we have to collect the single-electron Fourier transforms (\ref{sfour}), but now integrated over the left or the right dot area, i.e. we calculate $\tilde{\boldsymbol{\psi}}^{L(R)}_i(\mathbf{k})$ by putting $D_{L(R)}$ in the Fourier integral in Eq. (\ref{sfour}). Then, by proceeding as before, we obtain the valley index in the left $\mathcal{K}_L(t)$, and the right dot $\mathcal{K}_R(t)$. Note that naturally $\mathcal{K}_L(R)\in[-1,1]$ due to fact that electrons are evenly distributed between both dots.
To define two qubits the electrons must occupy both dots simultaneously. In other case, if both electrons occupied the same dot, the opposite unoccupied one, would have its qubit undefined.

\section{Double quantum dot}\label{sec:dqd}
\begin{figure}[bt]
	\center
	\includegraphics[width=8.8cm]{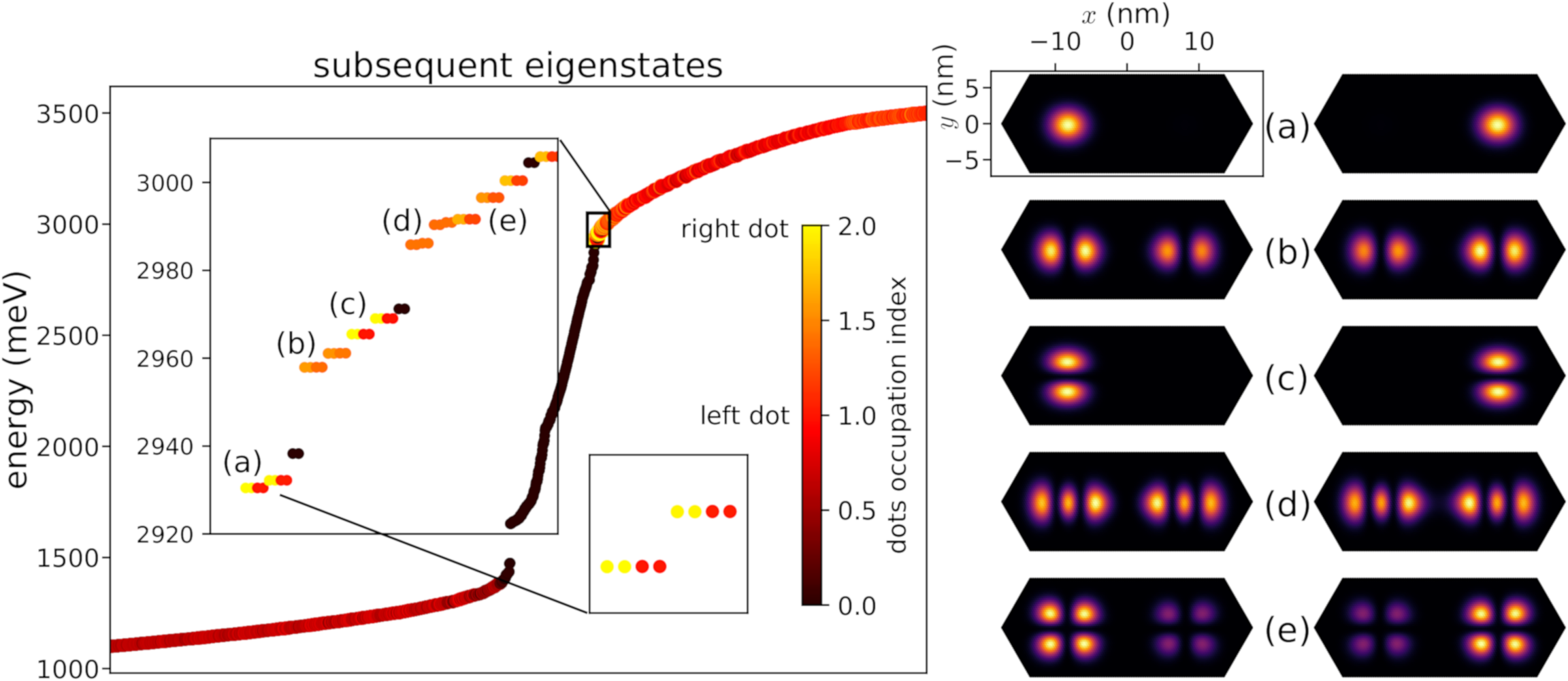}
	\caption{\label{fig:5} (top) Single-electron states represented by markings on the energy scale with color that express electron localization: outside the dots, near the flake edges forming the ``in-gap'' states (black markings), and localized in the left (red) or in the right dot (yellow). Insets collect subsets of states located in the vicinity of the CB edge. (bottom) Electron densities for states from groups (a-e) marked in the left inset.}
\end{figure}

Knowing the double-dot confinement potential we calculate a set of eigenstates of the Hamiltonian in Eq. (\ref{ham1e}) with the confinement potential energy $-|e|\phi$. These eigenstates will be further used to build two-electron basis states (of the Slater determinant form given in Eq. (\ref{slater})) for the considered two-qubit system. In Fig.~\ref{fig:5}, there are presented subsequent single-electron eigenstates for the double-dot potential from Fig.~\ref{fig:4} albeit with a bit higher barrier between the dots: $V_B = -1700$~mV, while other voltages remain the same as in Fig.~\ref{fig:4}.
We assigned colors to the dot occupancy parameter, defined as
\begin{equation}
\mathcal{N}=\int_{D_L}d^2r|\boldsymbol{\psi}_i(\mathbf{r})|^2+2\int_{D_R}d^2r|\boldsymbol{\psi}_i(\mathbf{r})|^2 \in [0,2],
\end{equation}
with $\boldsymbol{\psi}_i$ being the $i$-th eigenstate, and $D_L$ indicating the area over the left dot (but without including the edge), while $D_R$ --over the right one. Red markings mean that the electron is located in the left dot ($\mathcal{N}\simeq 1$), yellow -- in the right dot ($\mathcal{N}\simeq 2$), orange -- the electron is evenly spread between both dots ($\mathcal{N}\simeq 1.5$), and finally, black means that it is outside the dots ($\mathcal{N}\simeq 0$). The latter means that the electron is in a state located at the edge of the flake, forming the so-called edge state. These edge states marked as black in Fig. \ref{fig:5}) are forbidden for the electron confined within the dot, thus creating a band gap visible in Fig.~\ref{fig:5}(left)---in the case of an infinite flake these states would disappear. It is also known that those states are not present in torus geometry \cite{MB2, MB3}, equivalent to periodic boundary conditions, and their presence in open boundary conditions does not affect confined states in gate-defined regions in any way. We chose to keep open boundary conditions here due to our eight gate geometry that ensures decoupling of qubits from the edge.

In the inset of Fig. \ref{fig:5}, we present several lowest states derived from the CB minimum at $\pm K$ points. Some of them form characteristic yellow-red arrangements of four states (e.g. group (a) and (c)), meaning that for the electron located in the left dot (red markings) we have two spin-orbit-split doublets (spin-valley subspace), and the same for the right dot (yellow markings). This gives eight states in total (see the small inset in Fig.~\ref{fig:5}(left)). 

\begin{figure}[t]
	\center
	\includegraphics[width=7.8cm]{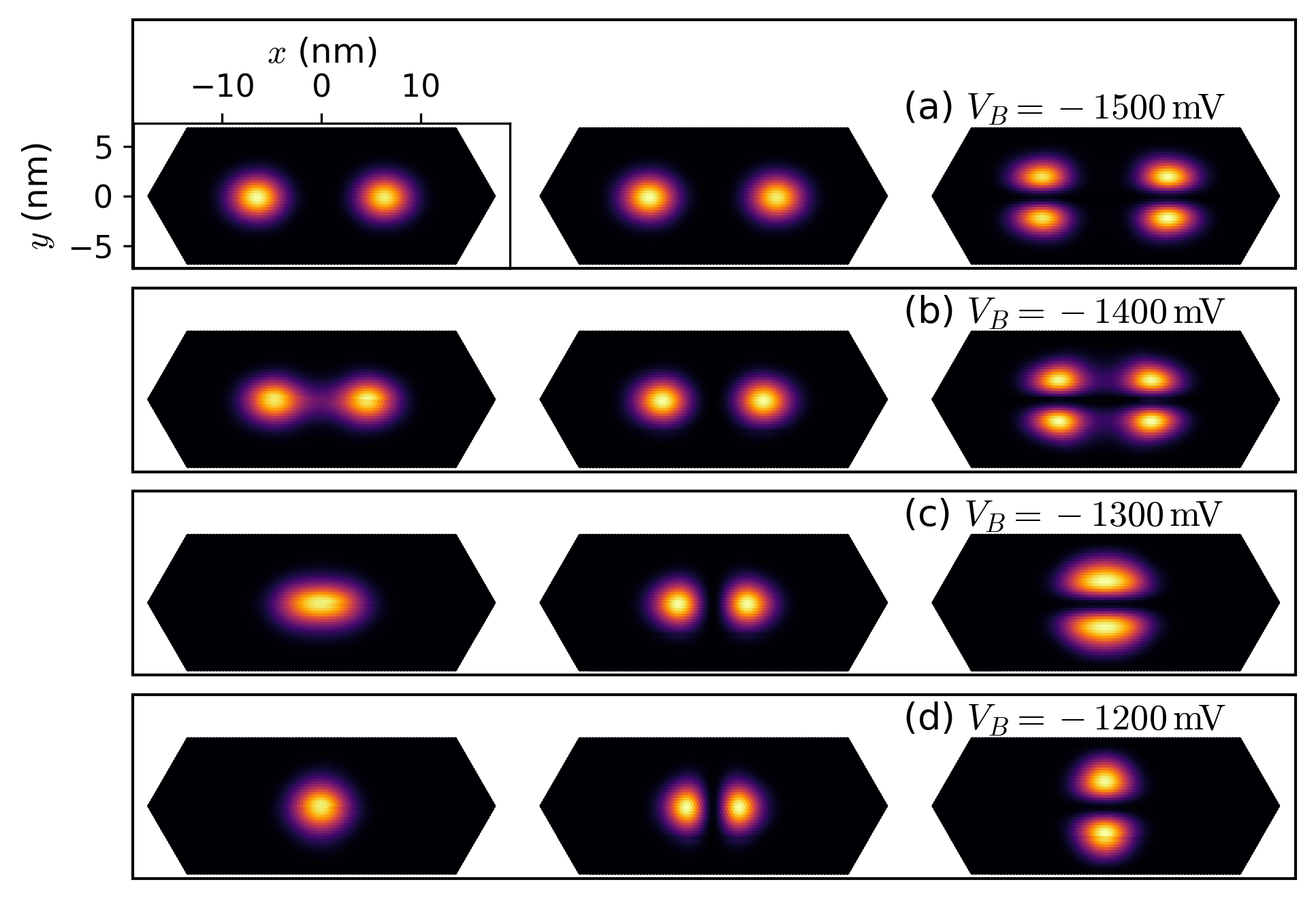}
	\caption{\label{fig:6} Single-electron eigenstates for lower interdot barriers: (a) $V_B=-1500$~mV,  (b) $-1400$, (c) $-1300$, and $(d) -1200$. Lowering the barrier between the dots makes the eigenstates spatially (anti)symmetric.}
\end{figure}

On the other hand, some states are formed with symmetric densities and are marked in orange in Fig. \ref{fig:5}, e.g. groups b) and d). Electron densities of states from several groups, marked by (a-e) in the left inset, are presented on the right side of Fig.~\ref{fig:5}.
What is noteworthy is that bringing the dots closer to each other by modulating $V_B$, and thus lowering the interdot barrier, symmetrizes the eigenstates. They become spatially symmetric or antisymmetric, with the latter moving up on the energy scale. Electron densities for dots brought closer to each other are presented in Fig.~\ref{fig:6}.
Left and center columns contain the lowest symmetric and antisymmetric states, respectively.

We now utilize the single-electron basis found for $V_B=-1300$~mV. The value of $V_B$ was chosen so as to get spatially (anti)symmetric states. This is motivated by the fact that we want to study and control interactions between two electrons in close proximity. To do so, we must set up a basis of antisymmetric Slater determinants given by Eq.~(\ref{ci}).
We take first 80 states from the CB minimum. In so defined a basis, we construct a matrix representing the two-electron Hamiltonian from Eq. (\ref{h2e}) with the Coulomb interaction via the configuration-interaction, filled by single-electron eigenenergies $\mathcal{E}_i+\mathcal{E}_j=\mathcal{E}_{ij}$ and Coulomb two-electron matrix elements $V_{ijkl}$ as in Eq. (\ref{vijkl}). We solve the eigenproblem by the exact diagonalization of the constructed $H_{2e}$ matrix. The key parameter that controls coupling between electrons is the barrier height controlled by $V_B$ voltage. 
\begin{figure}[b]
	\center
	\includegraphics[width=9cm]{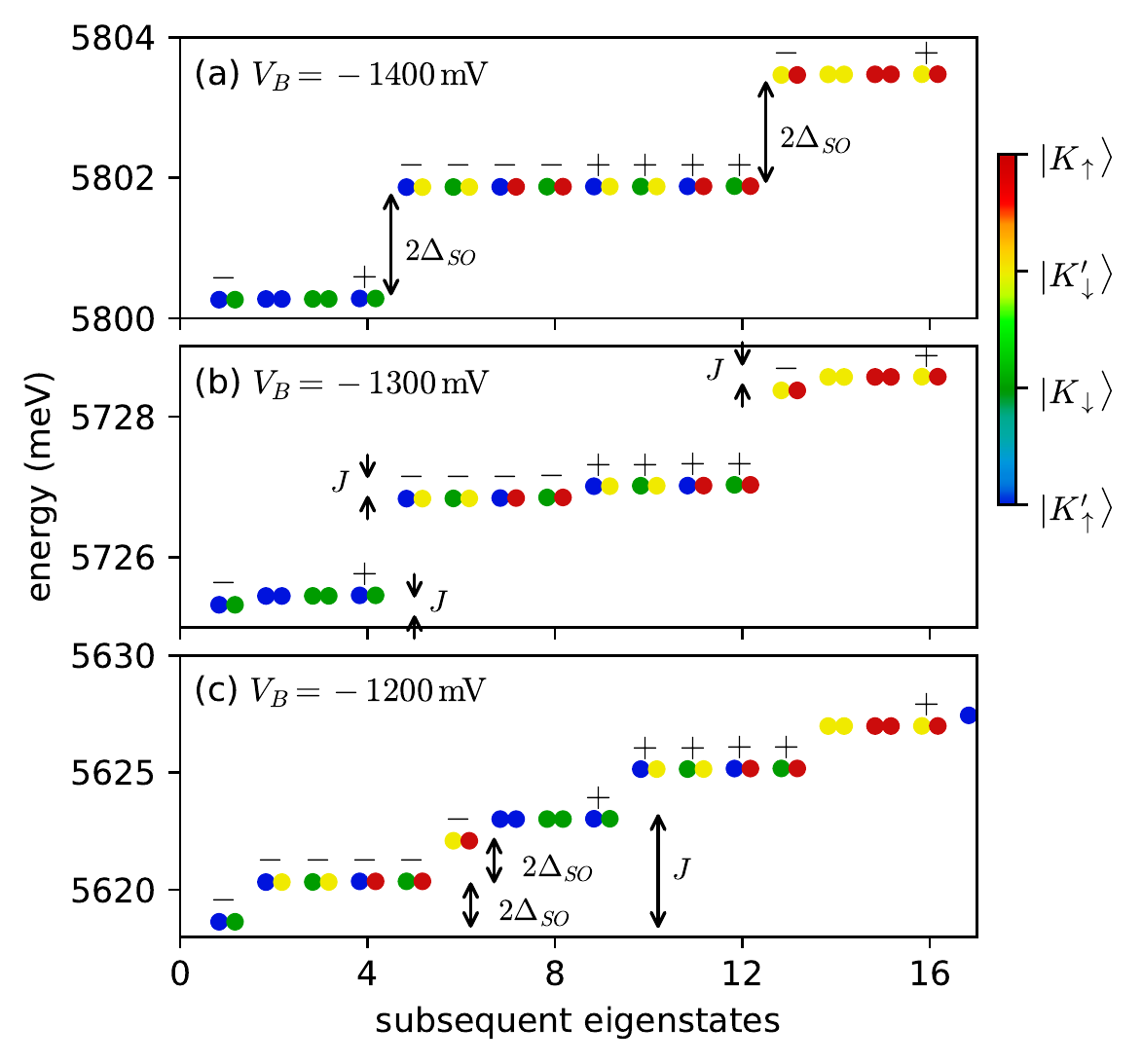}
	\caption{\label{fig:7} Two-electron states for different coupling between the dots, controlled via $V_B=V_3=V_7$ voltages, manifesting three regimes of the interplay between the exchange energy and spin-orbit splitting: a) $J\ll2\Delta_{SO}$, b) $J<2\Delta_{SO}$, c) $J>2\Delta_{SO}$. These 16 lowest states span the ground two-electron spin-valley degeneracy subspace. We identify them as built up mainly from the first four single-electron CB states: $|K'_\uparrow\rangle, |K_\downarrow\rangle, |K'_\downarrow\rangle, |K_\uparrow\rangle$, colored in blue, green, yellow and red, respectively. The $-$ or $+$ signs above the $\bullet\bullet$ symbol, that represents an eigenstate, denotes the singlet or the $T_0$-triplet combinations.}
\end{figure}

In Fig.~\ref{fig:7} there are presented first sixteen two-electron eigenstates spanning a spin- and valley-degenerated subspace \cite{rohling}, with the same (in the limit of decoupled dots) spatial state, being the ground state (with closer dots this state becomes a spatial symmetric-antisymmetric pair).
They are built up mainly (but not exactly, because in our exact calculations we employ the full Coulomb interaction) from the first four single-electron states from the CB edge (spin-valley degeneracy): $\{|K'_\uparrow\rangle, |K_\downarrow\rangle, |K'_\downarrow\rangle, |K_\uparrow\rangle \}$, split by the spin-orbit coupling energy $\Delta_{SO}$.
We identify them by calculating the total spin (see Appendix~\ref{apx:d}), and the total valley index $\mathcal{K}$, defined in Eq. (\ref{totk}).

The first four states form the singlet-triplet base: $\{|K'_\uparrow K_\downarrow\rangle\! -\!|K_\downarrow K'_\uparrow\rangle, |K'_\uparrow K'_\uparrow\rangle, |K'_\uparrow K_\downarrow\rangle\!+\!|K_\downarrow K'_\uparrow\rangle, |K_\downarrow K_\downarrow\rangle\}$ split by the spin-valley exchange energy $J$. Singlets ale denoted in Fig.~\ref{fig:7} by "$-$" sign above a pair of dots representing the given state, while $T_0$-triplets by "$+$".
The next eight states, four singlets and four $T_0$'s, are separated by the doubled single-electron spin-orbit splitting value ($2\Delta_{SO}$).
The next four states also form a singlet-triplet set, but made of the two upper states: $\{|K'_\downarrow\rangle, |K_\uparrow\rangle\}$.
By lowering $|V_B|$, from $V_B=-1400$~mV in Fig.~\ref{fig:7}(a) to $V_B=-1200$ in (c), we observe that states are gradually reorganized on the energy scale.
It is clear that $V_B$ controls the interdot exchange: for $V_B=-1400$~mV, $J\ll2\Delta_{SO}$, while for $V_B=-1200$ the exchange dominates: $J>2\Delta_{SO}$. 

\begin{figure}[t]
	\center
	\includegraphics[width=8.5cm]{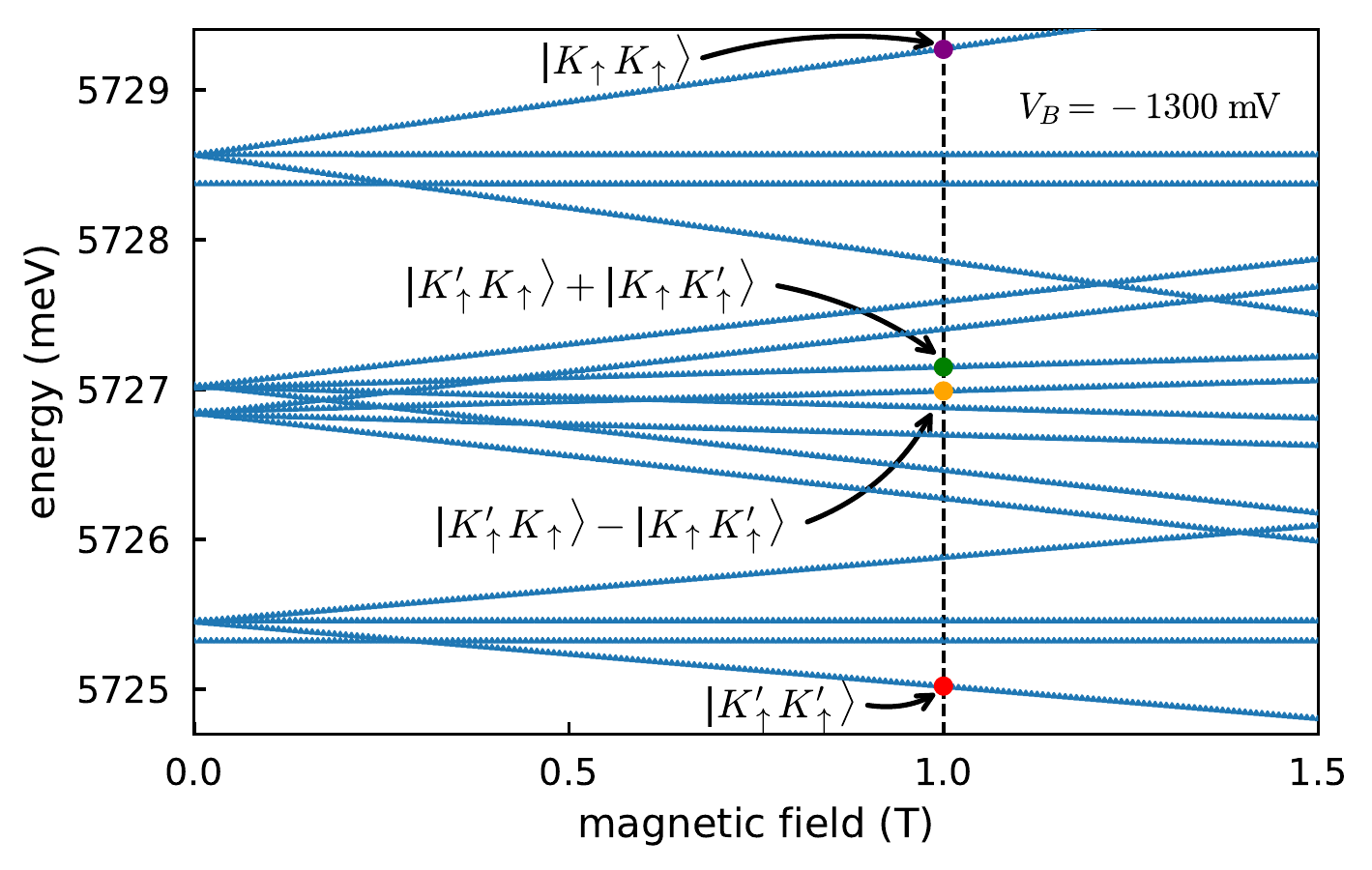}
	\caption{\label{fig:8} Evolution of the two-electron states from Fig.~\ref{fig:7}(b), i.e. for $V_B=-1300$, in an external magnetic field. Red, orange, green, and violet dots represent states that span the two-valley-qubit subspace (with the spin-up) for the applied perpendicular magnetic field $B_z=1$~T.}
\end{figure}

When analyzing two-electron states it is also sensible to examine the dependence of eigenenergies on the applied external magnetic field \cite{szafran}. To calculate the energy spectrum in such a case we add to the Hamiltonian in Eq. (\ref{ham1e}) the standard Zeeman term:
\begin{equation}
H_Z=\sum_{i\,\sigma \sigma'\alpha \beta }\gamma_Z\, \mathbf{B}\cdot\mathbf{s}_{\sigma\sigma'}\,\delta_{\alpha \beta}
\,{\hat{c}}^{\dagger }_{i\alpha \sigma }{\hat{c}}_{i\alpha \sigma '},
\end{equation}
with the magnetic field $\mathbf{B}$.
For $\gamma_Z=g_e\mu_B/2$
we arrive at the Zeeman energy $g\mu_B \mathbf{s} \cdot \mathbf{B}/2$. 
To also address the spatial effects related to the magnetic field we apply the so-called Peierls substitution \cite{hofst}. We multiply the hopping matrix, by the additional factor $t_{ij}\rightarrow \tilde{t}_{ij}= t_{ij}\exp\left( \imath\theta_B\right)$ in the Hamiltonian in Eq. (\ref{ham1e}). Now the vector potential $\mathbf{A}$ (we use the Landau gauge, $\mathbf{A}=[0,B_zx,0]^T$ for the perpendicular magnetic field $\mathbf{B} = [0,0,B_z]^T$) enters Eq.~(\ref{ham1e}) via the Peierls phase $\theta_B$, calculated as the path integral between neighboring nodes: 
$\theta_B=e/\hbar \int\mathbf{A}\cdot d\mathbf{r}$.

The most noticeable result of applying a magnetic field is the splitting, introduced between levels with opposite total spin and/or valley index. Note that in the first (1-4) and the last (13-16) four states (numbering as in Fig.~\ref{fig:7}(a)), singlet $S$ and triplet $T_0$ have opposite spin and valley index, meaning that their energy is almost constant in the magnetic field. Two triplet $T_+$ and $T_-$ states are split in B: the upper pair (14, 15) stronger than the lower one (2, 3). In the middle-eight states we have $S$ and $T_0$ composed of pairs with opposite spins or opposite valleys, which manifests in different $B$-field dependence, with $g$-factors that are different for spin and valley indices, i.e. in MoS$_2$ $g_v>g_s$ \cite{mynjp}.

\section{Valley manipulation and exchange}
\subsection{Single-qubit operations}\label{sec:sqop}
So far we have described the calculation method to determine system wavefunction evolution under applied external electrostatic potential via local gating, as well as to determine the dressed eigenspectrum of the two-electron system. Moreover, we described a method for calculating valley isospin in each dot separately (left or right). Now we are ready to define a two-qubit system based on the valley degree of freedom in each dot: left and right. 
To disentangle the valley degree of both electrons from their spins we apply an external magnetic field, arriving at a work subspace of, let us say, spin-up states: $\{|K'_\uparrow\rangle, |K_\uparrow\rangle \}$ for each electron. This magnetic field also enables us to set the specific frequency of valley transitions within the given spin subspace \cite{mynjp}. 
This way, the lowest (red dot) state in Fig.~\ref{fig:8}, $|K'_\uparrow K'_\uparrow\rangle$ represents the $|00\rangle$ two-qubit state, with $\mathcal{K}_L=\mathcal{K}_R=-1$, while the linear combination of the singlet and triplet (orange and green dots in Fig.~\ref{fig:8}), with $+$ or $-$ sign, gives $|01\rangle$ ($\mathcal{K}_L=-1$, $\mathcal{K}_R=1$) or $|10\rangle$ states ($\mathcal{K}_L=1$, $\mathcal{K}_R=-1$). The violet dot represents $|11\rangle$ state, with $\mathcal{K}_L=\mathcal{K}_R=1$.
Remind that we define qubits not on indistinguishable electrons, which are described by an antisymmetric wave function, but on the valley index of localized, spatially separated dots.
%{\color{red} What would be physical description of state after $\sqrt{SWAP}$ gate, e.g. $1/2[(|01>+|10>) + i(|01>-|10>)]$)?}

Thanks to the fact that qubits are localized in different dots we can easily manipulate them electrically, addressing their valley indices separately, by applying an oscillating voltage to the local gates in each dot separately \cite{mymos2}. Now we will show how it is done in our two-qubit device. We assume that initially the system is in the singlet state within the spin-up subspace, meaning that the valley index in each dot is zero and each qubit is in the equally-weighted superposition of $K$ and $K'$ states (orange or green dots in Fig.~\ref{fig:8}).

\begin{figure}[htb]
	\center
	\includegraphics[width=8.9cm]{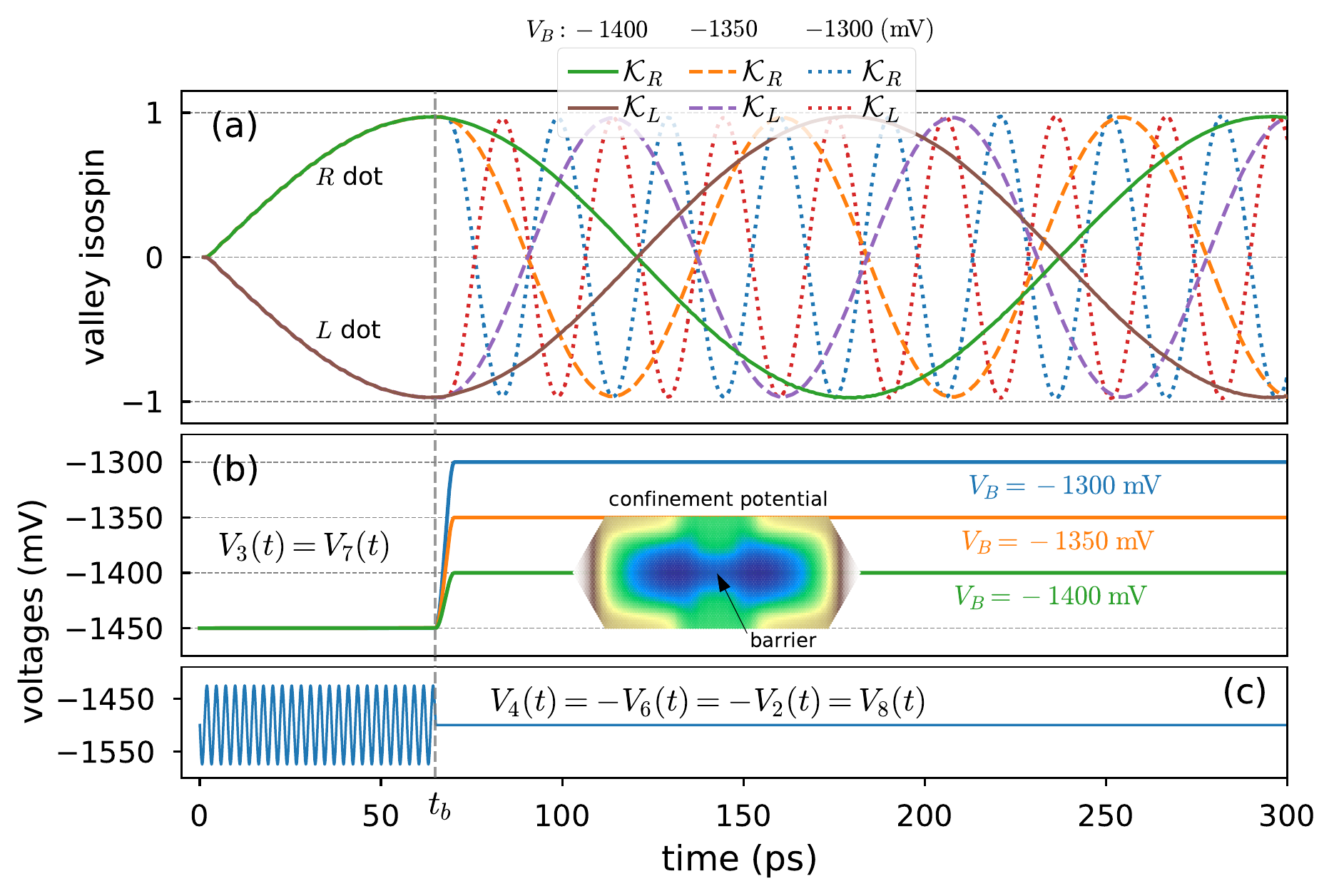}
	\caption{\label{fig:9} (a) Two-qubit system evolution. In the first step, at $0<t<t_b=64$~ps, modulated voltages at nearby gates (c) induce the left (right) valley-qubit rotations, depicted by the brown (green) curve, after which we arrive at opposite valley states: $\mathcal{K}_L=-1$ and $\mathcal{K}_R=1$. In the second step, for $t>t_b$, we reduce the interdot barrier by rising $V_B$ voltage, and observe the valley exchange between the dots,  which can be seen as the (two-qubit) valley-SWAP operation. (b)~Raising $V_B$ up to three different levels at $t_b=64$~ps reduces the interdot barrier, couples the dots, and induces the valley-exchange with three different exchange periods, as in (a).}
\end{figure}
We start the time evolution governed by Eq. (\ref{timeq}) and turn on oscillating voltages on $G_4$ and $G_6$ gates next to the left dot: $V_4(t)=-V_6(t)=V_{ac}\sin(\omega t)$, $V_{ac}=75$~mV. The pumping frequency $\omega=1/T$, where $T = 2.37$~ps (see Fig.~\ref{fig:9}(c)), is tuned to the energy splitting that equals spin-orbit splitting $2\Delta_{SO}=1.75$~meV for $B_z=0$. However, for $B_z=1$~T the energy splitting (energy difference between the red and green dots in Fig.~\ref{fig:8}) is a bit larger, and equals about $1.95$~meV, due to additional valley-Zeeman splitting \cite{mynjp} between these two valley states, which at the same time are the qubit basis states.
This way, by modulating the confinement potential in the left dot area, we induce a transition between the valley states of the left qubit, and thus rotate $|\mathcal{K}_L\rangle$ qubit \cite{mymos2}. 
Valley index calculated for the left dot $\mathcal{K}_L$ (i.e. defined in the same way as the total valley index in Eq. (\ref{totk}), but calculated for the Fourier transforms integrated over the left dot) is depicted in Fig.~\ref{fig:9}(a) by a brown curve. One can observe a gradual decrease from $0$ to $-1$, meaning that after time $t_b=64$~ps the left qubit is in $K'$ state. Similar situation happens in the right dot, where the potential is also modulated by an additional oscillating voltage applied to gates $G_2$ and $G_8$ (forming potential of the right dot) albeit in antiphase: $V_2(t)=-V_8(t)=-V_{ac}\sin(\omega t)$ (see Fig.~\ref{fig:9}(c)). The aforementioned modulation also results in an inter-valley transition in the right dot (green curve), and finally at $t_b$ the right qubit is in the $K$ state with the valley index equal $1$: $|\mathcal{K}_R=1\rangle$. The voltages in the dots oscillate in antiphase with the purpose to obtain antiparallel qubits at the end of this evolution step. 

\subsection{Two-qubit operation}
During the previous step (for $0<t<t_b$), the barrier between the dots was set to be high ($V_3(t)=V_7(t)=-1450$~mV), ensuring that no valley isospin exchange between left and right dots occurs. We have shown how to perform single qubit operations on individual qubits through manipulating voltages applied to nearby gates. To fulfill the universality criterion \cite{univer} we also need to implement any two-qubit operation. The simplest one, that naturally emerges in a system of two isospins, is their exchange, or SWAP using the quantum information language. It is known that the $\sqrt{\mbox{SWAP}}$ gate, which performs half of a two-qubit swap, is universal in a sense that any multi-qubit gate can be constructed from only $\sqrt{\mbox{SWAP}}$ and single-qubit gates \cite{swap}.

Now at time $t_b=64$~ps we lower the interdot barrier by raising (negative) voltage $V_B$, as shown in Fig.~\ref{fig:9}(b), and thus begin swapping the valley isospins between the dots. The modulation of voltages on gates $G_2$, $G_8$, and $G_4$, $G_6$ is now turned off. In Fig.~\ref{fig:9}(a) we observe that valley indices between dots $\mathcal{K}_L$ and $\mathcal{K}_R$ exchange their values. What is characteristic, this process is faster when the barrier, controlled by $V_B$, is lower. In other words, barrier height controls the coupling between the dots. We have performed simulations for three different barrier heights. For $V_B=-1400$~mV (brown and green curves) the swap completes in $T_\mathrm{SWAP}=230$~ps, for $-1350$~mV (violet and orange) in $94$~ps, while for $-1300$~mV (red and blue) it takes $30$~ps. The exchange time translates to singlet-triplet energy difference $J=h/T_\mathrm{SWAP}$ which agrees with the exchange energies $20$, $45$, and $125$~$\mu\mathrm{eV}$, respectively, obtained from the two-electron eigenspectrum presented in Fig.~\ref{fig:7}. Thus, the $\sqrt{\mbox{SWAP}}$ gate timings, which last half of the $T_\mathrm{SWAP}$, are relatively short---we can perform a full operation cycle within less than $100$~ps. This time can be tuned precisely by adjusting the interdot barrier height via $V_B=V_3=V_7$ voltages.

We have assumed the singlet state as the starting state of our simulation, but this is not the only option. With magnetic field $B_z=1$~T the lowest state is the polarized triplet $|K'_\uparrow K'_\uparrow\rangle$ (see Fig.~\ref{fig:8}). We could also assume this state as the starting one. Then, to observe the exchange, in the first stage of operation we have to rotate the valley in only one dot. As a result, we would also get two opposite valleys, and then be able to observe their exchange.

\section{Single-shot readout via Pauli blockade}
To get a complete physical implementation of the quantum computer, apart from single-qubit operations and swapping, we need the ability to initialize the state of the qubits as well as a qubit-specific measurement capability. Among numerous spin initialization and readout methods in gated quantum dots (QDs), the most common approach is to employ the Pauli spin-blockade mechanism \cite{pauli1, pauli2}.

In the following paragraphs, we will show that the Pauli blockade effect can be extended in our setting to the valley degree of freedom, similarly as it is done for carbon nanotubes \cite{nt1,nt2}. In Pauli blockade, a double quantum dot containing two electrons in total is tuned to the transition between two charge-states: (1,1) with one electron in each dot and (0,2) with both electrons in the right dot. 
This transition involves the electron tunnelling from the left to the right dot. If we properly tune the voltage bias between the dots, we set a blockade for the (1,1)-valley-triplet state, for which transition to an energetically accessible (0,2)-valley-singlet state is forbidden. It is crucial that (0,2)-valley-triplet state which would not block is sufficiently separated in energy scale by (0,2)-singlet-triplet energy difference which in our case is about $5$~meV. On the other hand, (1,1)-valley-singlet state is in this regime unblocked.

\begin{figure}[t]
	\center
	\includegraphics[width=8cm]{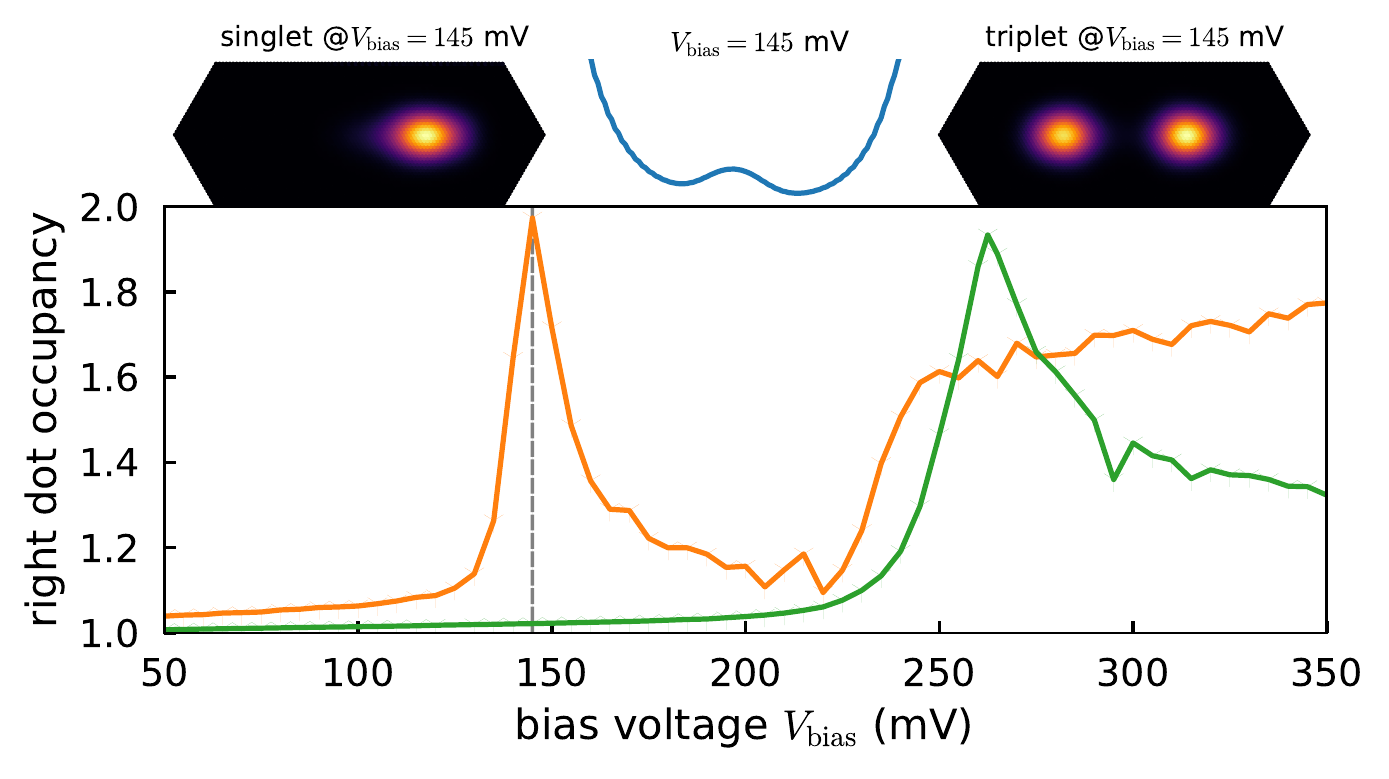}
	\caption{\label{fig:10} Valley-Pauli-blockade. The right dot occupancy at various biasing voltages, calculated for the singlet (orange) and the triplet (green) initial state, shows resonance (that means increased transition between the dots) at different bias. This way, proper biasing (i.e. $V_\mathrm{bias}=145$~mV) enables electron-transfer in the singlet state while blocks in the triplet state.}
\end{figure}
Cases discussed above are presented in Fig.~\ref{fig:10}. To observe the blockade we add a positive offset voltage $V_\mathrm{bias}>0$ between the dots, by raising $V_1$ voltage: $\tilde{V}_1(t)=V_1+V_\mathrm{bias}$. Then we stimulate an electron-transition from the left dot to the right, lower one (see the blue potential profile in Fig.~\ref{fig:10}), by applying an additional oscillating voltage to gate $G_5$: $\tilde{V}_5(t)=V_5+V_\mathrm{stim}\sin(\omega_\mathrm{stim}t)$, where $V_\mathrm{stim}=5$~mV, and $2\pi/\omega_\mathrm{stim}=5$~ps.
One should know, that when simulating the blockade, we took slightly different single-electron eigenstates, obtained for $V_1=-1550$~mV and $V_B=-1300$~mV, while other voltages remained the same as in Fig.~\ref{fig:4}. This step aimed to add states with asymmetric densities (located majorly in the right dot) to the basis.
%Moreover, presented results were calculated for zero $B_z$ magnetic field.

After turning on $V_\mathrm{stim}\sin(\omega_\mathrm{stim}t)$, we gradually increased $V_\mathrm{bias}$ and calculated time-evolutions for each of the considered $V_\mathrm{bias}$ values, and for two different initial two-electron states: singlet $|K'_\uparrow K_\downarrow\rangle-|K_\downarrow K'_\uparrow\rangle$, and triplet $|K'_\uparrow K'_\uparrow\rangle$. After a few dozen picoseconds, we observe that for the singlet initial state and $V_\mathrm{bias}=145$~mV the ``left electron'' transfers completely to the right dot (non-blocked state), while for the triplet state the occupancy of both dots is almost unchanged (blocked state). The resulting electron densities for the singlet (triplet) state are presented at the top left (right) of Fig.~\ref{fig:10}. One should note, that the occupancy of the right dot, calculated as the total electron density $\rho(\mathbf{r})$ (defined in Appendix~\ref{apx:d}) integrated over the right dot, does not remain constant after the transfer. In fact, the density oscillates over time as the electron goes back and forth between the dots, and the occupancy plot, presented in Fig.~\ref{fig:10}, shows their maximum values over time.
It is now clear that for the singlet state (orange curve) and for the triplet state (green curve) the resonant transitions occurs at different biasing: $V^S_\mathrm{bias}=145$~mV and $V^T_\mathrm{bias}=265$~mV, respectively.
Thus at $V_\mathrm{bias}=V^S_\mathrm{bias}$, we observe the Pauli valley-blockade for the electron density defining the left valley-qubit. The estimated blockade fidelity is about $97.5$\%.

Therefore, using the valley-blockade one can set up or read out the left valley-qubit state. However, we should keep in mind that the spin degree of freedom makes the (spin-valley) singlet-triplet subspace extend to 16 states (not just $1+3$). This complicates proper setting up the Pauli blockade. To resolve valley from spin we had to apply a magnetic field, then using proper transition frequencies, as in Fig.~\ref{fig:8}, of the single-valley-qubit operations ensures that we stay within the given spin (let's say up) subspace. But how to initialize or check that two qubits are in the desired valley state \textit{together} with the given spin? Magnetic field enriches blockade operation \cite{pauli3}. If we take a look at the Fig.~\ref{fig:8} we will see that for $B_z=1$~T the analyzed triplet $|K'_\uparrow K'_\uparrow\rangle$, and singlet $|K'_\uparrow K_\downarrow\rangle-|K_\downarrow K'_\uparrow\rangle$ are the two lowest states. By proper adjusting the left lead potential, i.e. to energy lower than the third state $|K'_\uparrow K_\downarrow\rangle+|K_\downarrow K'_\uparrow\rangle$ (unwanted triplet), we allow only these two (1,1) states to populate the double dot.
This way we ensure that only $|K_\uparrow K_\uparrow\rangle$ is blocked, thus implementing the valley-qubit initialization.

\section{Summary}
In the following work we have studied a two-electron system in TMDC's gate-defined double quantum dot from the point of view of valley-qubit implementation. Utilizing the configuration-interaction method, the realistic theory of Coulomb interaction, and the time-dependent Schr\"{o}dinger equation coupled with the Poisson equation, that models realistic dielectric environment, we were able to describe the proposed nanodevice with an eight-gate geometry and time-modulated electric potentials.

By performing numerical simulations, we have shown how one can obtain single- and two-qubit gates in the valley two-qubit system by electrically controlling the state of the electrons, and the interdot coupling, in a static magnetic field. First, we explained each qubit-rotations (single-qubit operations) in the left or the right dot, controlled separately by the local gates. Then, we coupled both qubits getting the valley-SWAP (two-qubit operation). Finally, we discussed how to set-up the valley-Pauli-blockade to implement the valley-qubit initialization and readout.
In this way, we have obtained a physical scheme  to realize universal quantum computation based on valley isospin in the gate-defined TMDCs double quantum dots.

In our theoretical description of the two-electron system, we used the exact configuration-interaction method that gives a basis of states dressed in an interaction between the electrons. The Coulomb interaction as well as the confinement potential are modelled realistically including atomic matrix elements, the screening effect by nearby dielectric layers, and voltages applied to the control gates layout with geometry inspired by experiments. Variable control voltages modulate the confinement potential leading to non-trivial device operation calculated using the time-dependent Schr\"{o}dinger equation, solved in the configuration-interaction basis self-consistently with the Poisson equation.

\section{Acknowledgements}
Authors would like to thank Dariusz \.{Z}ebrowski and Pawe\l{} Potasz for invaluable discussions. This work has  been supported by National Science Centre, under Grant No. 2016/20/S/ST3/00141. TW acknowledges financial support from Polish Ministry of Sience and Higher Education via Grant No. D\textbackslash 2015\textbackslash 002645. MB acknowledges financial support from National Science Center (NCN), Poland, Maestro Grant No. 2014/14/A/ST3/00654.
This research was supported in part by PL-Grid Infrastructure, Wroc\l{}aw Center for Networking and Supercomputing and Compute Canada.

\appendix
\section{Two-electron Hamiltonian}\label{apx:a}
Here we present ab explicit derivation of the formula for Hamiltonian~(\ref{h2e}):
\begin{align*}
&\sum_{i<j,k<l}\tilde{|ij\rangle}\tilde{\langle ij|}H_{2e}\tilde{|kl\rangle}\tilde{\langle kl|}=\\
&=\sum_{i<j,k<l}\tilde{|ij\rangle}\tilde{\langle kl|}\left[\mathcal{E}_{ij}(\delta_{ik}\delta_{jl}-\delta_{il}\delta_{jk}) + \tilde{\langle ij|}\bar{V}_C\tilde{|kl\rangle} \right]=
\end{align*}
\begin{align*}
&\begin{aligned}
=\sum_{i<j,k<l}d^\dagger_i d^\dagger_j\tilde{|0\rangle}&\tilde{\langle 0|}d_k d_l \Big[\mathcal{E}_{ij}\delta_{ik}\delta_{jl}+\\
&+\langle ij| \bar{V}_C|kl\rangle-\langle ij|\bar{V}_C|lk\rangle \Big]=\\
\end{aligned}\\
&\begin{aligned}
=\sum_{i<j} d^\dagger_i d^\dagger_j\tilde{|0\rangle}&\tilde{\langle 0|}d_i d_j\,\mathcal{E}_{ij}+\\ 
&+\sum_{i<j,k<l} d^\dagger_i d^\dagger_j\tilde{|0\rangle}\tilde{\langle 0|}d_k d_l \left(V_{ijkl}-V_{ijlk}\right).
\end{aligned}
\end{align*}

The two-electron energy is simply the sum of the single-electron-state energies: $\mathcal{E}_{ij}=\mathcal{E}_i+\mathcal{E}_j$.
Note that, firstly: $\delta_{il}\delta_{jk}$ vanishes since $i<j,\,k<l$. Secondly, we have $\tilde{\langle ij|}\bar{V}_C\tilde{|kl\rangle}= \langle ij| \bar{V}_C|kl\rangle-\langle ij|\bar{V}_C|lk\rangle$, so we introduce the abbreviation $\langle ij| \bar{V}_C|kl\rangle \equiv V_{ijkl}$. 

\section{Atomic Coulomb elements}\label{apx:b}
In this section we present numerical values for the atomic Coulomb-matrix elements. We use the following numbering of orbitals: $\alpha=1,2,3$ for $d_{z^2}$, $d_{xy}$, $d_{x^2-y^2}$ Mo orbitals respectively.
\begin{table}[b]
	\begin{tabular}{c|c|l|}
		\hline
		group \# &
		\multicolumn{2}{c|}{one-center integrals $\langle ss|V^0_C|ss\rangle$} \\ \hline
		1 & 14.43 & 
		$\langle s^{\alpha}s^{\alpha}|V^0_C|s^{\alpha}s^{\alpha}\rangle$, $\alpha=1,2,3$\\ 
		\hline
		2 & 13.63 & 
		$\langle s^{2}s^{3}|V^0_C|s^{2}s^{3}\rangle$, 
		$\langle s^{3}s^{2}|V^0_C|s^{3}s^{2}\rangle$\\ 
		\hline
		3 & 12.86 & 
		\begin{tabular}[c]{@{}l@{}}
			$\langle s^{1}s^{2}|V^0_C|s^{1}s^{2}\rangle$, 
			$\langle s^{2}s^{1}|V^0_C|s^{2}s^{1}\rangle$,\\
			$\langle s^{1}s^{3}|V^0_C|s^{1}s^{3}\rangle$, 
			$\langle s^{3}s^{1}|V^0_C|s^{3}s^{1}\rangle$
		\end{tabular}\\ 
		\hline
		7 & 0.79 & 
		\begin{tabular}[c]{@{}l@{}}
			$\langle s^{1}s^{2}|V^0_C|s^{2}s^{1}\rangle$, 
			$\langle s^{2}s^{1}|V^0_C|s^{1}s^{2}\rangle$,\\
			$\langle s^{1}s^{3}|V^0_C|s^{3}s^{1}\rangle$, 
			$\langle s^{3}s^{1}|V^0_C|s^{1}s^{3}\rangle$,\\
			$\langle s^{1}s^{1}|V^0_C|s^{2}s^{2}\rangle$, 
			$\langle s^{2}s^{2}|V^0_C|s^{1}s^{1}\rangle$,\\
			$\langle s^{1}s^{1}|V^0_C|s^{3}s^{3}\rangle$, 
			$\langle s^{3}s^{3}|V^0_C|s^{1}s^{1}\rangle$
		\end{tabular}\\ 
		\hline
		8 & 0.41 & 
		\begin{tabular}[c]{@{}l@{}}
			$\langle s^{2}s^{3}|V^0_C|s^{3}s^{2}\rangle$, 
			$\langle s^{3}s^{2}|V^0_C|s^{2}s^{3}\rangle$,\\
			$\langle s^{2}s^{2}|V^0_C|s^{3}s^{3}\rangle$, 
			$\langle s^{3}s^{3}|V^0_C|s^{2}s^{2}\rangle$
		\end{tabular}\\ 
		\hline
		&
		\multicolumn{2}{c|}{two-center integrals $\langle sp|V^0_C|sp\rangle$} \\ \hline
		4 & 4.71 & 
		\begin{tabular}[c]{@{}l@{}}
			$\langle s^{2}p^{2}|V^0_C|s^{2}p^{2}\rangle$, 
			$\langle s^{3}p^{3}|V^0_C|s^{3}p^{3}\rangle$,\\
			$\langle s^{2}p^{3}|V^0_C|s^{2}p^{3}\rangle$, 
			$\langle s^{3}p^{2}|V^0_C|s^{3}p^{2}\rangle$
		\end{tabular}\\ 
		\hline
		5 & 4.55 & 
		\begin{tabular}[c]{@{}l@{}}
			$\langle s^{1}p^{2}|V^0_C|s^{1}p^{2}\rangle$, 
			$\langle s^{2}p^{1}|V^0_C|s^{2}p^{1}\rangle$,\\
			$\langle s^{1}p^{3}|V^0_C|s^{1}p^{3}\rangle$, 
			$\langle s^{3}p^{1}|V^0_C|s^{3}p^{1}\rangle$
		\end{tabular}\\ 
		\hline
		6 & 4.42 & 
		$\langle s^{1}p^{1}|V^0_C|s^{1}p^{1}\rangle$\\ 
		\hline
	\end{tabular}
	\caption{\label{tab:1} The largest Coulomb atomic integrals with the listed largest energies from each group (eV). We include only elements with their energy value larger than $0.3$~eV, while the others are taken classically, as interaction between two point-charges. Note that each two-center integral occurs six times, one for each $R_{1..6}$ direction.}
\end{table}
\begin{figure}[t]
	\center
	\includegraphics[width=8.5cm]{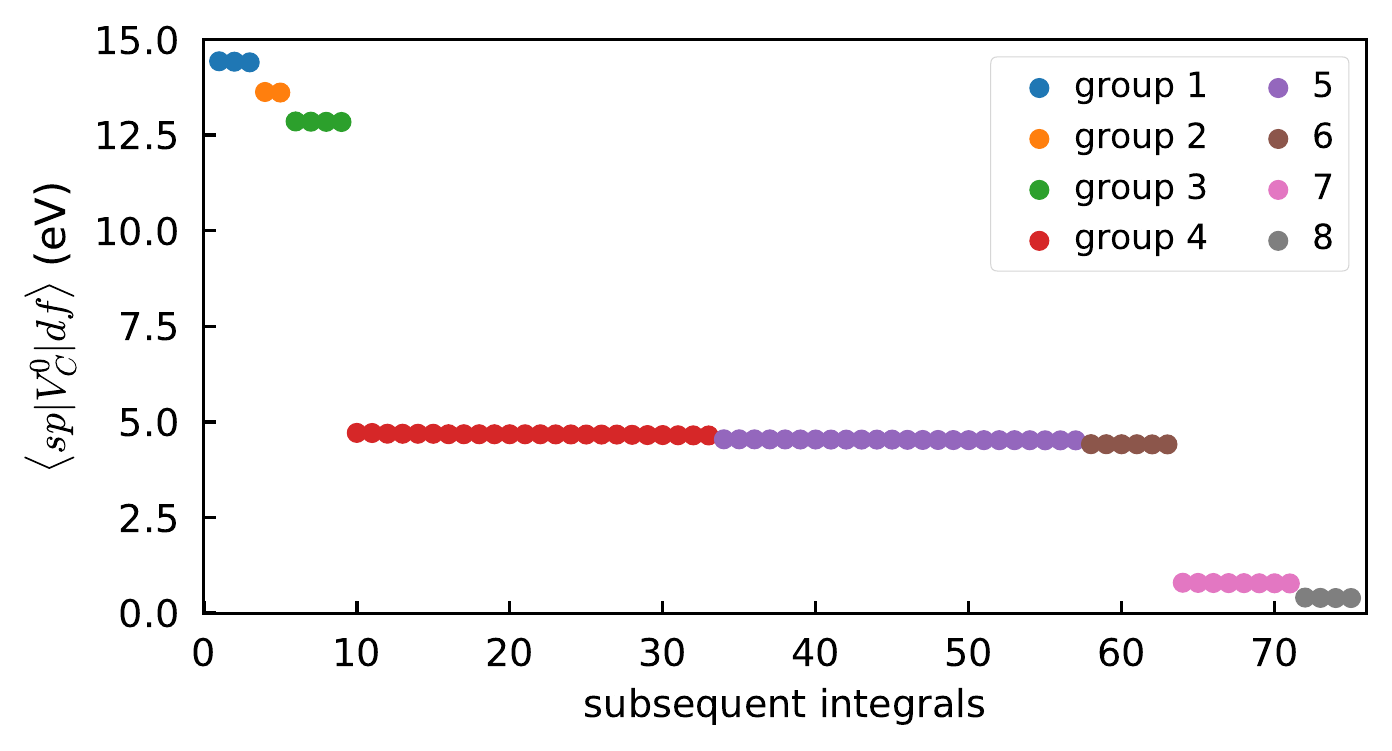}
	\caption{\label{fig:11} Interaction energy values for subsequent groups of integrals as listed in the Table~\ref{tab:1}. }
\end{figure}

When calculating one- ($\mathbf{r}_s=\mathbf{r}_p$) and two-center ($\mathbf{r}_s\neq\mathbf{r}_p$) integrals listed in the Table~\ref{tab:1},
the 3D Coulomb interaction in vacuum
$V^0_{C}(r)=\frac{|e|^2}{4\pi\varepsilon_0 r}$
was taken. Then, the obtained values were renormalized by the $\mathrm{rn}(|\mathbf{r}_s-\mathbf{r}_p|)$ function defined in Appendix~\ref{apx:c}, and this way we estimate $\mathcal{V}_{spdf}$ values, defined in Eq. (\ref{vspdf}), for the \textit{screened} Coulomb interaction:
\begin{equation}
\langle s^{\alpha_s}p^{\alpha_p}|\bar{V}_C| s^{\alpha_d}p^{\alpha_f}\rangle\simeq\frac{\langle s^{\alpha_s}p^{\alpha_p}|V^0_C| s^{\alpha_d}p^{\alpha_f}\rangle}{\kappa\,\mathrm{rn}(|\mathbf{r}_s-\mathbf{r}_p|)}.
\end{equation}
This way we include the screening effects of the dielectric environment (nearby insulators) and the monolayer flake. 
The calculated integrals, shown in Fig.~\ref{fig:11} and listed in Table~\ref{tab:1}, do not take into account screening, and can be applied to other structures with, for example, different insulating layers.

\section{Coulomb effective interaction}\label{apx:c}
Let us now introduce the screened Coulomb potential used in our model. We checked three different approaches to describe the screened Coulomb interaction in 2D structures, and verified them with an exact numerical potential calculated for the our structure.

In the simplest, naive, approach at the flake level we take the $\varepsilon$ as the average from two neighboring materials: $\kappa=(\varepsilon_\mathrm{hBN}+\varepsilon_{\mathrm{SiO}_2})/2=4.5$ ($\varepsilon_0$). \cite{eff2}
We assume the following dielectric constants: $\varepsilon_\mathrm{hBN}=5.1$ (for the top-layer of hexagonal boron nitride) \cite{eff1}, and $\varepsilon_{\mathrm{SiO}_2}=3.9$ (for the bottom-layer of quartz). However, the following potential derived by Keldysh \cite{eff3,eff4} to model the Coulomb interaction in thin semiconductor layer embedded between top and bottom layers with given permittivity, is much more accurate:
\begin{equation}
V_K(r)=\frac{|e|^2}{4\pi\varepsilon_0}\frac{\pi}{2r_0}\left[H_0\left(\frac{\kappa r}{r_0}\right)-Y_0\left(\frac{\kappa r}{r_0}\right)\right],
\end{equation}
with the zero-order Struve and the second-kind Bessel functions. Another potential $V_\mathrm{TYD}$, by Tuan, Yang and Dery \cite{eff5}, was  introduced for better modeling of the Coulomb interaction in TMDC monolayers. 
In addition to information about the permeability of adjacent dielectric layers, it takes into account the values of polarizabilities $\chi_+$ of the central Mo atomic sheet, and $\chi_-$ for the top and bottom S (chalcogen) sheets. We took the following parameters for the Keldysh and TYD potentials: $\kappa=(\varepsilon_t+\varepsilon_b)/2=(\varepsilon_\mathrm{hBN}+\varepsilon_{\mathrm{SiO}_2})/2=4.5$~$(\varepsilon_0)$, $r_0=7.5d=4.875$~nm, $l_{+}=2\pi\chi_+=5.6d=3.64$~nm, $l_{-}=2\pi\chi_-= 5d=3.25$~nm ($d=0.65$~nm for MoS2).
See [\onlinecite{eff5}] for a detailed analysis of fitting parameters to these models.

We compare all of these model potentials with the standard 3D Coulomb potential $V_C(r)=V^0_C(r)/\kappa=\frac{|e|^2}{4\pi\varepsilon_0}\frac{1}{\kappa r}$,
and potential $V_N(r)$ calculated numerically via the Poisson equation for the space-dependent permittivity:
\begin{equation}
\kappa(z)=\begin{cases}
\varepsilon_\mathrm{hBN} & \mathrm{for}\quad z > d/2,\\
1 & \mathrm{for}\quad -d/2 < z < d/2,\\
\varepsilon_{\mathrm{SiO}_2} & \mathrm{for}\quad z < -d/2,
\end{cases}
\end{equation}
and average 2D polarizability for the MoS$_2$ monolayer taken as $\chi=0.55$~nm, which screens electron charge in the monolayer itself, weakening it by the factor $\varepsilon=1+\frac{4\pi\chi}{d}\simeq 12$ \cite{eff6,eff7}.

\begin{figure}[b]
	\center
	\includegraphics[width=8cm]{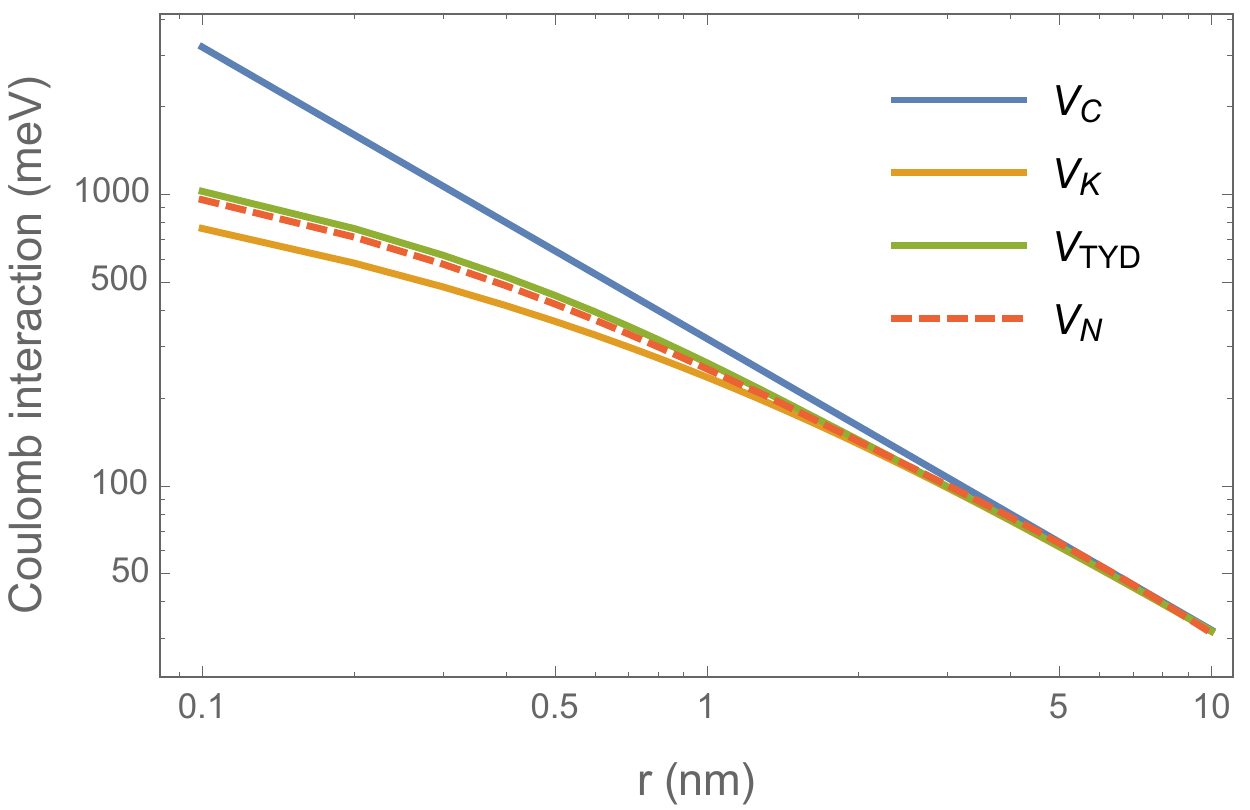}
	\caption{\label{fig:12} The screened Coulomb interaction captured by several model potentials, and compared with the exact numerical result $V_N(r)$. Comparison between the Coulomb $V_{C}(r)$ and best $V_\mathrm{TYD}(r)$ potential gives a correction for screening, described by the $\mathrm{rn}(r)$ function.}
\end{figure}

Comparison of all the potentials is presented in Fig.~\ref{fig:12}.
We can see that Keldysh, TYD, and numerical potentials give qualitatively comparable results, while the bare Coulomb potential overestimates interaction especially at small distances ($r<1$~nm).  
The closest to numerical are results for TYD, and we take exactly that potential to describe the electron-electron interaction in our model: 
\begin{equation}
\bar{V}_C(r)=V_\mathrm{TYD}(r)\simeq\frac{V^0_C(r)}{\kappa\,\mathrm{rn}(r)}.
\end{equation}

Now we are able to define function that renormalizes the bare Coulomb potential to describe realistic screening:
$\mathrm{rn}(r)=V_C(r)/V_\mathrm{TYD}(r)$.
The exact analytical formula for the TYD potential is rather complicated, thus, for our convenience, we approximated $\mathrm{rn}(r)$ using the following formula: $\mathrm{rn}(r)\simeq\frac{r+0.2}{r-0.01}$, $r>0.01$ ($r$ is in nm). This means that, e.g. for $r=0.319$~nm we have to divide the energies obtained from the Monte-Carlo (listed in Table~\ref{tab:1}) by $1.68$ (times $\kappa$).
On the other hand, for interaction at the same node, for one-center integrals, we would divide by about 3 (assuming the average radius of the Mo atom). However, if we take into account the fact that for atomic length scales the screening naturally decreases (in contrary to a simple model of dielectric slabs), 
then for $r\sim d$ the real epsilon is (roughly speaking) about twice as small \cite{eff6}. Hence, for interaction on the node itself, we assumed $1.5$ instead of $3$: $\mathrm{rn}(r\simeq 0)=1.5$.

\section{Two-electron density and total spin}\label{apx:d}
Let us remind that the full two-electron wavefunction is expanded in the configuration-interaction-basis as
\begin{equation}
\boldsymbol{\Psi}(\mathbf{r}_1,\mathbf{r}_2)=\sum_{i<j}d_{ij}\, 
\boldsymbol{\Psi}_{ij}(\mathbf{r}_1,\mathbf{r}_2).
\end{equation}
Thus, expansion for the total electron density reads:
\begin{align*}
\begin{aligned}
&\rho(\mathbf{r})=\\
&=\int\! d\mathbf{r}'|\boldsymbol{\Psi}(\mathbf{r},\mathbf{r}')|^2+ |\boldsymbol{\Psi}(\mathbf{r}'\!,\mathbf{r})|^2=2\!\int\! d\mathbf{r}'|\boldsymbol{\Psi}(\mathbf{r},\mathbf{r}')|^2=\\
&=2\!\sum_{i<j,k<l} d^\ast_{ij}d_{kl} \int\!d\mathbf{r}'\boldsymbol{\Psi}^\dag_{ij}(\mathbf{r},\mathbf{r}') 
\boldsymbol{\Psi}_{kl}(\mathbf{r},\mathbf{r}')=\\
&=\!\sum_{i<j,k<l} d^\ast_{ij}d_{kl}
\left\{\boldsymbol{\psi}_i^\dag(\mathbf{r})\boldsymbol{\psi}_k(\mathbf{r})\delta_{jl}-\boldsymbol{\psi}_i^\dag(\mathbf{r})\boldsymbol{\psi}_l(\mathbf{r})\delta_{jk}\right.\\
&\qquad\qquad\left.-\boldsymbol{\psi}_j^\dag(\mathbf{r})\boldsymbol{\psi}_k(\mathbf{r})\delta_{il}
+\boldsymbol{\psi}_j^\dag(\mathbf{r})\boldsymbol{\psi}_l(\mathbf{r})\delta_{ik}\right\},
\end{aligned}
\end{align*}
where for clarity we omitted the $\otimes$ symbol. Due to the antisymmetry constraint $\Psi_{ij}^{\sigma_i\alpha_i\sigma_j\alpha_j}(\mathbf{r}'\!,\mathbf{r})= -\Psi_{ij}^{\sigma_j\alpha_j\sigma_i\alpha_i}(\mathbf{r},\mathbf{r}')$ we get
$|\boldsymbol{\Psi}(\mathbf{r}'\!,\mathbf{r})|^2= |\boldsymbol{\Psi}(\mathbf{r},\mathbf{r}')|^2$. Note that obviously one has
$\int\!d\mathbf{r}\,\rho(\mathbf{r})=2$.
The total spin is calculated in the same way:
\begin{equation}
\begin{split}
&\langle\sigma_z\rangle=\langle\boldsymbol{\Psi}|\boldsymbol{\sigma}_z\otimes\mathbf{1}+ \mathbf{1}\otimes\boldsymbol{\sigma}_z|\boldsymbol{\Psi}\rangle=
2\langle\boldsymbol{\Psi}|\boldsymbol{\sigma}_z\otimes\mathbf{1}|\boldsymbol{\Psi}\rangle =\\
&=2\!\sum_{i<j,k<l} d^\ast_{ij}d_{kl} 
\int\! d\mathbf{r}\,d\mathbf{r}'\boldsymbol{\Psi}^\dag_{ij}(\mathbf{r},\mathbf{r}') \boldsymbol{\sigma}_z\otimes\mathbf{1}\boldsymbol{\Psi}_{kl}(\mathbf{r},\mathbf{r}'),
\end{split}
\end{equation}
with $\boldsymbol{\sigma}_z\equiv\sigma_z\otimes\mathbf{1}_3$, and the identity matrix $\mathbf{1}\equiv\mathbf{1}_6$.
By defining $\sigma_{ijkl}=\langle i|\sigma_z|k\rangle\delta_{jl}+\langle j|\sigma_z|l\rangle\delta_{ik}$,
with e.g. $\langle i|\sigma_z|k\rangle=\int\! d\mathbf{r}\,
\boldsymbol{\psi}_i^\dag(\mathbf{r})\boldsymbol{\sigma}_z \boldsymbol{\psi}_k(\mathbf{r})$,
we can write formula for the total spin compactly:
\begin{equation}
    \langle\sigma_z\rangle=\sum_{i<j,k<l} d^\ast_{ij}d_{kl}\left\{\sigma_{ijkl}-\sigma_{ijlk}\right\}.
\end{equation}

\bibliography{valleyswap}

\begin{thebibliography}{119}%
\makeatletter
\providecommand \@ifxundefined [1]{%
 \@ifx{#1\undefined}
}%
\providecommand \@ifnum [1]{%
 \ifnum #1\expandafter \@firstoftwo
 \else \expandafter \@secondoftwo
 \fi
}%
\providecommand \@ifx [1]{%
 \ifx #1\expandafter \@firstoftwo
 \else \expandafter \@secondoftwo
 \fi
}%
\providecommand \natexlab [1]{#1}%
\providecommand \enquote  [1]{``#1''}%
\providecommand \bibnamefont  [1]{#1}%
\providecommand \bibfnamefont [1]{#1}%
\providecommand \citenamefont [1]{#1}%
\providecommand \href@noop [0]{\@secondoftwo}%
\providecommand \href [0]{\begingroup \@sanitize@url \@href}%
\providecommand \@href[1]{\@@startlink{#1}\@@href}%
\providecommand \@@href[1]{\endgroup#1\@@endlink}%
\providecommand \@sanitize@url [0]{\catcode `\\12\catcode `\$12\catcode
  `\&12\catcode `\#12\catcode `\^12\catcode `\_12\catcode `\%12\relax}%
\providecommand \@@startlink[1]{}%
\providecommand \@@endlink[0]{}%
\providecommand \url  [0]{\begingroup\@sanitize@url \@url }%
\providecommand \@url [1]{\endgroup\@href {#1}{\urlprefix }}%
\providecommand \urlprefix  [0]{URL }%
\providecommand \Eprint [0]{\href }%
\providecommand \doibase [0]{https://doi.org/}%
\providecommand \selectlanguage [0]{\@gobble}%
\providecommand \bibinfo  [0]{\@secondoftwo}%
\providecommand \bibfield  [0]{\@secondoftwo}%
\providecommand \translation [1]{[#1]}%
\providecommand \BibitemOpen [0]{}%
\providecommand \bibitemStop [0]{}%
\providecommand \bibitemNoStop [0]{.\EOS\space}%
\providecommand \EOS [0]{\spacefactor3000\relax}%
\providecommand \BibitemShut  [1]{\csname bibitem#1\endcsname}%
\let\auto@bib@innerbib\@empty
%</preamble>
\bibitem [{\citenamefont {Gambetta}\ \emph {et~al.}(2017)\citenamefont
  {Gambetta}, \citenamefont {Chow},\ and\ \citenamefont {Steffen}}]{scq}%
  \BibitemOpen
  \bibfield  {author} {\bibinfo {author} {\bibfnamefont {J.~M.}\ \bibnamefont
  {Gambetta}}, \bibinfo {author} {\bibfnamefont {J.~M.}\ \bibnamefont {Chow}},\
  and\ \bibinfo {author} {\bibfnamefont {M.}~\bibnamefont {Steffen}},\
  }\bibfield  {title} {\bibinfo {title} {Building logical qubits in a
  superconducting quantum computing system},\ }\href@noop {} {\bibfield
  {journal} {\bibinfo  {journal} {npj Quantum Information}\ }\textbf {\bibinfo
  {volume} {3}},\ \bibinfo {pages} {1} (\bibinfo {year} {2017})},\ \bibinfo
  {note} {article number: 2}\BibitemShut {NoStop}%
\bibitem [{\citenamefont {Kjaergaard}\ \emph {et~al.}(2020)\citenamefont
  {Kjaergaard}, \citenamefont {Schwartz}, \citenamefont {Braumüller},
  \citenamefont {Krantz}, \citenamefont {Wang}, \citenamefont {Gustavsson},\
  and\ \citenamefont {Oliver}}]{scq1}%
  \BibitemOpen
  \bibfield  {author} {\bibinfo {author} {\bibfnamefont {M.}~\bibnamefont
  {Kjaergaard}}, \bibinfo {author} {\bibfnamefont {M.~E.}\ \bibnamefont
  {Schwartz}}, \bibinfo {author} {\bibfnamefont {J.}~\bibnamefont
  {Braumüller}}, \bibinfo {author} {\bibfnamefont {P.}~\bibnamefont {Krantz}},
  \bibinfo {author} {\bibfnamefont {J.~I.-J.}\ \bibnamefont {Wang}}, \bibinfo
  {author} {\bibfnamefont {S.}~\bibnamefont {Gustavsson}},\ and\ \bibinfo
  {author} {\bibfnamefont {W.~D.}\ \bibnamefont {Oliver}},\ }\bibfield  {title}
  {\bibinfo {title} {Superconducting qubits: Current state of play},\ }\href
  {https://doi.org/10.1146/annurev-conmatphys-031119-050605} {\bibfield
  {journal} {\bibinfo  {journal} {Annual Review of Condensed Matter Physics}\
  }\textbf {\bibinfo {volume} {11}},\ \bibinfo {pages} {369} (\bibinfo {year}
  {2020})}\BibitemShut {NoStop}%
\bibitem [{\citenamefont {Arute}\ \emph {et~al.}(2019)\citenamefont {Arute},
  \citenamefont {Arya}, \citenamefont {Babbush}, \citenamefont {Bacon},
  \citenamefont {Bardin}, \citenamefont {Barends}, \citenamefont {Biswas},
  \citenamefont {Boixo}, \citenamefont {Brandao}, \citenamefont {Buell} \emph
  {et~al.}}]{google}%
  \BibitemOpen
  \bibfield  {author} {\bibinfo {author} {\bibfnamefont {F.}~\bibnamefont
  {Arute}}, \bibinfo {author} {\bibfnamefont {K.}~\bibnamefont {Arya}},
  \bibinfo {author} {\bibfnamefont {R.}~\bibnamefont {Babbush}}, \bibinfo
  {author} {\bibfnamefont {D.}~\bibnamefont {Bacon}}, \bibinfo {author}
  {\bibfnamefont {J.~C.}\ \bibnamefont {Bardin}}, \bibinfo {author}
  {\bibfnamefont {R.}~\bibnamefont {Barends}}, \bibinfo {author} {\bibfnamefont
  {R.}~\bibnamefont {Biswas}}, \bibinfo {author} {\bibfnamefont
  {S.}~\bibnamefont {Boixo}}, \bibinfo {author} {\bibfnamefont {F.~G.}\
  \bibnamefont {Brandao}}, \bibinfo {author} {\bibfnamefont {D.~A.}\
  \bibnamefont {Buell}}, \emph {et~al.},\ }\bibfield  {title} {\bibinfo {title}
  {Quantum supremacy using a programmable superconducting processor},\
  }\href@noop {} {\bibfield  {journal} {\bibinfo  {journal} {Nature}\ }\textbf
  {\bibinfo {volume} {574}},\ \bibinfo {pages} {505} (\bibinfo {year}
  {2019})}\BibitemShut {NoStop}%
\bibitem [{\citenamefont {Roffe}(2019)}]{qec}%
  \BibitemOpen
  \bibfield  {author} {\bibinfo {author} {\bibfnamefont {J.}~\bibnamefont
  {Roffe}},\ }\bibfield  {title} {\bibinfo {title} {Quantum error correction:
  an introductory guide},\ }\href@noop {} {\bibfield  {journal} {\bibinfo
  {journal} {Contemporary Physics}\ }\textbf {\bibinfo {volume} {60}},\
  \bibinfo {pages} {226} (\bibinfo {year} {2019})}\BibitemShut {NoStop}%
\bibitem [{\citenamefont {Zwanenburg}\ \emph {et~al.}(2013)\citenamefont
  {Zwanenburg}, \citenamefont {Dzurak}, \citenamefont {Morello}, \citenamefont
  {Simmons}, \citenamefont {Hollenberg}, \citenamefont {Klimeck}, \citenamefont
  {Rogge}, \citenamefont {Coppersmith},\ and\ \citenamefont
  {Eriksson}}]{Zwanenburg_Eriksson_2013}%
  \BibitemOpen
  \bibfield  {author} {\bibinfo {author} {\bibfnamefont {F.~A.}\ \bibnamefont
  {Zwanenburg}}, \bibinfo {author} {\bibfnamefont {A.~S.}\ \bibnamefont
  {Dzurak}}, \bibinfo {author} {\bibfnamefont {A.}~\bibnamefont {Morello}},
  \bibinfo {author} {\bibfnamefont {M.~Y.}\ \bibnamefont {Simmons}}, \bibinfo
  {author} {\bibfnamefont {L.~C.~L.}\ \bibnamefont {Hollenberg}}, \bibinfo
  {author} {\bibfnamefont {G.}~\bibnamefont {Klimeck}}, \bibinfo {author}
  {\bibfnamefont {S.}~\bibnamefont {Rogge}}, \bibinfo {author} {\bibfnamefont
  {S.~N.}\ \bibnamefont {Coppersmith}},\ and\ \bibinfo {author} {\bibfnamefont
  {M.~A.}\ \bibnamefont {Eriksson}},\ }\bibfield  {title} {\bibinfo {title}
  {Silicon quantum electronics},\ }\href
  {https://doi.org/10.1103/RevModPhys.85.961} {\bibfield  {journal} {\bibinfo
  {journal} {Rev. Mod. Phys.}\ }\textbf {\bibinfo {volume} {85}},\ \bibinfo
  {pages} {961} (\bibinfo {year} {2013})}\BibitemShut {NoStop}%
\bibitem [{\citenamefont {Kim}\ \emph {et~al.}(2014)\citenamefont {Kim},
  \citenamefont {Shi}, \citenamefont {Simmons}, \citenamefont {Ward},
  \citenamefont {Prance}, \citenamefont {Koh}, \citenamefont {Gamble},
  \citenamefont {Savage}, \citenamefont {Lagally}, \citenamefont {Friesen},
  \citenamefont {Coppersmith},\ and\ \citenamefont
  {Eriksson}}]{Kim_Eriksson_2014}%
  \BibitemOpen
  \bibfield  {author} {\bibinfo {author} {\bibfnamefont {D.}~\bibnamefont
  {Kim}}, \bibinfo {author} {\bibfnamefont {Z.}~\bibnamefont {Shi}}, \bibinfo
  {author} {\bibfnamefont {C.~B.}\ \bibnamefont {Simmons}}, \bibinfo {author}
  {\bibfnamefont {D.~R.}\ \bibnamefont {Ward}}, \bibinfo {author}
  {\bibfnamefont {J.~R.}\ \bibnamefont {Prance}}, \bibinfo {author}
  {\bibfnamefont {T.~S.}\ \bibnamefont {Koh}}, \bibinfo {author} {\bibfnamefont
  {J.~K.}\ \bibnamefont {Gamble}}, \bibinfo {author} {\bibfnamefont {D.~E.}\
  \bibnamefont {Savage}}, \bibinfo {author} {\bibfnamefont {M.~G.}\
  \bibnamefont {Lagally}}, \bibinfo {author} {\bibfnamefont {M.}~\bibnamefont
  {Friesen}}, \bibinfo {author} {\bibfnamefont {S.~N.}\ \bibnamefont
  {Coppersmith}},\ and\ \bibinfo {author} {\bibfnamefont {M.~A.}\ \bibnamefont
  {Eriksson}},\ }\bibfield  {title} {\bibinfo {title} {Quantum control and
  process tomography of a semiconductor quantum dot hybrid qubit},\ }\href
  {https://doi.org/10.1038/nature13407} {\bibfield  {journal} {\bibinfo
  {journal} {Nature}\ }\textbf {\bibinfo {volume} {511}},\ \bibinfo {pages}
  {70} (\bibinfo {year} {2014})}\BibitemShut {NoStop}%
\bibitem [{\citenamefont {Zajac}\ \emph {et~al.}(2018)\citenamefont {Zajac},
  \citenamefont {Sigillito}, \citenamefont {Russ}, \citenamefont {Borjans},
  \citenamefont {Taylor}, \citenamefont {Burkard},\ and\ \citenamefont
  {Petta}}]{Zajac_Petta_2018}%
  \BibitemOpen
  \bibfield  {author} {\bibinfo {author} {\bibfnamefont {D.~M.}\ \bibnamefont
  {Zajac}}, \bibinfo {author} {\bibfnamefont {A.~J.}\ \bibnamefont
  {Sigillito}}, \bibinfo {author} {\bibfnamefont {M.}~\bibnamefont {Russ}},
  \bibinfo {author} {\bibfnamefont {F.}~\bibnamefont {Borjans}}, \bibinfo
  {author} {\bibfnamefont {J.~M.}\ \bibnamefont {Taylor}}, \bibinfo {author}
  {\bibfnamefont {G.}~\bibnamefont {Burkard}},\ and\ \bibinfo {author}
  {\bibfnamefont {J.~R.}\ \bibnamefont {Petta}},\ }\bibfield  {title} {\bibinfo
  {title} {Resonantly driven cnot gate for electron spins},\ }\href
  {https://doi.org/10.1126/science.aao5965} {\bibfield  {journal} {\bibinfo
  {journal} {Science}\ }\textbf {\bibinfo {volume} {359}},\ \bibinfo {pages}
  {439} (\bibinfo {year} {2018})}\BibitemShut {NoStop}%
\bibitem [{\citenamefont {Watson}\ \emph {et~al.}(2018)\citenamefont {Watson},
  \citenamefont {Philips}, \citenamefont {Kawakami}, \citenamefont {Ward},
  \citenamefont {Scarlino}, \citenamefont {Veldhorst}, \citenamefont {Savage},
  \citenamefont {Lagally}, \citenamefont {Friesen}, \citenamefont
  {Coppersmith}, \citenamefont {Eriksson},\ and\ \citenamefont
  {Vandersypen}}]{Watson_Vandersypen_2018}%
  \BibitemOpen
  \bibfield  {author} {\bibinfo {author} {\bibfnamefont {T.~F.}\ \bibnamefont
  {Watson}}, \bibinfo {author} {\bibfnamefont {S.~G.~J.}\ \bibnamefont
  {Philips}}, \bibinfo {author} {\bibfnamefont {E.}~\bibnamefont {Kawakami}},
  \bibinfo {author} {\bibfnamefont {D.~R.}\ \bibnamefont {Ward}}, \bibinfo
  {author} {\bibfnamefont {P.}~\bibnamefont {Scarlino}}, \bibinfo {author}
  {\bibfnamefont {M.}~\bibnamefont {Veldhorst}}, \bibinfo {author}
  {\bibfnamefont {D.~E.}\ \bibnamefont {Savage}}, \bibinfo {author}
  {\bibfnamefont {M.~G.}\ \bibnamefont {Lagally}}, \bibinfo {author}
  {\bibfnamefont {M.}~\bibnamefont {Friesen}}, \bibinfo {author} {\bibfnamefont
  {S.~N.}\ \bibnamefont {Coppersmith}}, \bibinfo {author} {\bibfnamefont
  {M.~A.}\ \bibnamefont {Eriksson}},\ and\ \bibinfo {author} {\bibfnamefont
  {L.~M.~K.}\ \bibnamefont {Vandersypen}},\ }\bibfield  {title} {\bibinfo
  {title} {A programmable two-qubit quantum processor in silicon},\ }\href
  {https://doi.org/10.1038/nature25766} {\bibfield  {journal} {\bibinfo
  {journal} {Nature}\ }\textbf {\bibinfo {volume} {555}},\ \bibinfo {pages}
  {633} (\bibinfo {year} {2018})}\BibitemShut {NoStop}%
\bibitem [{\citenamefont {Mi}\ \emph {et~al.}(2018{\natexlab{a}})\citenamefont
  {Mi}, \citenamefont {Benito}, \citenamefont {Putz}, \citenamefont {Zajac},
  \citenamefont {Taylor}, \citenamefont {Burkard},\ and\ \citenamefont
  {Petta}}]{Mi_Petta_2018}%
  \BibitemOpen
  \bibfield  {author} {\bibinfo {author} {\bibfnamefont {X.}~\bibnamefont
  {Mi}}, \bibinfo {author} {\bibfnamefont {M.}~\bibnamefont {Benito}}, \bibinfo
  {author} {\bibfnamefont {S.}~\bibnamefont {Putz}}, \bibinfo {author}
  {\bibfnamefont {D.~M.}\ \bibnamefont {Zajac}}, \bibinfo {author}
  {\bibfnamefont {J.~M.}\ \bibnamefont {Taylor}}, \bibinfo {author}
  {\bibfnamefont {G.}~\bibnamefont {Burkard}},\ and\ \bibinfo {author}
  {\bibfnamefont {J.~R.}\ \bibnamefont {Petta}},\ }\bibfield  {title} {\bibinfo
  {title} {A coherent spin--photon interface in silicon},\ }\href
  {https://doi.org/10.1038/nature25769} {\bibfield  {journal} {\bibinfo
  {journal} {Nature}\ }\textbf {\bibinfo {volume} {555}},\ \bibinfo {pages}
  {599} (\bibinfo {year} {2018}{\natexlab{a}})}\BibitemShut {NoStop}%
\bibitem [{\citenamefont {Landig}\ \emph {et~al.}(2018)\citenamefont {Landig},
  \citenamefont {Koski}, \citenamefont {Scarlino}, \citenamefont {Mendes},
  \citenamefont {Blais}, \citenamefont {Reichl}, \citenamefont {Wegscheider},
  \citenamefont {Wallraff}, \citenamefont {Ensslin},\ and\ \citenamefont
  {Ihn}}]{Landig_Ihn_2018}%
  \BibitemOpen
  \bibfield  {author} {\bibinfo {author} {\bibfnamefont {A.~J.}\ \bibnamefont
  {Landig}}, \bibinfo {author} {\bibfnamefont {J.~V.}\ \bibnamefont {Koski}},
  \bibinfo {author} {\bibfnamefont {P.}~\bibnamefont {Scarlino}}, \bibinfo
  {author} {\bibfnamefont {U.~C.}\ \bibnamefont {Mendes}}, \bibinfo {author}
  {\bibfnamefont {A.}~\bibnamefont {Blais}}, \bibinfo {author} {\bibfnamefont
  {C.}~\bibnamefont {Reichl}}, \bibinfo {author} {\bibfnamefont
  {W.}~\bibnamefont {Wegscheider}}, \bibinfo {author} {\bibfnamefont
  {A.}~\bibnamefont {Wallraff}}, \bibinfo {author} {\bibfnamefont
  {K.}~\bibnamefont {Ensslin}},\ and\ \bibinfo {author} {\bibfnamefont
  {T.}~\bibnamefont {Ihn}},\ }\bibfield  {title} {\bibinfo {title} {Coherent
  spin--photon coupling using a resonant exchange qubit},\ }\href
  {https://doi.org/10.1038/s41586-018-0365-y} {\bibfield  {journal} {\bibinfo
  {journal} {Nature}\ }\textbf {\bibinfo {volume} {560}},\ \bibinfo {pages}
  {179} (\bibinfo {year} {2018})}\BibitemShut {NoStop}%
\bibitem [{\citenamefont {Zheng}\ \emph {et~al.}(2019)\citenamefont {Zheng},
  \citenamefont {Samkharadze}, \citenamefont {Noordam}, \citenamefont {Kalhor},
  \citenamefont {Brousse}, \citenamefont {Sammak}, \citenamefont {Scappucci},\
  and\ \citenamefont {Vandersypen}}]{Zheng_Vandersypen_2019}%
  \BibitemOpen
  \bibfield  {author} {\bibinfo {author} {\bibfnamefont {G.}~\bibnamefont
  {Zheng}}, \bibinfo {author} {\bibfnamefont {N.}~\bibnamefont {Samkharadze}},
  \bibinfo {author} {\bibfnamefont {M.~L.}\ \bibnamefont {Noordam}}, \bibinfo
  {author} {\bibfnamefont {N.}~\bibnamefont {Kalhor}}, \bibinfo {author}
  {\bibfnamefont {D.}~\bibnamefont {Brousse}}, \bibinfo {author} {\bibfnamefont
  {A.}~\bibnamefont {Sammak}}, \bibinfo {author} {\bibfnamefont
  {G.}~\bibnamefont {Scappucci}},\ and\ \bibinfo {author} {\bibfnamefont
  {L.~M.~K.}\ \bibnamefont {Vandersypen}},\ }\bibfield  {title} {\bibinfo
  {title} {Rapid gate-based spin read-out in silicon using an on-chip
  resonator},\ }\href {https://doi.org/10.1038/s41565-019-0488-9} {\bibfield
  {journal} {\bibinfo  {journal} {Nature Nanotechnology}\ }\textbf {\bibinfo
  {volume} {14}},\ \bibinfo {pages} {742} (\bibinfo {year} {2019})}\BibitemShut
  {NoStop}%
\bibitem [{\citenamefont {Petit}\ \emph {et~al.}(2020)\citenamefont {Petit},
  \citenamefont {Eenink}, \citenamefont {Russ}, \citenamefont {Lawrie},
  \citenamefont {Hendrickx}, \citenamefont {Philips}, \citenamefont {Clarke},
  \citenamefont {Vandersypen},\ and\ \citenamefont
  {Veldhorst}}]{Petit_Veldhorst_2020}%
  \BibitemOpen
  \bibfield  {author} {\bibinfo {author} {\bibfnamefont {L.}~\bibnamefont
  {Petit}}, \bibinfo {author} {\bibfnamefont {H.~G.~J.}\ \bibnamefont
  {Eenink}}, \bibinfo {author} {\bibfnamefont {M.}~\bibnamefont {Russ}},
  \bibinfo {author} {\bibfnamefont {W.~I.~L.}\ \bibnamefont {Lawrie}}, \bibinfo
  {author} {\bibfnamefont {N.~W.}\ \bibnamefont {Hendrickx}}, \bibinfo {author}
  {\bibfnamefont {S.~G.~J.}\ \bibnamefont {Philips}}, \bibinfo {author}
  {\bibfnamefont {J.~S.}\ \bibnamefont {Clarke}}, \bibinfo {author}
  {\bibfnamefont {L.~M.~K.}\ \bibnamefont {Vandersypen}},\ and\ \bibinfo
  {author} {\bibfnamefont {M.}~\bibnamefont {Veldhorst}},\ }\bibfield  {title}
  {\bibinfo {title} {Universal quantum logic in hot silicon qubits},\ }\href
  {https://doi.org/10.1038/s41586-020-2170-7} {\bibfield  {journal} {\bibinfo
  {journal} {Nature}\ }\textbf {\bibinfo {volume} {580}},\ \bibinfo {pages}
  {355} (\bibinfo {year} {2020})}\BibitemShut {NoStop}%
\bibitem [{\citenamefont {Korm\'anyos}\ \emph
  {et~al.}(2014{\natexlab{a}})\citenamefont {Korm\'anyos}, \citenamefont
  {Z\'olyomi}, \citenamefont {Drummond},\ and\ \citenamefont {Burkard}}]{prx}%
  \BibitemOpen
  \bibfield  {author} {\bibinfo {author} {\bibfnamefont {A.}~\bibnamefont
  {Korm\'anyos}}, \bibinfo {author} {\bibfnamefont {V.}~\bibnamefont
  {Z\'olyomi}}, \bibinfo {author} {\bibfnamefont {N.~D.}\ \bibnamefont
  {Drummond}},\ and\ \bibinfo {author} {\bibfnamefont {G.}~\bibnamefont
  {Burkard}},\ }\bibfield  {title} {\bibinfo {title} {Spin-orbit coupling,
  quantum dots, and qubits in monolayer transition metal dichalcogenides},\
  }\href {https://doi.org/10.1103/PhysRevX.4.011034} {\bibfield  {journal}
  {\bibinfo  {journal} {Phys. Rev. X}\ }\textbf {\bibinfo {volume} {4}},\
  \bibinfo {pages} {011034} (\bibinfo {year} {2014}{\natexlab{a}})}\BibitemShut
  {NoStop}%
\bibitem [{\citenamefont {Ko\ifmmode~\acute{s}\else \'{s}\fi{}mider}\ \emph
  {et~al.}(2013)\citenamefont {Ko\ifmmode~\acute{s}\else \'{s}\fi{}mider},
  \citenamefont {Gonz\'alez},\ and\ \citenamefont {Fern\'andez-Rossier}}]{kos}%
  \BibitemOpen
  \bibfield  {author} {\bibinfo {author} {\bibfnamefont {K.}~\bibnamefont
  {Ko\ifmmode~\acute{s}\else \'{s}\fi{}mider}}, \bibinfo {author}
  {\bibfnamefont {J.~W.}\ \bibnamefont {Gonz\'alez}},\ and\ \bibinfo {author}
  {\bibfnamefont {J.}~\bibnamefont {Fern\'andez-Rossier}},\ }\bibfield  {title}
  {\bibinfo {title} {Large spin splitting in the conduction band of transition
  metal dichalcogenide monolayers},\ }\href
  {https://doi.org/10.1103/PhysRevB.88.245436} {\bibfield  {journal} {\bibinfo
  {journal} {Phys. Rev. B}\ }\textbf {\bibinfo {volume} {88}},\ \bibinfo
  {pages} {245436} (\bibinfo {year} {2013})}\BibitemShut {NoStop}%
\bibitem [{\citenamefont {Wu}\ \emph {et~al.}(2011)\citenamefont {Wu},
  \citenamefont {Lue},\ and\ \citenamefont {Chang}}]{valq0}%
  \BibitemOpen
  \bibfield  {author} {\bibinfo {author} {\bibfnamefont {G.~Y.}\ \bibnamefont
  {Wu}}, \bibinfo {author} {\bibfnamefont {N.-Y.}\ \bibnamefont {Lue}},\ and\
  \bibinfo {author} {\bibfnamefont {L.}~\bibnamefont {Chang}},\ }\bibfield
  {title} {\bibinfo {title} {Graphene quantum dots for valley-based quantum
  computing: A feasibility study},\ }\href
  {https://doi.org/10.1103/PhysRevB.84.195463} {\bibfield  {journal} {\bibinfo
  {journal} {Phys. Rev. B}\ }\textbf {\bibinfo {volume} {84}},\ \bibinfo
  {pages} {195463} (\bibinfo {year} {2011})}\BibitemShut {NoStop}%
\bibitem [{\citenamefont {Wu}\ \emph {et~al.}(2016{\natexlab{a}})\citenamefont
  {Wu}, \citenamefont {Tong}, \citenamefont {Liu}, \citenamefont {Yu},\ and\
  \citenamefont {Yao}}]{valq}%
  \BibitemOpen
  \bibfield  {author} {\bibinfo {author} {\bibfnamefont {Y.}~\bibnamefont
  {Wu}}, \bibinfo {author} {\bibfnamefont {Q.}~\bibnamefont {Tong}}, \bibinfo
  {author} {\bibfnamefont {G.-B.}\ \bibnamefont {Liu}}, \bibinfo {author}
  {\bibfnamefont {H.}~\bibnamefont {Yu}},\ and\ \bibinfo {author}
  {\bibfnamefont {W.}~\bibnamefont {Yao}},\ }\bibfield  {title} {\bibinfo
  {title} {Spin-valley qubit in nanostructures of monolayer semiconductors:
  Optical control and hyperfine interaction},\ }\href
  {https://doi.org/10.1103/PhysRevB.93.045313} {\bibfield  {journal} {\bibinfo
  {journal} {Phys. Rev. B}\ }\textbf {\bibinfo {volume} {93}},\ \bibinfo
  {pages} {045313} (\bibinfo {year} {2016}{\natexlab{a}})}\BibitemShut
  {NoStop}%
\bibitem [{\citenamefont {Paw\l{}owski}\ \emph {et~al.}(2018)\citenamefont
  {Paw\l{}owski}, \citenamefont {\ifmmode~\dot{Z}\else \.{Z}\fi{}ebrowski},\
  and\ \citenamefont {Bednarek}}]{mymos2}%
  \BibitemOpen
  \bibfield  {author} {\bibinfo {author} {\bibfnamefont {J.}~\bibnamefont
  {Paw\l{}owski}}, \bibinfo {author} {\bibfnamefont {D.}~\bibnamefont
  {\ifmmode~\dot{Z}\else \.{Z}\fi{}ebrowski}},\ and\ \bibinfo {author}
  {\bibfnamefont {S.}~\bibnamefont {Bednarek}},\ }\bibfield  {title} {\bibinfo
  {title} {Valley qubit in a gated ${\mathbf{mos}}_{2}$ monolayer quantum
  dot},\ }\href {https://doi.org/10.1103/PhysRevB.97.155412} {\bibfield
  {journal} {\bibinfo  {journal} {Phys. Rev. B}\ }\textbf {\bibinfo {volume}
  {97}},\ \bibinfo {pages} {155412} (\bibinfo {year} {2018})}\BibitemShut
  {NoStop}%
\bibitem [{\citenamefont {Rohling}\ and\ \citenamefont
  {Burkard}(2012)}]{rohling}%
  \BibitemOpen
  \bibfield  {author} {\bibinfo {author} {\bibfnamefont {N.}~\bibnamefont
  {Rohling}}\ and\ \bibinfo {author} {\bibfnamefont {G.}~\bibnamefont
  {Burkard}},\ }\bibfield  {title} {\bibinfo {title} {Universal quantum
  computing with spin and valley states},\ }\href
  {https://doi.org/10.1088/1367-2630/14/8/083008} {\bibfield  {journal}
  {\bibinfo  {journal} {New Journal of Physics}\ }\textbf {\bibinfo {volume}
  {14}},\ \bibinfo {pages} {083008} (\bibinfo {year} {2012})}\BibitemShut
  {NoStop}%
\bibitem [{\citenamefont {Paw{\l}owski}(2019)}]{mynjp}%
  \BibitemOpen
  \bibfield  {author} {\bibinfo {author} {\bibfnamefont {J.}~\bibnamefont
  {Paw{\l}owski}},\ }\bibfield  {title} {\bibinfo {title} {Spin-valley system
  in a gated {MoS}2-monolayer quantum dot},\ }\href@noop {} {\bibfield
  {journal} {\bibinfo  {journal} {New Journal of Physics}\ }\textbf {\bibinfo
  {volume} {21}},\ \bibinfo {pages} {123029} (\bibinfo {year}
  {2019})}\BibitemShut {NoStop}%
\bibitem [{\citenamefont {Goh}\ \emph {et~al.}(2020{\natexlab{a}})\citenamefont
  {Goh}, \citenamefont {Bussolotti}, \citenamefont {Lau}, \citenamefont
  {Kotekar-Patil}, \citenamefont {Ooi},\ and\ \citenamefont {Chee}}]{kotekar}%
  \BibitemOpen
  \bibfield  {author} {\bibinfo {author} {\bibfnamefont {K.~E.~J.}\
  \bibnamefont {Goh}}, \bibinfo {author} {\bibfnamefont {F.}~\bibnamefont
  {Bussolotti}}, \bibinfo {author} {\bibfnamefont {C.~S.}\ \bibnamefont {Lau}},
  \bibinfo {author} {\bibfnamefont {D.}~\bibnamefont {Kotekar-Patil}}, \bibinfo
  {author} {\bibfnamefont {Z.~E.}\ \bibnamefont {Ooi}},\ and\ \bibinfo {author}
  {\bibfnamefont {J.~Y.}\ \bibnamefont {Chee}},\ }\bibfield  {title} {\bibinfo
  {title} {Toward valley-coupled spin qubits},\ }\href@noop {} {\bibfield
  {journal} {\bibinfo  {journal} {Advanced Quantum Technologies}\ }\textbf
  {\bibinfo {volume} {3}},\ \bibinfo {pages} {1900123} (\bibinfo {year}
  {2020}{\natexlab{a}})}\BibitemShut {NoStop}%
\bibitem [{\citenamefont {Laird}\ \emph {et~al.}(2013)\citenamefont {Laird},
  \citenamefont {Pei},\ and\ \citenamefont {Kouwenhoven}}]{tubes}%
  \BibitemOpen
  \bibfield  {author} {\bibinfo {author} {\bibfnamefont {E.~A.}\ \bibnamefont
  {Laird}}, \bibinfo {author} {\bibfnamefont {F.}~\bibnamefont {Pei}},\ and\
  \bibinfo {author} {\bibfnamefont {L.~P.}\ \bibnamefont {Kouwenhoven}},\
  }\bibfield  {title} {\bibinfo {title} {A valley--spin qubit in a carbon
  nanotube},\ }\href@noop {} {\bibfield  {journal} {\bibinfo  {journal} {Nature
  Nanotechnology}\ }\textbf {\bibinfo {volume} {8}},\ \bibinfo {pages} {565}
  (\bibinfo {year} {2013})}\BibitemShut {NoStop}%
\bibitem [{\citenamefont {Pei}\ \emph {et~al.}(2012)\citenamefont {Pei},
  \citenamefont {Laird}, \citenamefont {Steele},\ and\ \citenamefont
  {Kouwenhoven}}]{nt2}%
  \BibitemOpen
  \bibfield  {author} {\bibinfo {author} {\bibfnamefont {F.}~\bibnamefont
  {Pei}}, \bibinfo {author} {\bibfnamefont {E.~A.}\ \bibnamefont {Laird}},
  \bibinfo {author} {\bibfnamefont {G.~A.}\ \bibnamefont {Steele}},\ and\
  \bibinfo {author} {\bibfnamefont {L.~P.}\ \bibnamefont {Kouwenhoven}},\
  }\bibfield  {title} {\bibinfo {title} {Valley--spin blockade and spin
  resonance in carbon nanotubes},\ }\href@noop {} {\bibfield  {journal}
  {\bibinfo  {journal} {Nature Nanotechnology}\ }\textbf {\bibinfo {volume}
  {7}},\ \bibinfo {pages} {630} (\bibinfo {year} {2012})}\BibitemShut {NoStop}%
\bibitem [{\citenamefont {Gong}\ \emph {et~al.}(2013)\citenamefont {Gong},
  \citenamefont {Liu}, \citenamefont {Yu}, \citenamefont {Xiao}, \citenamefont
  {Cui}, \citenamefont {Xu},\ and\ \citenamefont {Yao}}]{bil}%
  \BibitemOpen
  \bibfield  {author} {\bibinfo {author} {\bibfnamefont {Z.}~\bibnamefont
  {Gong}}, \bibinfo {author} {\bibfnamefont {G.-B.}\ \bibnamefont {Liu}},
  \bibinfo {author} {\bibfnamefont {H.}~\bibnamefont {Yu}}, \bibinfo {author}
  {\bibfnamefont {D.}~\bibnamefont {Xiao}}, \bibinfo {author} {\bibfnamefont
  {X.}~\bibnamefont {Cui}}, \bibinfo {author} {\bibfnamefont {X.}~\bibnamefont
  {Xu}},\ and\ \bibinfo {author} {\bibfnamefont {W.}~\bibnamefont {Yao}},\
  }\bibfield  {title} {\bibinfo {title} {Magnetoelectric effects and
  valley-controlled spin quantum gates in transition metal dichalcogenide
  bilayers},\ }\href@noop {} {\bibfield  {journal} {\bibinfo  {journal} {Nature
  communications}\ }\textbf {\bibinfo {volume} {4}},\ \bibinfo {pages} {1}
  (\bibinfo {year} {2013})},\ \bibinfo {note} {article number:
  2053}\BibitemShut {NoStop}%
\bibitem [{\citenamefont {Ciccarino}\ \emph {et~al.}(2019)\citenamefont
  {Ciccarino}, \citenamefont {Chakraborty}, \citenamefont {Englund},\ and\
  \citenamefont {Narang}}]{bil2}%
  \BibitemOpen
  \bibfield  {author} {\bibinfo {author} {\bibfnamefont {C.~J.}\ \bibnamefont
  {Ciccarino}}, \bibinfo {author} {\bibfnamefont {C.}~\bibnamefont
  {Chakraborty}}, \bibinfo {author} {\bibfnamefont {D.~R.}\ \bibnamefont
  {Englund}},\ and\ \bibinfo {author} {\bibfnamefont {P.}~\bibnamefont
  {Narang}},\ }\bibfield  {title} {\bibinfo {title} {Carrier dynamics and
  spin–valley–layer effects in bilayer transition metal dichalcogenides},\
  }\href {https://doi.org/10.1039/C8FD00159F} {\bibfield  {journal} {\bibinfo
  {journal} {Faraday Discuss.}\ }\textbf {\bibinfo {volume} {214}},\ \bibinfo
  {pages} {175} (\bibinfo {year} {2019})}\BibitemShut {NoStop}%
\bibitem [{\citenamefont {Eich}\ \emph {et~al.}(2018)\citenamefont {Eich},
  \citenamefont {Herman}, \citenamefont {Pisoni}, \citenamefont {Overweg},
  \citenamefont {Kurzmann}, \citenamefont {Lee}, \citenamefont {Rickhaus},
  \citenamefont {Watanabe}, \citenamefont {Taniguchi}, \citenamefont {Sigrist},
  \citenamefont {Ihn},\ and\ \citenamefont {Ensslin}}]{bil1}%
  \BibitemOpen
  \bibfield  {author} {\bibinfo {author} {\bibfnamefont {M.}~\bibnamefont
  {Eich}}, \bibinfo {author} {\bibfnamefont {F.~c.~v.}\ \bibnamefont {Herman}},
  \bibinfo {author} {\bibfnamefont {R.}~\bibnamefont {Pisoni}}, \bibinfo
  {author} {\bibfnamefont {H.}~\bibnamefont {Overweg}}, \bibinfo {author}
  {\bibfnamefont {A.}~\bibnamefont {Kurzmann}}, \bibinfo {author}
  {\bibfnamefont {Y.}~\bibnamefont {Lee}}, \bibinfo {author} {\bibfnamefont
  {P.}~\bibnamefont {Rickhaus}}, \bibinfo {author} {\bibfnamefont
  {K.}~\bibnamefont {Watanabe}}, \bibinfo {author} {\bibfnamefont
  {T.}~\bibnamefont {Taniguchi}}, \bibinfo {author} {\bibfnamefont
  {M.}~\bibnamefont {Sigrist}}, \bibinfo {author} {\bibfnamefont
  {T.}~\bibnamefont {Ihn}},\ and\ \bibinfo {author} {\bibfnamefont
  {K.}~\bibnamefont {Ensslin}},\ }\bibfield  {title} {\bibinfo {title} {Spin
  and valley states in gate-defined bilayer graphene quantum dots},\ }\href
  {https://doi.org/10.1103/PhysRevX.8.031023} {\bibfield  {journal} {\bibinfo
  {journal} {Phys. Rev. X}\ }\textbf {\bibinfo {volume} {8}},\ \bibinfo {pages}
  {031023} (\bibinfo {year} {2018})}\BibitemShut {NoStop}%
\bibitem [{\citenamefont {Penthorn}\ \emph {et~al.}(2019)\citenamefont
  {Penthorn}, \citenamefont {Schoenfield}, \citenamefont {Rooney},
  \citenamefont {Edge},\ and\ \citenamefont {Jiang}}]{svq1}%
  \BibitemOpen
  \bibfield  {author} {\bibinfo {author} {\bibfnamefont {N.~E.}\ \bibnamefont
  {Penthorn}}, \bibinfo {author} {\bibfnamefont {J.~S.}\ \bibnamefont
  {Schoenfield}}, \bibinfo {author} {\bibfnamefont {J.~D.}\ \bibnamefont
  {Rooney}}, \bibinfo {author} {\bibfnamefont {L.~F.}\ \bibnamefont {Edge}},\
  and\ \bibinfo {author} {\bibfnamefont {H.}~\bibnamefont {Jiang}},\ }\bibfield
   {title} {\bibinfo {title} {Two-axis quantum control of a fast valley qubit
  in silicon},\ }\href@noop {} {\bibfield  {journal} {\bibinfo  {journal} {npj
  Quantum Information}\ }\textbf {\bibinfo {volume} {5}},\ \bibinfo {pages} {1}
  (\bibinfo {year} {2019})},\ \bibinfo {note} {article number: 94}\BibitemShut
  {NoStop}%
\bibitem [{\citenamefont {Mi}\ \emph {et~al.}(2018{\natexlab{b}})\citenamefont
  {Mi}, \citenamefont {Kohler},\ and\ \citenamefont {Petta}}]{svq2}%
  \BibitemOpen
  \bibfield  {author} {\bibinfo {author} {\bibfnamefont {X.}~\bibnamefont
  {Mi}}, \bibinfo {author} {\bibfnamefont {S.}~\bibnamefont {Kohler}},\ and\
  \bibinfo {author} {\bibfnamefont {J.~R.}\ \bibnamefont {Petta}},\ }\bibfield
  {title} {\bibinfo {title} {Landau-zener interferometry of valley-orbit states
  in si/sige double quantum dots},\ }\href
  {https://doi.org/10.1103/PhysRevB.98.161404} {\bibfield  {journal} {\bibinfo
  {journal} {Phys. Rev. B}\ }\textbf {\bibinfo {volume} {98}},\ \bibinfo
  {pages} {161404} (\bibinfo {year} {2018}{\natexlab{b}})}\BibitemShut
  {NoStop}%
\bibitem [{\citenamefont {Settnes}\ \emph {et~al.}(2016)\citenamefont
  {Settnes}, \citenamefont {Power}, \citenamefont {Brandbyge},\ and\
  \citenamefont {Jauho}}]{valq1}%
  \BibitemOpen
  \bibfield  {author} {\bibinfo {author} {\bibfnamefont {M.}~\bibnamefont
  {Settnes}}, \bibinfo {author} {\bibfnamefont {S.~R.}\ \bibnamefont {Power}},
  \bibinfo {author} {\bibfnamefont {M.}~\bibnamefont {Brandbyge}},\ and\
  \bibinfo {author} {\bibfnamefont {A.-P.}\ \bibnamefont {Jauho}},\ }\bibfield
  {title} {\bibinfo {title} {Graphene nanobubbles as valley filters and beam
  splitters},\ }\href {https://doi.org/10.1103/PhysRevLett.117.276801}
  {\bibfield  {journal} {\bibinfo  {journal} {Phys. Rev. Lett.}\ }\textbf
  {\bibinfo {volume} {117}},\ \bibinfo {pages} {276801} (\bibinfo {year}
  {2016})}\BibitemShut {NoStop}%
\bibitem [{\citenamefont {Schaibley}\ \emph {et~al.}(2016)\citenamefont
  {Schaibley}, \citenamefont {Yu}, \citenamefont {Clark}, \citenamefont
  {Rivera}, \citenamefont {Ross}, \citenamefont {Seyler}, \citenamefont {Yao},\
  and\ \citenamefont {Xu}}]{valq2}%
  \BibitemOpen
  \bibfield  {author} {\bibinfo {author} {\bibfnamefont {J.~R.}\ \bibnamefont
  {Schaibley}}, \bibinfo {author} {\bibfnamefont {H.}~\bibnamefont {Yu}},
  \bibinfo {author} {\bibfnamefont {G.}~\bibnamefont {Clark}}, \bibinfo
  {author} {\bibfnamefont {P.}~\bibnamefont {Rivera}}, \bibinfo {author}
  {\bibfnamefont {J.~S.}\ \bibnamefont {Ross}}, \bibinfo {author}
  {\bibfnamefont {K.~L.}\ \bibnamefont {Seyler}}, \bibinfo {author}
  {\bibfnamefont {W.}~\bibnamefont {Yao}},\ and\ \bibinfo {author}
  {\bibfnamefont {X.}~\bibnamefont {Xu}},\ }\bibfield  {title} {\bibinfo
  {title} {Valleytronics in 2d materials},\ }\href@noop {} {\bibfield
  {journal} {\bibinfo  {journal} {Nature Reviews Materials}\ }\textbf {\bibinfo
  {volume} {1}},\ \bibinfo {pages} {1} (\bibinfo {year} {2016})}\BibitemShut
  {NoStop}%
\bibitem [{\citenamefont {Sekera}\ \emph {et~al.}(2017)\citenamefont {Sekera},
  \citenamefont {Bruder}, \citenamefont {Mele},\ and\ \citenamefont
  {Tiwari}}]{filter}%
  \BibitemOpen
  \bibfield  {author} {\bibinfo {author} {\bibfnamefont {T.}~\bibnamefont
  {Sekera}}, \bibinfo {author} {\bibfnamefont {C.}~\bibnamefont {Bruder}},
  \bibinfo {author} {\bibfnamefont {E.~J.}\ \bibnamefont {Mele}},\ and\
  \bibinfo {author} {\bibfnamefont {R.~P.}\ \bibnamefont {Tiwari}},\ }\bibfield
   {title} {\bibinfo {title} {Switchable valley filter based on a graphene
  $p\ensuremath{-}n$ junction in a magnetic field},\ }\href
  {https://doi.org/10.1103/PhysRevB.95.205431} {\bibfield  {journal} {\bibinfo
  {journal} {Phys. Rev. B}\ }\textbf {\bibinfo {volume} {95}},\ \bibinfo
  {pages} {205431} (\bibinfo {year} {2017})}\BibitemShut {NoStop}%
\bibitem [{\citenamefont {Scuri}\ \emph {et~al.}(2020)\citenamefont {Scuri},
  \citenamefont {Andersen}, \citenamefont {Zhou}, \citenamefont {Wild},
  \citenamefont {Sung}, \citenamefont {Gelly}, \citenamefont {B\'erub\'e},
  \citenamefont {Heo}, \citenamefont {Shao}, \citenamefont {Joe}, \citenamefont
  {Mier~Valdivia}, \citenamefont {Taniguchi}, \citenamefont {Watanabe},
  \citenamefont {Lon\ifmmode~\check{c}\else \v{c}\fi{}ar}, \citenamefont {Kim},
  \citenamefont {Lukin},\ and\ \citenamefont {Park}}]{twist}%
  \BibitemOpen
  \bibfield  {author} {\bibinfo {author} {\bibfnamefont {G.}~\bibnamefont
  {Scuri}}, \bibinfo {author} {\bibfnamefont {T.~I.}\ \bibnamefont {Andersen}},
  \bibinfo {author} {\bibfnamefont {Y.}~\bibnamefont {Zhou}}, \bibinfo {author}
  {\bibfnamefont {D.~S.}\ \bibnamefont {Wild}}, \bibinfo {author}
  {\bibfnamefont {J.}~\bibnamefont {Sung}}, \bibinfo {author} {\bibfnamefont
  {R.~J.}\ \bibnamefont {Gelly}}, \bibinfo {author} {\bibfnamefont
  {D.}~\bibnamefont {B\'erub\'e}}, \bibinfo {author} {\bibfnamefont
  {H.}~\bibnamefont {Heo}}, \bibinfo {author} {\bibfnamefont {L.}~\bibnamefont
  {Shao}}, \bibinfo {author} {\bibfnamefont {A.~Y.}\ \bibnamefont {Joe}},
  \bibinfo {author} {\bibfnamefont {A.~M.}\ \bibnamefont {Mier~Valdivia}},
  \bibinfo {author} {\bibfnamefont {T.}~\bibnamefont {Taniguchi}}, \bibinfo
  {author} {\bibfnamefont {K.}~\bibnamefont {Watanabe}}, \bibinfo {author}
  {\bibfnamefont {M.}~\bibnamefont {Lon\ifmmode~\check{c}\else \v{c}\fi{}ar}},
  \bibinfo {author} {\bibfnamefont {P.}~\bibnamefont {Kim}}, \bibinfo {author}
  {\bibfnamefont {M.~D.}\ \bibnamefont {Lukin}},\ and\ \bibinfo {author}
  {\bibfnamefont {H.}~\bibnamefont {Park}},\ }\bibfield  {title} {\bibinfo
  {title} {Electrically tunable valley dynamics in twisted
  ${\mathrm{wse}}_{2}/{\mathrm{wse}}_{2}$ bilayers},\ }\href
  {https://doi.org/10.1103/PhysRevLett.124.217403} {\bibfield  {journal}
  {\bibinfo  {journal} {Phys. Rev. Lett.}\ }\textbf {\bibinfo {volume} {124}},\
  \bibinfo {pages} {217403} (\bibinfo {year} {2020})}\BibitemShut {NoStop}%
\bibitem [{\citenamefont {Xiao}\ \emph {et~al.}(2012)\citenamefont {Xiao},
  \citenamefont {Liu}, \citenamefont {Feng}, \citenamefont {Xu},\ and\
  \citenamefont {Yao}}]{Xiao_Yao_2012}%
  \BibitemOpen
  \bibfield  {author} {\bibinfo {author} {\bibfnamefont {D.}~\bibnamefont
  {Xiao}}, \bibinfo {author} {\bibfnamefont {G.-B.}\ \bibnamefont {Liu}},
  \bibinfo {author} {\bibfnamefont {W.}~\bibnamefont {Feng}}, \bibinfo {author}
  {\bibfnamefont {X.}~\bibnamefont {Xu}},\ and\ \bibinfo {author}
  {\bibfnamefont {W.}~\bibnamefont {Yao}},\ }\bibfield  {title} {\bibinfo
  {title} {{Coupled Spin and Valley Physics in Monolayers of
  ${\mathrm{MoS}}_{2}$ and Other Group-VI Dichalcogenides}},\ }\href
  {https://doi.org/10.1103/PhysRevLett.108.196802} {\bibfield  {journal}
  {\bibinfo  {journal} {Phys. Rev. Lett.}\ }\textbf {\bibinfo {volume} {108}},\
  \bibinfo {pages} {196802} (\bibinfo {year} {2012})}\BibitemShut {NoStop}%
\bibitem [{\citenamefont {Cao}\ \emph {et~al.}(2012)\citenamefont {Cao},
  \citenamefont {Wang}, \citenamefont {Han}, \citenamefont {Ye}, \citenamefont
  {Zhu}, \citenamefont {Shi}, \citenamefont {Niu}, \citenamefont {Tan},
  \citenamefont {Wang}, \citenamefont {Liu},\ and\ \citenamefont
  {Feng}}]{Cao_Feng_2012}%
  \BibitemOpen
  \bibfield  {author} {\bibinfo {author} {\bibfnamefont {T.}~\bibnamefont
  {Cao}}, \bibinfo {author} {\bibfnamefont {G.}~\bibnamefont {Wang}}, \bibinfo
  {author} {\bibfnamefont {W.}~\bibnamefont {Han}}, \bibinfo {author}
  {\bibfnamefont {H.}~\bibnamefont {Ye}}, \bibinfo {author} {\bibfnamefont
  {C.}~\bibnamefont {Zhu}}, \bibinfo {author} {\bibfnamefont {J.}~\bibnamefont
  {Shi}}, \bibinfo {author} {\bibfnamefont {Q.}~\bibnamefont {Niu}}, \bibinfo
  {author} {\bibfnamefont {P.}~\bibnamefont {Tan}}, \bibinfo {author}
  {\bibfnamefont {E.}~\bibnamefont {Wang}}, \bibinfo {author} {\bibfnamefont
  {B.}~\bibnamefont {Liu}},\ and\ \bibinfo {author} {\bibfnamefont
  {J.}~\bibnamefont {Feng}},\ }\bibfield  {title} {\bibinfo {title}
  {{Valley-selective circular dichroism of monolayer molybdenum disulphide}},\
  }\href {https://doi.org/10.1038/ncomms1882} {\bibfield  {journal} {\bibinfo
  {journal} {Nature Communications}\ }\textbf {\bibinfo {volume} {3}},\
  \bibinfo {pages} {887} (\bibinfo {year} {2012})}\BibitemShut {NoStop}%
\bibitem [{\citenamefont {Mak}\ \emph {et~al.}(2012)\citenamefont {Mak},
  \citenamefont {He}, \citenamefont {Shan},\ and\ \citenamefont
  {Heinz}}]{Mak_Heinz_2012}%
  \BibitemOpen
  \bibfield  {author} {\bibinfo {author} {\bibfnamefont {K.~F.}\ \bibnamefont
  {Mak}}, \bibinfo {author} {\bibfnamefont {K.}~\bibnamefont {He}}, \bibinfo
  {author} {\bibfnamefont {J.}~\bibnamefont {Shan}},\ and\ \bibinfo {author}
  {\bibfnamefont {T.~F.}\ \bibnamefont {Heinz}},\ }\bibfield  {title} {\bibinfo
  {title} {{Control of valley polarization in monolayer MoS2 by optical
  helicity}},\ }\href {https://doi.org/10.1038/nnano.2012.96} {\bibfield
  {journal} {\bibinfo  {journal} {Nature Nanotechnology}\ }\textbf {\bibinfo
  {volume} {7}},\ \bibinfo {pages} {494} (\bibinfo {year} {2012})}\BibitemShut
  {NoStop}%
\bibitem [{\citenamefont {Zeng}\ \emph {et~al.}(2012)\citenamefont {Zeng},
  \citenamefont {Dai}, \citenamefont {Yao}, \citenamefont {Xiao},\ and\
  \citenamefont {Cui}}]{Zeng_Cui_2012}%
  \BibitemOpen
  \bibfield  {author} {\bibinfo {author} {\bibfnamefont {H.}~\bibnamefont
  {Zeng}}, \bibinfo {author} {\bibfnamefont {J.}~\bibnamefont {Dai}}, \bibinfo
  {author} {\bibfnamefont {W.}~\bibnamefont {Yao}}, \bibinfo {author}
  {\bibfnamefont {D.}~\bibnamefont {Xiao}},\ and\ \bibinfo {author}
  {\bibfnamefont {X.}~\bibnamefont {Cui}},\ }\bibfield  {title} {\bibinfo
  {title} {{Valley polarization in MoS2 monolayers by optical pumping}},\
  }\href {https://doi.org/10.1038/nnano.2012.95} {\bibfield  {journal}
  {\bibinfo  {journal} {Nature Nanotechnology}\ }\textbf {\bibinfo {volume}
  {7}},\ \bibinfo {pages} {490} (\bibinfo {year} {2012})}\BibitemShut {NoStop}%
\bibitem [{\citenamefont {Wang}\ \emph
  {et~al.}(2018{\natexlab{a}})\citenamefont {Wang}, \citenamefont {Chernikov},
  \citenamefont {Glazov}, \citenamefont {Heinz}, \citenamefont {Marie},
  \citenamefont {Amand},\ and\ \citenamefont {Urbaszek}}]{Wang_Urbaszek_2018}%
  \BibitemOpen
  \bibfield  {author} {\bibinfo {author} {\bibfnamefont {G.}~\bibnamefont
  {Wang}}, \bibinfo {author} {\bibfnamefont {A.}~\bibnamefont {Chernikov}},
  \bibinfo {author} {\bibfnamefont {M.~M.}\ \bibnamefont {Glazov}}, \bibinfo
  {author} {\bibfnamefont {T.~F.}\ \bibnamefont {Heinz}}, \bibinfo {author}
  {\bibfnamefont {X.}~\bibnamefont {Marie}}, \bibinfo {author} {\bibfnamefont
  {T.}~\bibnamefont {Amand}},\ and\ \bibinfo {author} {\bibfnamefont
  {B.}~\bibnamefont {Urbaszek}},\ }\bibfield  {title} {\bibinfo {title}
  {{Colloquium: Excitons in atomically thin transition metal
  dichalcogenides}},\ }\href {https://doi.org/10.1103/RevModPhys.90.021001}
  {\bibfield  {journal} {\bibinfo  {journal} {Rev. Mod. Phys.}\ }\textbf
  {\bibinfo {volume} {90}},\ \bibinfo {pages} {021001} (\bibinfo {year}
  {2018}{\natexlab{a}})}\BibitemShut {NoStop}%
\bibitem [{\citenamefont {Baimuratov}\ and\ \citenamefont
  {H{\"o}gele}(2020)}]{baimuratov}%
  \BibitemOpen
  \bibfield  {author} {\bibinfo {author} {\bibfnamefont {A.~S.}\ \bibnamefont
  {Baimuratov}}\ and\ \bibinfo {author} {\bibfnamefont {A.}~\bibnamefont
  {H{\"o}gele}},\ }\bibfield  {title} {\bibinfo {title} {Valley-selective
  energy transfer between quantum dots in atomically thin semiconductors},\
  }\href@noop {} {\bibfield  {journal} {\bibinfo  {journal} {Scientific
  Reports}\ }\textbf {\bibinfo {volume} {10}},\ \bibinfo {pages} {1} (\bibinfo
  {year} {2020})}\BibitemShut {NoStop}%
\bibitem [{\citenamefont {Jones}\ \emph {et~al.}(2013)\citenamefont {Jones},
  \citenamefont {Yu}, \citenamefont {Ghimire}, \citenamefont {Wu},
  \citenamefont {Aivazian}, \citenamefont {Ross}, \citenamefont {Zhao},
  \citenamefont {Yan}, \citenamefont {Mandrus}, \citenamefont {Xiao},
  \citenamefont {Yao},\ and\ \citenamefont {Xu}}]{Jones_Xu_2013}%
  \BibitemOpen
  \bibfield  {author} {\bibinfo {author} {\bibfnamefont {A.~M.}\ \bibnamefont
  {Jones}}, \bibinfo {author} {\bibfnamefont {H.}~\bibnamefont {Yu}}, \bibinfo
  {author} {\bibfnamefont {N.~J.}\ \bibnamefont {Ghimire}}, \bibinfo {author}
  {\bibfnamefont {S.}~\bibnamefont {Wu}}, \bibinfo {author} {\bibfnamefont
  {G.}~\bibnamefont {Aivazian}}, \bibinfo {author} {\bibfnamefont {J.~S.}\
  \bibnamefont {Ross}}, \bibinfo {author} {\bibfnamefont {B.}~\bibnamefont
  {Zhao}}, \bibinfo {author} {\bibfnamefont {J.}~\bibnamefont {Yan}}, \bibinfo
  {author} {\bibfnamefont {D.~G.}\ \bibnamefont {Mandrus}}, \bibinfo {author}
  {\bibfnamefont {D.}~\bibnamefont {Xiao}}, \bibinfo {author} {\bibfnamefont
  {W.}~\bibnamefont {Yao}},\ and\ \bibinfo {author} {\bibfnamefont
  {X.}~\bibnamefont {Xu}},\ }\bibfield  {title} {\bibinfo {title} {{Optical
  generation of excitonic valley coherence in monolayer WSe2}},\ }\href
  {https://doi.org/10.1038/nnano.2013.151} {\bibfield  {journal} {\bibinfo
  {journal} {Nature Nanotechnology}\ }\textbf {\bibinfo {volume} {8}},\
  \bibinfo {pages} {634} (\bibinfo {year} {2013})}\BibitemShut {NoStop}%
\bibitem [{\citenamefont {Wang}\ \emph {et~al.}(2016)\citenamefont {Wang},
  \citenamefont {Marie}, \citenamefont {Liu}, \citenamefont {Amand},
  \citenamefont {Robert}, \citenamefont {Cadiz}, \citenamefont {Renucci},\ and\
  \citenamefont {Urbaszek}}]{Wang_Urbaszek_2016}%
  \BibitemOpen
  \bibfield  {author} {\bibinfo {author} {\bibfnamefont {G.}~\bibnamefont
  {Wang}}, \bibinfo {author} {\bibfnamefont {X.}~\bibnamefont {Marie}},
  \bibinfo {author} {\bibfnamefont {B.~L.}\ \bibnamefont {Liu}}, \bibinfo
  {author} {\bibfnamefont {T.}~\bibnamefont {Amand}}, \bibinfo {author}
  {\bibfnamefont {C.}~\bibnamefont {Robert}}, \bibinfo {author} {\bibfnamefont
  {F.}~\bibnamefont {Cadiz}}, \bibinfo {author} {\bibfnamefont
  {P.}~\bibnamefont {Renucci}},\ and\ \bibinfo {author} {\bibfnamefont
  {B.}~\bibnamefont {Urbaszek}},\ }\bibfield  {title} {\bibinfo {title}
  {{Control of Exciton Valley Coherence in Transition Metal Dichalcogenide
  Monolayers}},\ }\href {https://doi.org/10.1103/PhysRevLett.117.187401}
  {\bibfield  {journal} {\bibinfo  {journal} {Phys. Rev. Lett.}\ }\textbf
  {\bibinfo {volume} {117}},\ \bibinfo {pages} {187401} (\bibinfo {year}
  {2016})}\BibitemShut {NoStop}%
\bibitem [{\citenamefont {Hao}\ \emph {et~al.}(2016)\citenamefont {Hao},
  \citenamefont {Moody}, \citenamefont {Wu}, \citenamefont {Dass},
  \citenamefont {Xu}, \citenamefont {Chen}, \citenamefont {Sun}, \citenamefont
  {Li}, \citenamefont {Li}, \citenamefont {MacDonald},\ and\ \citenamefont
  {Li}}]{Hao_Li_2016}%
  \BibitemOpen
  \bibfield  {author} {\bibinfo {author} {\bibfnamefont {K.}~\bibnamefont
  {Hao}}, \bibinfo {author} {\bibfnamefont {G.}~\bibnamefont {Moody}}, \bibinfo
  {author} {\bibfnamefont {F.}~\bibnamefont {Wu}}, \bibinfo {author}
  {\bibfnamefont {C.~K.}\ \bibnamefont {Dass}}, \bibinfo {author}
  {\bibfnamefont {L.}~\bibnamefont {Xu}}, \bibinfo {author} {\bibfnamefont
  {C.-H.}\ \bibnamefont {Chen}}, \bibinfo {author} {\bibfnamefont
  {L.}~\bibnamefont {Sun}}, \bibinfo {author} {\bibfnamefont {M.-Y.}\
  \bibnamefont {Li}}, \bibinfo {author} {\bibfnamefont {L.-J.}\ \bibnamefont
  {Li}}, \bibinfo {author} {\bibfnamefont {A.~H.}\ \bibnamefont {MacDonald}},\
  and\ \bibinfo {author} {\bibfnamefont {X.}~\bibnamefont {Li}},\ }\bibfield
  {title} {\bibinfo {title} {{Direct measurement of exciton valley coherence in
  monolayer WSe2}},\ }\href {https://doi.org/10.1038/nphys3674} {\bibfield
  {journal} {\bibinfo  {journal} {Nature Physics}\ }\textbf {\bibinfo {volume}
  {12}},\ \bibinfo {pages} {677} (\bibinfo {year} {2016})}\BibitemShut
  {NoStop}%
\bibitem [{\citenamefont {Srivastava}\ \emph {et~al.}(2015)\citenamefont
  {Srivastava}, \citenamefont {Sidler}, \citenamefont {Allain}, \citenamefont
  {Lembke}, \citenamefont {Kis},\ and\ \citenamefont
  {Imamoglu}}]{Srivastava_Imamoglu_2015b}%
  \BibitemOpen
  \bibfield  {author} {\bibinfo {author} {\bibfnamefont {A.}~\bibnamefont
  {Srivastava}}, \bibinfo {author} {\bibfnamefont {M.}~\bibnamefont {Sidler}},
  \bibinfo {author} {\bibfnamefont {A.~V.}\ \bibnamefont {Allain}}, \bibinfo
  {author} {\bibfnamefont {D.~S.}\ \bibnamefont {Lembke}}, \bibinfo {author}
  {\bibfnamefont {A.}~\bibnamefont {Kis}},\ and\ \bibinfo {author}
  {\bibfnamefont {A.}~\bibnamefont {Imamoglu}},\ }\bibfield  {title} {\bibinfo
  {title} {{Valley Zeeman effect in elementary optical excitations of monolayer
  WSe2}},\ }\href {https://doi.org/10.1038/nphys3203} {\bibfield  {journal}
  {\bibinfo  {journal} {Nature Physics}\ }\textbf {\bibinfo {volume} {11}},\
  \bibinfo {pages} {141} (\bibinfo {year} {2015})}\BibitemShut {NoStop}%
\bibitem [{\citenamefont {Aivazian}\ \emph {et~al.}(2015)\citenamefont
  {Aivazian}, \citenamefont {Gong}, \citenamefont {Jones}, \citenamefont {Chu},
  \citenamefont {Yan}, \citenamefont {Mandrus}, \citenamefont {Zhang},
  \citenamefont {Cobden}, \citenamefont {Yao},\ and\ \citenamefont
  {Xu}}]{Aivazian_Xu_2015}%
  \BibitemOpen
  \bibfield  {author} {\bibinfo {author} {\bibfnamefont {G.}~\bibnamefont
  {Aivazian}}, \bibinfo {author} {\bibfnamefont {Z.}~\bibnamefont {Gong}},
  \bibinfo {author} {\bibfnamefont {A.~M.}\ \bibnamefont {Jones}}, \bibinfo
  {author} {\bibfnamefont {R.-L.}\ \bibnamefont {Chu}}, \bibinfo {author}
  {\bibfnamefont {J.}~\bibnamefont {Yan}}, \bibinfo {author} {\bibfnamefont
  {D.~G.}\ \bibnamefont {Mandrus}}, \bibinfo {author} {\bibfnamefont
  {C.}~\bibnamefont {Zhang}}, \bibinfo {author} {\bibfnamefont
  {D.}~\bibnamefont {Cobden}}, \bibinfo {author} {\bibfnamefont
  {W.}~\bibnamefont {Yao}},\ and\ \bibinfo {author} {\bibfnamefont
  {X.}~\bibnamefont {Xu}},\ }\bibfield  {title} {\bibinfo {title} {{Magnetic
  control of valley pseudospin in monolayer WSe2}},\ }\href
  {https://doi.org/10.1038/nphys3201} {\bibfield  {journal} {\bibinfo
  {journal} {Nature Physics}\ }\textbf {\bibinfo {volume} {11}},\ \bibinfo
  {pages} {148} (\bibinfo {year} {2015})}\BibitemShut {NoStop}%
\bibitem [{\citenamefont {Scrace}\ \emph {et~al.}(2015)\citenamefont {Scrace},
  \citenamefont {Tsai}, \citenamefont {Barman}, \citenamefont {Schweidenback},
  \citenamefont {Petrou}, \citenamefont {Kioseoglou}, \citenamefont {Ozfidan},
  \citenamefont {Korkusi\'nski},\ and\ \citenamefont
  {Hawrylak}}]{Scrace_Hawrylak_2015}%
  \BibitemOpen
  \bibfield  {author} {\bibinfo {author} {\bibfnamefont {T.}~\bibnamefont
  {Scrace}}, \bibinfo {author} {\bibfnamefont {Y.}~\bibnamefont {Tsai}},
  \bibinfo {author} {\bibfnamefont {B.}~\bibnamefont {Barman}}, \bibinfo
  {author} {\bibfnamefont {L.}~\bibnamefont {Schweidenback}}, \bibinfo {author}
  {\bibfnamefont {A.}~\bibnamefont {Petrou}}, \bibinfo {author} {\bibfnamefont
  {G.}~\bibnamefont {Kioseoglou}}, \bibinfo {author} {\bibfnamefont
  {I.}~\bibnamefont {Ozfidan}}, \bibinfo {author} {\bibfnamefont
  {M.}~\bibnamefont {Korkusi\'nski}},\ and\ \bibinfo {author} {\bibfnamefont
  {P.}~\bibnamefont {Hawrylak}},\ }\bibfield  {title} {\bibinfo {title}
  {{Magnetoluminescence and valley polarized state of a two-dimensional
  electron gas in WS2 monolayers}},\ }\href
  {https://doi.org/10.1038/nnano.2015.78} {\bibfield  {journal} {\bibinfo
  {journal} {Nat. Nano.}\ }\textbf {\bibinfo {volume} {10}},\ \bibinfo {pages}
  {603} (\bibinfo {year} {2015})}\BibitemShut {NoStop}%
\bibitem [{\citenamefont {Braz}\ \emph {et~al.}(2018)\citenamefont {Braz},
  \citenamefont {Amorim},\ and\ \citenamefont {Castro}}]{Braz_Castro_2018}%
  \BibitemOpen
  \bibfield  {author} {\bibinfo {author} {\bibfnamefont {J.~E.~H.}\
  \bibnamefont {Braz}}, \bibinfo {author} {\bibfnamefont {B.}~\bibnamefont
  {Amorim}},\ and\ \bibinfo {author} {\bibfnamefont {E.~V.}\ \bibnamefont
  {Castro}},\ }\bibfield  {title} {\bibinfo {title} {{Valley-polarized magnetic
  state in hole-doped monolayers of transition-metal dichalcogenides}},\ }\href
  {https://doi.org/10.1103/PhysRevB.98.161406} {\bibfield  {journal} {\bibinfo
  {journal} {Phys. Rev. B}\ }\textbf {\bibinfo {volume} {98}},\ \bibinfo
  {pages} {161406} (\bibinfo {year} {2018})}\BibitemShut {NoStop}%
\bibitem [{\citenamefont {Miserev}\ \emph {et~al.}(2019)\citenamefont
  {Miserev}, \citenamefont {Klinovaja},\ and\ \citenamefont
  {Loss}}]{Miserev_Loss_2019}%
  \BibitemOpen
  \bibfield  {author} {\bibinfo {author} {\bibfnamefont {D.}~\bibnamefont
  {Miserev}}, \bibinfo {author} {\bibfnamefont {J.}~\bibnamefont {Klinovaja}},\
  and\ \bibinfo {author} {\bibfnamefont {D.}~\bibnamefont {Loss}},\ }\bibfield
  {title} {\bibinfo {title} {{Exchange intervalley scattering and magnetic
  phase diagram of transition metal dichalcogenide monolayers}},\ }\href
  {https://doi.org/10.1103/PhysRevB.100.014428} {\bibfield  {journal} {\bibinfo
   {journal} {Phys. Rev. B}\ }\textbf {\bibinfo {volume} {100}},\ \bibinfo
  {pages} {014428} (\bibinfo {year} {2019})}\BibitemShut {NoStop}%
\bibitem [{\citenamefont {Mak}\ \emph {et~al.}(2013)\citenamefont {Mak},
  \citenamefont {He}, \citenamefont {Lee}, \citenamefont {Lee}, \citenamefont
  {Hone}, \citenamefont {Heinz},\ and\ \citenamefont {Shan}}]{Mak_Shan_2013}%
  \BibitemOpen
  \bibfield  {author} {\bibinfo {author} {\bibfnamefont {K.~F.}\ \bibnamefont
  {Mak}}, \bibinfo {author} {\bibfnamefont {K.}~\bibnamefont {He}}, \bibinfo
  {author} {\bibfnamefont {C.}~\bibnamefont {Lee}}, \bibinfo {author}
  {\bibfnamefont {G.~H.}\ \bibnamefont {Lee}}, \bibinfo {author} {\bibfnamefont
  {J.}~\bibnamefont {Hone}}, \bibinfo {author} {\bibfnamefont {T.~F.}\
  \bibnamefont {Heinz}},\ and\ \bibinfo {author} {\bibfnamefont
  {J.}~\bibnamefont {Shan}},\ }\bibfield  {title} {\bibinfo {title} {{Tightly
  bound trions in monolayer MoS2}},\ }\href {https://doi.org/10.1038/nmat3505}
  {\bibfield  {journal} {\bibinfo  {journal} {Nature Materials}\ }\textbf
  {\bibinfo {volume} {12}},\ \bibinfo {pages} {207} (\bibinfo {year}
  {2013})}\BibitemShut {NoStop}%
\bibitem [{\citenamefont {Jadczak}\ \emph {et~al.}(2017)\citenamefont
  {Jadczak}, \citenamefont {Kutrowska-Girzycka}, \citenamefont
  {Kapu{\'{s}}ci{\'{n}}ski}, \citenamefont {Huang}, \citenamefont
  {W{\'{o}}js},\ and\ \citenamefont {Bryja}}]{Jadczak_Bryja_2017}%
  \BibitemOpen
  \bibfield  {author} {\bibinfo {author} {\bibfnamefont {J.}~\bibnamefont
  {Jadczak}}, \bibinfo {author} {\bibfnamefont {J.}~\bibnamefont
  {Kutrowska-Girzycka}}, \bibinfo {author} {\bibfnamefont {P.}~\bibnamefont
  {Kapu{\'{s}}ci{\'{n}}ski}}, \bibinfo {author} {\bibfnamefont {Y.~S.}\
  \bibnamefont {Huang}}, \bibinfo {author} {\bibfnamefont {A.}~\bibnamefont
  {W{\'{o}}js}},\ and\ \bibinfo {author} {\bibfnamefont {L.}~\bibnamefont
  {Bryja}},\ }\bibfield  {title} {\bibinfo {title} {{Probing of free and
  localized excitons and trions in atomically thin {WSe}2, {WS}2, {MoSe}2 and
  {MoS}2 in photoluminescence and reflectivity experiments}},\ }\href
  {https://doi.org/10.1088/1361-6528/aa87d0} {\bibfield  {journal} {\bibinfo
  {journal} {Nanotechnology}\ }\textbf {\bibinfo {volume} {28}},\ \bibinfo
  {pages} {395702} (\bibinfo {year} {2017})}\BibitemShut {NoStop}%
\bibitem [{\citenamefont {Back}\ \emph {et~al.}(2017)\citenamefont {Back},
  \citenamefont {Sidler}, \citenamefont {Cotlet}, \citenamefont {Srivastava},
  \citenamefont {Takemura}, \citenamefont {Kroner},\ and\ \citenamefont
  {Imamo\ifmmode~\breve{g}\else \u{g}\fi{}lu}}]{Back_Imamoglu_2017}%
  \BibitemOpen
  \bibfield  {author} {\bibinfo {author} {\bibfnamefont {P.}~\bibnamefont
  {Back}}, \bibinfo {author} {\bibfnamefont {M.}~\bibnamefont {Sidler}},
  \bibinfo {author} {\bibfnamefont {O.}~\bibnamefont {Cotlet}}, \bibinfo
  {author} {\bibfnamefont {A.}~\bibnamefont {Srivastava}}, \bibinfo {author}
  {\bibfnamefont {N.}~\bibnamefont {Takemura}}, \bibinfo {author}
  {\bibfnamefont {M.}~\bibnamefont {Kroner}},\ and\ \bibinfo {author}
  {\bibfnamefont {A.}~\bibnamefont {Imamo\ifmmode~\breve{g}\else
  \u{g}\fi{}lu}},\ }\bibfield  {title} {\bibinfo {title} {{Giant
  Paramagnetism-Induced Valley Polarization of Electrons in Charge-Tunable
  Monolayer ${\mathrm{MoSe}}_{2}$}},\ }\href
  {https://doi.org/10.1103/PhysRevLett.118.237404} {\bibfield  {journal}
  {\bibinfo  {journal} {Phys. Rev. Lett.}\ }\textbf {\bibinfo {volume} {118}},\
  \bibinfo {pages} {237404} (\bibinfo {year} {2017})}\BibitemShut {NoStop}%
\bibitem [{\citenamefont {Roch}\ \emph {et~al.}(2019)\citenamefont {Roch},
  \citenamefont {Froehlicher}, \citenamefont {Leisgang}, \citenamefont {Makk},
  \citenamefont {Watanabe}, \citenamefont {Taniguchi},\ and\ \citenamefont
  {Warburton}}]{Roch_Warburton_2019}%
  \BibitemOpen
  \bibfield  {author} {\bibinfo {author} {\bibfnamefont {J.~G.}\ \bibnamefont
  {Roch}}, \bibinfo {author} {\bibfnamefont {G.}~\bibnamefont {Froehlicher}},
  \bibinfo {author} {\bibfnamefont {N.}~\bibnamefont {Leisgang}}, \bibinfo
  {author} {\bibfnamefont {P.}~\bibnamefont {Makk}}, \bibinfo {author}
  {\bibfnamefont {K.}~\bibnamefont {Watanabe}}, \bibinfo {author}
  {\bibfnamefont {T.}~\bibnamefont {Taniguchi}},\ and\ \bibinfo {author}
  {\bibfnamefont {R.~J.}\ \bibnamefont {Warburton}},\ }\bibfield  {title}
  {\bibinfo {title} {{Spin-polarized electrons in monolayer
  {MoS}$_{\textrm{2}}$}},\ }\href {https://doi.org/10.1038/s41565-019-0397-y}
  {\bibfield  {journal} {\bibinfo  {journal} {Nature Nanotechnology}\ }\textbf
  {\bibinfo {volume} {14}},\ \bibinfo {pages} {432} (\bibinfo {year}
  {2019})}\BibitemShut {NoStop}%
\bibitem [{\citenamefont {Jadczak}\ \emph {et~al.}(2019)\citenamefont
  {Jadczak}, \citenamefont {Bryja}, \citenamefont {Kutrowska-Girzycka},
  \citenamefont {Kapuscinski}, \citenamefont {Bieniek}, \citenamefont {Huang},\
  and\ \citenamefont {Hawrylak}}]{Jadczak_Hawrylak_2019}%
  \BibitemOpen
  \bibfield  {author} {\bibinfo {author} {\bibfnamefont {J.}~\bibnamefont
  {Jadczak}}, \bibinfo {author} {\bibfnamefont {L.}~\bibnamefont {Bryja}},
  \bibinfo {author} {\bibfnamefont {J.}~\bibnamefont {Kutrowska-Girzycka}},
  \bibinfo {author} {\bibfnamefont {P.}~\bibnamefont {Kapuscinski}}, \bibinfo
  {author} {\bibfnamefont {M.}~\bibnamefont {Bieniek}}, \bibinfo {author}
  {\bibfnamefont {Y.-S.}\ \bibnamefont {Huang}},\ and\ \bibinfo {author}
  {\bibfnamefont {P.}~\bibnamefont {Hawrylak}},\ }\bibfield  {title} {\bibinfo
  {title} {{Room temperature multi-phonon upconversion photoluminescence in
  monolayer semiconductor {WS}$_{\textrm{2}}$}},\ }\href
  {https://doi.org/10.1038/s41467-018-07994-1} {\bibfield  {journal} {\bibinfo
  {journal} {Nature Communications}\ }\textbf {\bibinfo {volume} {10}},\
  \bibinfo {pages} {107} (\bibinfo {year} {2019})}\BibitemShut {NoStop}%
\bibitem [{\citenamefont {Jadczak}\ \emph {et~al.}(2020)\citenamefont
  {Jadczak}, \citenamefont {Kutrowska-Girzycka}, \citenamefont {Bieniek},
  \citenamefont {Kazimierczuk}, \citenamefont {Kossacki}, \citenamefont
  {Schindler}, \citenamefont {Debus}, \citenamefont {Watanabe}, \citenamefont
  {Taniguchi}, \citenamefont {Ho}, \citenamefont {Wójs}, \citenamefont
  {Hawrylak},\ and\ \citenamefont {Bryja}}]{Jadczak_Bryja_2020}%
  \BibitemOpen
  \bibfield  {author} {\bibinfo {author} {\bibfnamefont {J.}~\bibnamefont
  {Jadczak}}, \bibinfo {author} {\bibfnamefont {J.~J.}\ \bibnamefont
  {Kutrowska-Girzycka}}, \bibinfo {author} {\bibfnamefont {M.}~\bibnamefont
  {Bieniek}}, \bibinfo {author} {\bibfnamefont {T.}~\bibnamefont
  {Kazimierczuk}}, \bibinfo {author} {\bibfnamefont {P.}~\bibnamefont
  {Kossacki}}, \bibinfo {author} {\bibfnamefont {J.~J.~J.}\ \bibnamefont
  {Schindler}}, \bibinfo {author} {\bibfnamefont {J.}~\bibnamefont {Debus}},
  \bibinfo {author} {\bibfnamefont {K.}~\bibnamefont {Watanabe}}, \bibinfo
  {author} {\bibfnamefont {T.}~\bibnamefont {Taniguchi}}, \bibinfo {author}
  {\bibfnamefont {C.-H.}\ \bibnamefont {Ho}}, \bibinfo {author} {\bibfnamefont
  {A.}~\bibnamefont {Wójs}}, \bibinfo {author} {\bibfnamefont
  {P.}~\bibnamefont {Hawrylak}},\ and\ \bibinfo {author} {\bibfnamefont
  {L.}~\bibnamefont {Bryja}},\ }\bibfield  {title} {\bibinfo {title} {Probing
  of negatively charged and neutral excitons in mos2/hbn and hbn/mos2/hbn van
  der waals heterostructures},\ }\href
  {http://iopscience.iop.org/article/10.1088/1361-6528/abd507} {\bibfield
  {journal} {\bibinfo  {journal} {Nanotechnology}\ } (\bibinfo {year}
  {2020})}\BibitemShut {NoStop}%
\bibitem [{\citenamefont {Wang}\ \emph
  {et~al.}(2018{\natexlab{b}})\citenamefont {Wang}, \citenamefont {De~Greve},
  \citenamefont {Jauregui}, \citenamefont {Sushko}, \citenamefont {High},
  \citenamefont {Zhou}, \citenamefont {Scuri}, \citenamefont {Taniguchi},
  \citenamefont {Watanabe}, \citenamefont {Lukin}, \citenamefont {Park},\ and\
  \citenamefont {Kim}}]{Wang_Kim_2018}%
  \BibitemOpen
  \bibfield  {author} {\bibinfo {author} {\bibfnamefont {K.}~\bibnamefont
  {Wang}}, \bibinfo {author} {\bibfnamefont {K.}~\bibnamefont {De~Greve}},
  \bibinfo {author} {\bibfnamefont {L.~A.}\ \bibnamefont {Jauregui}}, \bibinfo
  {author} {\bibfnamefont {A.}~\bibnamefont {Sushko}}, \bibinfo {author}
  {\bibfnamefont {A.}~\bibnamefont {High}}, \bibinfo {author} {\bibfnamefont
  {Y.}~\bibnamefont {Zhou}}, \bibinfo {author} {\bibfnamefont {G.}~\bibnamefont
  {Scuri}}, \bibinfo {author} {\bibfnamefont {T.}~\bibnamefont {Taniguchi}},
  \bibinfo {author} {\bibfnamefont {K.}~\bibnamefont {Watanabe}}, \bibinfo
  {author} {\bibfnamefont {M.~D.}\ \bibnamefont {Lukin}}, \bibinfo {author}
  {\bibfnamefont {H.}~\bibnamefont {Park}},\ and\ \bibinfo {author}
  {\bibfnamefont {P.}~\bibnamefont {Kim}},\ }\bibfield  {title} {\bibinfo
  {title} {Electrical control of charged carriers and excitons in atomically
  thin materials},\ }\href {https://doi.org/10.1038/s41565-017-0030-x}
  {\bibfield  {journal} {\bibinfo  {journal} {Nature Nanotechnology}\ }\textbf
  {\bibinfo {volume} {13}},\ \bibinfo {pages} {128} (\bibinfo {year}
  {2018}{\natexlab{b}})}\BibitemShut {NoStop}%
\bibitem [{\citenamefont {Brotons-Gisbert}\ \emph {et~al.}(2019)\citenamefont
  {Brotons-Gisbert}, \citenamefont {Branny}, \citenamefont {Kumar},
  \citenamefont {Picard}, \citenamefont {Proux}, \citenamefont {Gray},
  \citenamefont {Burch}, \citenamefont {Watanabe}, \citenamefont {Taniguchi},\
  and\ \citenamefont {Gerardot}}]{BrotonsGisbert_Gerardot_2019}%
  \BibitemOpen
  \bibfield  {author} {\bibinfo {author} {\bibfnamefont {M.}~\bibnamefont
  {Brotons-Gisbert}}, \bibinfo {author} {\bibfnamefont {A.}~\bibnamefont
  {Branny}}, \bibinfo {author} {\bibfnamefont {S.}~\bibnamefont {Kumar}},
  \bibinfo {author} {\bibfnamefont {R.}~\bibnamefont {Picard}}, \bibinfo
  {author} {\bibfnamefont {R.}~\bibnamefont {Proux}}, \bibinfo {author}
  {\bibfnamefont {M.}~\bibnamefont {Gray}}, \bibinfo {author} {\bibfnamefont
  {K.~S.}\ \bibnamefont {Burch}}, \bibinfo {author} {\bibfnamefont
  {K.}~\bibnamefont {Watanabe}}, \bibinfo {author} {\bibfnamefont
  {T.}~\bibnamefont {Taniguchi}},\ and\ \bibinfo {author} {\bibfnamefont
  {B.~D.}\ \bibnamefont {Gerardot}},\ }\bibfield  {title} {\bibinfo {title}
  {Coulomb blockade in an atomically thin quantum dot coupled to a tunable
  {Fermi} reservoir},\ }\href {https://doi.org/10.1038/s41565-019-0402-5}
  {\bibfield  {journal} {\bibinfo  {journal} {Nature Nanotechnology}\ }\textbf
  {\bibinfo {volume} {14}},\ \bibinfo {pages} {442} (\bibinfo {year}
  {2019})}\BibitemShut {NoStop}%
\bibitem [{\citenamefont {Klinovaja}\ and\ \citenamefont
  {Loss}(2013)}]{klinovaja}%
  \BibitemOpen
  \bibfield  {author} {\bibinfo {author} {\bibfnamefont {J.}~\bibnamefont
  {Klinovaja}}\ and\ \bibinfo {author} {\bibfnamefont {D.}~\bibnamefont
  {Loss}},\ }\bibfield  {title} {\bibinfo {title} {Spintronics in mos${}_{2}$
  monolayer quantum wires},\ }\href
  {https://doi.org/10.1103/PhysRevB.88.075404} {\bibfield  {journal} {\bibinfo
  {journal} {Phys. Rev. B}\ }\textbf {\bibinfo {volume} {88}},\ \bibinfo
  {pages} {075404} (\bibinfo {year} {2013})}\BibitemShut {NoStop}%
\bibitem [{\citenamefont {Korm\'anyos}\ \emph
  {et~al.}(2014{\natexlab{b}})\citenamefont {Korm\'anyos}, \citenamefont
  {Z\'olyomi}, \citenamefont {Drummond},\ and\ \citenamefont
  {Burkard}}]{Kormanyos_Burkard_2014}%
  \BibitemOpen
  \bibfield  {author} {\bibinfo {author} {\bibfnamefont {A.}~\bibnamefont
  {Korm\'anyos}}, \bibinfo {author} {\bibfnamefont {V.}~\bibnamefont
  {Z\'olyomi}}, \bibinfo {author} {\bibfnamefont {N.~D.}\ \bibnamefont
  {Drummond}},\ and\ \bibinfo {author} {\bibfnamefont {G.}~\bibnamefont
  {Burkard}},\ }\bibfield  {title} {\bibinfo {title} {{Spin-Orbit Coupling,
  Quantum Dots, and Qubits in Monolayer Transition Metal Dichalcogenides}},\
  }\href {https://doi.org/10.1103/PhysRevX.4.011034} {\bibfield  {journal}
  {\bibinfo  {journal} {Phys. Rev. X}\ }\textbf {\bibinfo {volume} {4}},\
  \bibinfo {pages} {011034} (\bibinfo {year} {2014}{\natexlab{b}})}\BibitemShut
  {NoStop}%
\bibitem [{\citenamefont {Liu}\ \emph {et~al.}(2014{\natexlab{a}})\citenamefont
  {Liu}, \citenamefont {Pang}, \citenamefont {Yao},\ and\ \citenamefont
  {Yao}}]{Liu_Yao_2014}%
  \BibitemOpen
  \bibfield  {author} {\bibinfo {author} {\bibfnamefont {G.-B.}\ \bibnamefont
  {Liu}}, \bibinfo {author} {\bibfnamefont {H.}~\bibnamefont {Pang}}, \bibinfo
  {author} {\bibfnamefont {Y.}~\bibnamefont {Yao}},\ and\ \bibinfo {author}
  {\bibfnamefont {W.}~\bibnamefont {Yao}},\ }\bibfield  {title} {\bibinfo
  {title} {Intervalley coupling by quantum dot confinement potentials in
  monolayer transition metal dichalcogenides},\ }\href
  {https://doi.org/10.1088/1367-2630/16/10/105011} {\bibfield  {journal}
  {\bibinfo  {journal} {New Journal of Physics}\ }\textbf {\bibinfo {volume}
  {16}},\ \bibinfo {pages} {105011} (\bibinfo {year}
  {2014}{\natexlab{a}})}\BibitemShut {NoStop}%
\bibitem [{\citenamefont {Pavlovi{\'c}}\ and\ \citenamefont
  {Peeters}(2015)}]{Pavlovic_Peeters_2015}%
  \BibitemOpen
  \bibfield  {author} {\bibinfo {author} {\bibfnamefont {S.}~\bibnamefont
  {Pavlovi{\'c}}}\ and\ \bibinfo {author} {\bibfnamefont {F.~M.}\ \bibnamefont
  {Peeters}},\ }\bibfield  {title} {\bibinfo {title} {Electronic properties of
  triangular and hexagonal {MoS}$_2$ quantum dots},\ }\href
  {https://doi.org/10.1103/PhysRevB.91.155410} {\bibfield  {journal} {\bibinfo
  {journal} {Physical Review B}\ }\textbf {\bibinfo {volume} {91}},\ \bibinfo
  {pages} {155410} (\bibinfo {year} {2015})}\BibitemShut {NoStop}%
\bibitem [{\citenamefont {Wu}\ \emph {et~al.}(2016{\natexlab{b}})\citenamefont
  {Wu}, \citenamefont {Tong}, \citenamefont {Liu}, \citenamefont {Yu},\ and\
  \citenamefont {Yao}}]{Wu_Yao_2016}%
  \BibitemOpen
  \bibfield  {author} {\bibinfo {author} {\bibfnamefont {Y.}~\bibnamefont
  {Wu}}, \bibinfo {author} {\bibfnamefont {Q.}~\bibnamefont {Tong}}, \bibinfo
  {author} {\bibfnamefont {G.-B.}\ \bibnamefont {Liu}}, \bibinfo {author}
  {\bibfnamefont {H.}~\bibnamefont {Yu}},\ and\ \bibinfo {author}
  {\bibfnamefont {W.}~\bibnamefont {Yao}},\ }\bibfield  {title} {\bibinfo
  {title} {Spin-valley qubit in nanostructures of monolayer semiconductors:
  Optical control and hyperfine interaction},\ }\href
  {https://doi.org/10.1103/PhysRevB.93.045313} {\bibfield  {journal} {\bibinfo
  {journal} {Phys. Rev. B}\ }\textbf {\bibinfo {volume} {93}},\ \bibinfo
  {pages} {045313} (\bibinfo {year} {2016}{\natexlab{b}})}\BibitemShut
  {NoStop}%
\bibitem [{\citenamefont {Dias}\ \emph {et~al.}(2016)\citenamefont {Dias},
  \citenamefont {Fu}, \citenamefont {Villegas-Lelovsky},\ and\ \citenamefont
  {Qu}}]{Dias_Qu_2016}%
  \BibitemOpen
  \bibfield  {author} {\bibinfo {author} {\bibfnamefont {A.~C.}\ \bibnamefont
  {Dias}}, \bibinfo {author} {\bibfnamefont {J.}~\bibnamefont {Fu}}, \bibinfo
  {author} {\bibfnamefont {L.}~\bibnamefont {Villegas-Lelovsky}},\ and\
  \bibinfo {author} {\bibfnamefont {F.}~\bibnamefont {Qu}},\ }\bibfield
  {title} {\bibinfo {title} {Robust effective {Zeeman} energy in monolayer
  {MoS} $_{\textrm{2}}$ quantum dots},\ }\href
  {https://doi.org/10.1088/0953-8984/28/37/375803} {\bibfield  {journal}
  {\bibinfo  {journal} {Journal of Physics: Condensed Matter}\ }\textbf
  {\bibinfo {volume} {28}},\ \bibinfo {pages} {375803} (\bibinfo {year}
  {2016})}\BibitemShut {NoStop}%
\bibitem [{\citenamefont {Brooks}\ and\ \citenamefont
  {Burkard}(2017)}]{Brooks_Burkard_2017}%
  \BibitemOpen
  \bibfield  {author} {\bibinfo {author} {\bibfnamefont {M.}~\bibnamefont
  {Brooks}}\ and\ \bibinfo {author} {\bibfnamefont {G.}~\bibnamefont
  {Burkard}},\ }\bibfield  {title} {\bibinfo {title} {Spin-degenerate regimes
  for single quantum dots in transition metal dichalcogenide monolayers},\
  }\href {https://doi.org/10.1103/PhysRevB.95.245411} {\bibfield  {journal}
  {\bibinfo  {journal} {Phys. Rev. B}\ }\textbf {\bibinfo {volume} {95}},\
  \bibinfo {pages} {245411} (\bibinfo {year} {2017})}\BibitemShut {NoStop}%
\bibitem [{\citenamefont {Qu}\ \emph {et~al.}(2017)\citenamefont {Qu},
  \citenamefont {Dias}, \citenamefont {Fu}, \citenamefont {Villegas-Lelovsky},\
  and\ \citenamefont {Azevedo}}]{Qu_Azevedo_2017}%
  \BibitemOpen
  \bibfield  {author} {\bibinfo {author} {\bibfnamefont {F.}~\bibnamefont
  {Qu}}, \bibinfo {author} {\bibfnamefont {A.~C.}\ \bibnamefont {Dias}},
  \bibinfo {author} {\bibfnamefont {J.}~\bibnamefont {Fu}}, \bibinfo {author}
  {\bibfnamefont {L.}~\bibnamefont {Villegas-Lelovsky}},\ and\ \bibinfo
  {author} {\bibfnamefont {D.~L.}\ \bibnamefont {Azevedo}},\ }\bibfield
  {title} {\bibinfo {title} {Tunable spin and valley dependent magneto-optical
  absorption in molybdenum disulfide quantum dots},\ }\href
  {https://doi.org/10.1038/srep41044} {\bibfield  {journal} {\bibinfo
  {journal} {Scientific Reports}\ }\textbf {\bibinfo {volume} {7}},\ \bibinfo
  {pages} {41044} (\bibinfo {year} {2017})}\BibitemShut {NoStop}%
\bibitem [{\citenamefont {Sz{\'e}chenyi}\ \emph {et~al.}(2018)\citenamefont
  {Sz{\'e}chenyi}, \citenamefont {Chirolli},\ and\ \citenamefont
  {P{\'a}lyi}}]{Szechenyi_Palyi_2018}%
  \BibitemOpen
  \bibfield  {author} {\bibinfo {author} {\bibfnamefont {G.}~\bibnamefont
  {Sz{\'e}chenyi}}, \bibinfo {author} {\bibfnamefont {L.}~\bibnamefont
  {Chirolli}},\ and\ \bibinfo {author} {\bibfnamefont {A.}~\bibnamefont
  {P{\'a}lyi}},\ }\bibfield  {title} {\bibinfo {title} {Impurity-assisted
  electric control of spin-valley qubits in monolayer {MoS} $_{\textrm{2}}$},\
  }\href {https://doi.org/10.1088/2053-1583/aab80e} {\bibfield  {journal}
  {\bibinfo  {journal} {2D Materials}\ }\textbf {\bibinfo {volume} {5}},\
  \bibinfo {pages} {035004} (\bibinfo {year} {2018})}\BibitemShut {NoStop}%
\bibitem [{\citenamefont {David}\ \emph {et~al.}(2018)\citenamefont {David},
  \citenamefont {Burkard},\ and\ \citenamefont
  {Korm{\'{a}}nyos}}]{David_Kormanyos_2018}%
  \BibitemOpen
  \bibfield  {author} {\bibinfo {author} {\bibfnamefont {A.}~\bibnamefont
  {David}}, \bibinfo {author} {\bibfnamefont {G.}~\bibnamefont {Burkard}},\
  and\ \bibinfo {author} {\bibfnamefont {A.}~\bibnamefont {Korm{\'{a}}nyos}},\
  }\bibfield  {title} {\bibinfo {title} {{Effective theory of monolayer {TMDC}
  double quantum dots}},\ }\href {https://doi.org/10.1088/2053-1583/aac17f}
  {\bibfield  {journal} {\bibinfo  {journal} {2D Materials}\ }\textbf {\bibinfo
  {volume} {5}},\ \bibinfo {pages} {035031} (\bibinfo {year}
  {2018})}\BibitemShut {NoStop}%
\bibitem [{\citenamefont {Chen}\ \emph {et~al.}(2018)\citenamefont {Chen},
  \citenamefont {Li},\ and\ \citenamefont {Peeters}}]{Chen_Peeters_2018}%
  \BibitemOpen
  \bibfield  {author} {\bibinfo {author} {\bibfnamefont {Q.}~\bibnamefont
  {Chen}}, \bibinfo {author} {\bibfnamefont {L.~L.}\ \bibnamefont {Li}},\ and\
  \bibinfo {author} {\bibfnamefont {F.~M.}\ \bibnamefont {Peeters}},\
  }\bibfield  {title} {\bibinfo {title} {Magnetic field dependence of
  electronic properties of {MoS}$_2$ quantum dots with different edges},\
  }\href {https://doi.org/10.1103/PhysRevB.97.085437} {\bibfield  {journal}
  {\bibinfo  {journal} {Physical Review B}\ }\textbf {\bibinfo {volume} {97}},\
  \bibinfo {pages} {085437} (\bibinfo {year} {2018})}\BibitemShut {NoStop}%
\bibitem [{\citenamefont {Chen}\ and\ \citenamefont {Wu}(2020)}]{Chen_Wu_2020}%
  \BibitemOpen
  \bibfield  {author} {\bibinfo {author} {\bibfnamefont {F.-W.}\ \bibnamefont
  {Chen}}\ and\ \bibinfo {author} {\bibfnamefont {Y.-S.~G.}\ \bibnamefont
  {Wu}},\ }\bibfield  {title} {\bibinfo {title} {{Theory of field-modulated
  spin valley orbital pseudospin physics}},\ }\href
  {https://doi.org/10.1103/PhysRevResearch.2.013076} {\bibfield  {journal}
  {\bibinfo  {journal} {Phys. Rev. Research}\ }\textbf {\bibinfo {volume}
  {2}},\ \bibinfo {pages} {013076} (\bibinfo {year} {2020})}\BibitemShut
  {NoStop}%
\bibitem [{\citenamefont {Brooks}\ and\ \citenamefont
  {Burkard}(2020)}]{Brooks_Burkard_2020}%
  \BibitemOpen
  \bibfield  {author} {\bibinfo {author} {\bibfnamefont {M.}~\bibnamefont
  {Brooks}}\ and\ \bibinfo {author} {\bibfnamefont {G.}~\bibnamefont
  {Burkard}},\ }\bibfield  {title} {\bibinfo {title} {{Electric dipole spin
  resonance of two-dimensional semiconductor spin qubits}},\ }\href
  {https://doi.org/10.1103/PhysRevB.101.035204} {\bibfield  {journal} {\bibinfo
   {journal} {Phys. Rev. B}\ }\textbf {\bibinfo {volume} {101}},\ \bibinfo
  {pages} {035204} (\bibinfo {year} {2020})}\BibitemShut {NoStop}%
\bibitem [{\citenamefont {Song}\ \emph {et~al.}(2015)\citenamefont {Song},
  \citenamefont {Liu}, \citenamefont {Mosallanejad}, \citenamefont {You},
  \citenamefont {Han}, \citenamefont {Chen}, \citenamefont {Li}, \citenamefont
  {Cao}, \citenamefont {Xiao}, \citenamefont {Guo},\ and\ \citenamefont
  {Guo}}]{Song_Guo_2015}%
  \BibitemOpen
  \bibfield  {author} {\bibinfo {author} {\bibfnamefont {X.-X.}\ \bibnamefont
  {Song}}, \bibinfo {author} {\bibfnamefont {D.}~\bibnamefont {Liu}}, \bibinfo
  {author} {\bibfnamefont {V.}~\bibnamefont {Mosallanejad}}, \bibinfo {author}
  {\bibfnamefont {J.}~\bibnamefont {You}}, \bibinfo {author} {\bibfnamefont
  {T.-Y.}\ \bibnamefont {Han}}, \bibinfo {author} {\bibfnamefont {D.-T.}\
  \bibnamefont {Chen}}, \bibinfo {author} {\bibfnamefont {H.-O.}\ \bibnamefont
  {Li}}, \bibinfo {author} {\bibfnamefont {G.}~\bibnamefont {Cao}}, \bibinfo
  {author} {\bibfnamefont {M.}~\bibnamefont {Xiao}}, \bibinfo {author}
  {\bibfnamefont {G.-C.}\ \bibnamefont {Guo}},\ and\ \bibinfo {author}
  {\bibfnamefont {G.-P.}\ \bibnamefont {Guo}},\ }\bibfield  {title} {\bibinfo
  {title} {A gate defined quantum dot on the two-dimensional transition metal
  dichalcogenide semiconductor {WSe} $_2$},\ }\href
  {https://doi.org/10.1039/C5NR04961J} {\bibfield  {journal} {\bibinfo
  {journal} {Nanoscale}\ }\textbf {\bibinfo {volume} {7}},\ \bibinfo {pages}
  {16867} (\bibinfo {year} {2015})}\BibitemShut {NoStop}%
\bibitem [{\citenamefont {Zhang}\ \emph {et~al.}(2017)\citenamefont {Zhang},
  \citenamefont {Song}, \citenamefont {Luo}, \citenamefont {Deng},
  \citenamefont {Mosallanejad}, \citenamefont {Taniguchi}, \citenamefont
  {Watanabe}, \citenamefont {Li}, \citenamefont {Cao}, \citenamefont {Guo},
  \citenamefont {Nori},\ and\ \citenamefont {Guo}}]{Zhang_Guo_2017}%
  \BibitemOpen
  \bibfield  {author} {\bibinfo {author} {\bibfnamefont {Z.-Z.}\ \bibnamefont
  {Zhang}}, \bibinfo {author} {\bibfnamefont {X.-X.}\ \bibnamefont {Song}},
  \bibinfo {author} {\bibfnamefont {G.}~\bibnamefont {Luo}}, \bibinfo {author}
  {\bibfnamefont {G.-W.}\ \bibnamefont {Deng}}, \bibinfo {author}
  {\bibfnamefont {V.}~\bibnamefont {Mosallanejad}}, \bibinfo {author}
  {\bibfnamefont {T.}~\bibnamefont {Taniguchi}}, \bibinfo {author}
  {\bibfnamefont {K.}~\bibnamefont {Watanabe}}, \bibinfo {author}
  {\bibfnamefont {H.-O.}\ \bibnamefont {Li}}, \bibinfo {author} {\bibfnamefont
  {G.}~\bibnamefont {Cao}}, \bibinfo {author} {\bibfnamefont {G.-C.}\
  \bibnamefont {Guo}}, \bibinfo {author} {\bibfnamefont {F.}~\bibnamefont
  {Nori}},\ and\ \bibinfo {author} {\bibfnamefont {G.-P.}\ \bibnamefont
  {Guo}},\ }\bibfield  {title} {\bibinfo {title} {Electrotunable artificial
  molecules based on van der {Waals} heterostructures},\ }\href
  {https://doi.org/10.1126/sciadv.1701699} {\bibfield  {journal} {\bibinfo
  {journal} {Science Advances}\ }\textbf {\bibinfo {volume} {3}},\ \bibinfo
  {pages} {e1701699} (\bibinfo {year} {2017})}\BibitemShut {NoStop}%
\bibitem [{\citenamefont {Pisoni}\ \emph {et~al.}(2018)\citenamefont {Pisoni},
  \citenamefont {Korm\'anyos}, \citenamefont {Brooks}, \citenamefont {Lei},
  \citenamefont {Back}, \citenamefont {Eich}, \citenamefont {Overweg},
  \citenamefont {Lee}, \citenamefont {Rickhaus}, \citenamefont {Watanabe},
  \citenamefont {Taniguchi}, \citenamefont {Imamoglu}, \citenamefont {Burkard},
  \citenamefont {Ihn},\ and\ \citenamefont {Ensslin}}]{Pisoni_Ensslin_2018}%
  \BibitemOpen
  \bibfield  {author} {\bibinfo {author} {\bibfnamefont {R.}~\bibnamefont
  {Pisoni}}, \bibinfo {author} {\bibfnamefont {A.}~\bibnamefont {Korm\'anyos}},
  \bibinfo {author} {\bibfnamefont {M.}~\bibnamefont {Brooks}}, \bibinfo
  {author} {\bibfnamefont {Z.}~\bibnamefont {Lei}}, \bibinfo {author}
  {\bibfnamefont {P.}~\bibnamefont {Back}}, \bibinfo {author} {\bibfnamefont
  {M.}~\bibnamefont {Eich}}, \bibinfo {author} {\bibfnamefont {H.}~\bibnamefont
  {Overweg}}, \bibinfo {author} {\bibfnamefont {Y.}~\bibnamefont {Lee}},
  \bibinfo {author} {\bibfnamefont {P.}~\bibnamefont {Rickhaus}}, \bibinfo
  {author} {\bibfnamefont {K.}~\bibnamefont {Watanabe}}, \bibinfo {author}
  {\bibfnamefont {T.}~\bibnamefont {Taniguchi}}, \bibinfo {author}
  {\bibfnamefont {A.}~\bibnamefont {Imamoglu}}, \bibinfo {author}
  {\bibfnamefont {G.}~\bibnamefont {Burkard}}, \bibinfo {author} {\bibfnamefont
  {T.}~\bibnamefont {Ihn}},\ and\ \bibinfo {author} {\bibfnamefont
  {K.}~\bibnamefont {Ensslin}},\ }\bibfield  {title} {\bibinfo {title}
  {{Interactions and Magnetotransport through Spin-Valley Coupled Landau Levels
  in Monolayer ${\mathrm{MoS}}_{2}$}},\ }\href
  {https://doi.org/10.1103/PhysRevLett.121.247701} {\bibfield  {journal}
  {\bibinfo  {journal} {Phys. Rev. Lett.}\ }\textbf {\bibinfo {volume} {121}},\
  \bibinfo {pages} {247701} (\bibinfo {year} {2018})}\BibitemShut {NoStop}%
\bibitem [{\citenamefont {Lau}\ \emph {et~al.}(2019)\citenamefont {Lau},
  \citenamefont {Chee}, \citenamefont {Thian}, \citenamefont {Kawai},
  \citenamefont {Deng}, \citenamefont {Wong}, \citenamefont {Ooi},
  \citenamefont {Lim},\ and\ \citenamefont {Goh}}]{Lau_Goh_2019}%
  \BibitemOpen
  \bibfield  {author} {\bibinfo {author} {\bibfnamefont {C.~S.}\ \bibnamefont
  {Lau}}, \bibinfo {author} {\bibfnamefont {J.~Y.}\ \bibnamefont {Chee}},
  \bibinfo {author} {\bibfnamefont {D.}~\bibnamefont {Thian}}, \bibinfo
  {author} {\bibfnamefont {H.}~\bibnamefont {Kawai}}, \bibinfo {author}
  {\bibfnamefont {J.}~\bibnamefont {Deng}}, \bibinfo {author} {\bibfnamefont
  {S.~L.}\ \bibnamefont {Wong}}, \bibinfo {author} {\bibfnamefont {Z.~E.}\
  \bibnamefont {Ooi}}, \bibinfo {author} {\bibfnamefont {Y.-F.}\ \bibnamefont
  {Lim}},\ and\ \bibinfo {author} {\bibfnamefont {K.~E.~J.}\ \bibnamefont
  {Goh}},\ }\bibfield  {title} {\bibinfo {title} {{Carrier control in 2D
  transition metal dichalcogenides with Al2O3 dielectric}},\ }\href
  {https://doi.org/10.1038/s41598-019-45392-9} {\bibfield  {journal} {\bibinfo
  {journal} {Scientific Reports}\ }\textbf {\bibinfo {volume} {9}},\ \bibinfo
  {pages} {8769} (\bibinfo {year} {2019})}\BibitemShut {NoStop}%
\bibitem [{\citenamefont {Davari}\ \emph
  {et~al.}(2020{\natexlab{a}})\citenamefont {Davari}, \citenamefont {Stacy},
  \citenamefont {Mercado}, \citenamefont {Tull}, \citenamefont {Basnet},
  \citenamefont {Pandey}, \citenamefont {Watanabe}, \citenamefont {Taniguchi},
  \citenamefont {Hu},\ and\ \citenamefont {Churchill}}]{Davari_Churchill_2020}%
  \BibitemOpen
  \bibfield  {author} {\bibinfo {author} {\bibfnamefont {S.}~\bibnamefont
  {Davari}}, \bibinfo {author} {\bibfnamefont {J.}~\bibnamefont {Stacy}},
  \bibinfo {author} {\bibfnamefont {A.}~\bibnamefont {Mercado}}, \bibinfo
  {author} {\bibfnamefont {J.}~\bibnamefont {Tull}}, \bibinfo {author}
  {\bibfnamefont {R.}~\bibnamefont {Basnet}}, \bibinfo {author} {\bibfnamefont
  {K.}~\bibnamefont {Pandey}}, \bibinfo {author} {\bibfnamefont
  {K.}~\bibnamefont {Watanabe}}, \bibinfo {author} {\bibfnamefont
  {T.}~\bibnamefont {Taniguchi}}, \bibinfo {author} {\bibfnamefont
  {J.}~\bibnamefont {Hu}},\ and\ \bibinfo {author} {\bibfnamefont
  {H.}~\bibnamefont {Churchill}},\ }\bibfield  {title} {\bibinfo {title}
  {{Gate-Defined Accumulation-Mode Quantum Dots in Monolayer and Bilayer
  ${\mathrm{W}\mathrm{Se}}_{2}$}},\ }\href
  {https://doi.org/10.1103/PhysRevApplied.13.054058} {\bibfield  {journal}
  {\bibinfo  {journal} {Phys. Rev. Applied}\ }\textbf {\bibinfo {volume}
  {13}},\ \bibinfo {pages} {054058} (\bibinfo {year}
  {2020}{\natexlab{a}})}\BibitemShut {NoStop}%
\bibitem [{\citenamefont {Goh}\ \emph {et~al.}(2020{\natexlab{b}})\citenamefont
  {Goh}, \citenamefont {Bussolotti}, \citenamefont {Lau}, \citenamefont
  {Kotekar-Patil}, \citenamefont {Ooi},\ and\ \citenamefont
  {Chee}}]{Goh_Yee_2020}%
  \BibitemOpen
  \bibfield  {author} {\bibinfo {author} {\bibfnamefont {K.~E.~J.}\
  \bibnamefont {Goh}}, \bibinfo {author} {\bibfnamefont {F.}~\bibnamefont
  {Bussolotti}}, \bibinfo {author} {\bibfnamefont {C.~S.}\ \bibnamefont {Lau}},
  \bibinfo {author} {\bibfnamefont {D.}~\bibnamefont {Kotekar-Patil}}, \bibinfo
  {author} {\bibfnamefont {Z.~E.}\ \bibnamefont {Ooi}},\ and\ \bibinfo {author}
  {\bibfnamefont {J.~Y.}\ \bibnamefont {Chee}},\ }\bibfield  {title} {\bibinfo
  {title} {Toward valley-coupled spin qubits},\ }\href
  {https://doi.org/https://doi.org/10.1002/qute.201900123} {\bibfield
  {journal} {\bibinfo  {journal} {Advanced Quantum Technologies}\ }\textbf
  {\bibinfo {volume} {3}},\ \bibinfo {pages} {1900123} (\bibinfo {year}
  {2020}{\natexlab{b}})}\BibitemShut {NoStop}%
\bibitem [{\citenamefont {Chirolli}\ \emph {et~al.}(2019)\citenamefont
  {Chirolli}, \citenamefont {Prada}, \citenamefont {Guinea}, \citenamefont
  {Rold{\'a}n},\ and\ \citenamefont {San-Jose}}]{Chirolli_SanJose_2019}%
  \BibitemOpen
  \bibfield  {author} {\bibinfo {author} {\bibfnamefont {L.}~\bibnamefont
  {Chirolli}}, \bibinfo {author} {\bibfnamefont {E.}~\bibnamefont {Prada}},
  \bibinfo {author} {\bibfnamefont {F.}~\bibnamefont {Guinea}}, \bibinfo
  {author} {\bibfnamefont {R.}~\bibnamefont {Rold{\'a}n}},\ and\ \bibinfo
  {author} {\bibfnamefont {P.}~\bibnamefont {San-Jose}},\ }\bibfield  {title}
  {\bibinfo {title} {Strain-induced bound states in transition-metal
  dichalcogenide bubbles},\ }\href {https://doi.org/10.1088/2053-1583/ab0113}
  {\bibfield  {journal} {\bibinfo  {journal} {2D Materials}\ }\textbf {\bibinfo
  {volume} {6}},\ \bibinfo {pages} {025010} (\bibinfo {year}
  {2019})}\BibitemShut {NoStop}%
\bibitem [{\citenamefont {Bieniek}\ \emph {et~al.}(2020)\citenamefont
  {Bieniek}, \citenamefont {Szulakowska},\ and\ \citenamefont
  {Hawrylak}}]{MB2}%
  \BibitemOpen
  \bibfield  {author} {\bibinfo {author} {\bibfnamefont {M.}~\bibnamefont
  {Bieniek}}, \bibinfo {author} {\bibfnamefont {L.}~\bibnamefont
  {Szulakowska}},\ and\ \bibinfo {author} {\bibfnamefont {P.}~\bibnamefont
  {Hawrylak}},\ }\bibfield  {title} {\bibinfo {title} {Effect of valley, spin,
  and band nesting on the electronic properties of gated quantum dots in a
  single layer of transition metal dichalcogenides},\ }\href
  {https://doi.org/10.1103/PhysRevB.101.035401} {\bibfield  {journal} {\bibinfo
   {journal} {Phys. Rev. B}\ }\textbf {\bibinfo {volume} {101}},\ \bibinfo
  {pages} {035401} (\bibinfo {year} {2020})}\BibitemShut {NoStop}%
\bibitem [{\citenamefont {Szulakowska}\ \emph {et~al.}(2020)\citenamefont
  {Szulakowska}, \citenamefont {Cygorek}, \citenamefont {Bieniek},\ and\
  \citenamefont {Hawrylak}}]{MB3}%
  \BibitemOpen
  \bibfield  {author} {\bibinfo {author} {\bibfnamefont {L.}~\bibnamefont
  {Szulakowska}}, \bibinfo {author} {\bibfnamefont {M.}~\bibnamefont
  {Cygorek}}, \bibinfo {author} {\bibfnamefont {M.}~\bibnamefont {Bieniek}},\
  and\ \bibinfo {author} {\bibfnamefont {P.}~\bibnamefont {Hawrylak}},\
  }\bibfield  {title} {\bibinfo {title} {Valley- and spin-polarized
  broken-symmetry states of interacting electrons in gated
  $\mathrm{Mo}{\mathrm{s}}_{2}$ quantum dots},\ }\href
  {https://doi.org/10.1103/PhysRevB.102.245410} {\bibfield  {journal} {\bibinfo
   {journal} {Phys. Rev. B}\ }\textbf {\bibinfo {volume} {102}},\ \bibinfo
  {pages} {245410} (\bibinfo {year} {2020})}\BibitemShut {NoStop}%
\bibitem [{\citenamefont {Ciorga}\ \emph {et~al.}(2000)\citenamefont {Ciorga},
  \citenamefont {Sachrajda}, \citenamefont {Hawrylak}, \citenamefont {Gould},
  \citenamefont {Zawadzki}, \citenamefont {Jullian}, \citenamefont {Feng},\
  and\ \citenamefont {Wasilewski}}]{gated_haw}%
  \BibitemOpen
  \bibfield  {author} {\bibinfo {author} {\bibfnamefont {M.}~\bibnamefont
  {Ciorga}}, \bibinfo {author} {\bibfnamefont {A.~S.}\ \bibnamefont
  {Sachrajda}}, \bibinfo {author} {\bibfnamefont {P.}~\bibnamefont {Hawrylak}},
  \bibinfo {author} {\bibfnamefont {C.}~\bibnamefont {Gould}}, \bibinfo
  {author} {\bibfnamefont {P.}~\bibnamefont {Zawadzki}}, \bibinfo {author}
  {\bibfnamefont {S.}~\bibnamefont {Jullian}}, \bibinfo {author} {\bibfnamefont
  {Y.}~\bibnamefont {Feng}},\ and\ \bibinfo {author} {\bibfnamefont
  {Z.}~\bibnamefont {Wasilewski}},\ }\bibfield  {title} {\bibinfo {title}
  {Addition spectrum of a lateral dot from coulomb and spin-blockade
  spectroscopy},\ }\href {https://doi.org/10.1103/PhysRevB.61.R16315}
  {\bibfield  {journal} {\bibinfo  {journal} {Phys. Rev. B}\ }\textbf {\bibinfo
  {volume} {61}},\ \bibinfo {pages} {R16315} (\bibinfo {year}
  {2000})}\BibitemShut {NoStop}%
\bibitem [{\citenamefont {Bayer}\ \emph {et~al.}(2001)\citenamefont {Bayer},
  \citenamefont {Hawrylak}, \citenamefont {Hinzer}, \citenamefont {Fafard},
  \citenamefont {Korkusinski}, \citenamefont {Wasilewski}, \citenamefont
  {Stern},\ and\ \citenamefont {Forchel}}]{dgd_haw1}%
  \BibitemOpen
  \bibfield  {author} {\bibinfo {author} {\bibfnamefont {M.}~\bibnamefont
  {Bayer}}, \bibinfo {author} {\bibfnamefont {P.}~\bibnamefont {Hawrylak}},
  \bibinfo {author} {\bibfnamefont {K.}~\bibnamefont {Hinzer}}, \bibinfo
  {author} {\bibfnamefont {S.}~\bibnamefont {Fafard}}, \bibinfo {author}
  {\bibfnamefont {M.}~\bibnamefont {Korkusinski}}, \bibinfo {author}
  {\bibfnamefont {Z.}~\bibnamefont {Wasilewski}}, \bibinfo {author}
  {\bibfnamefont {O.}~\bibnamefont {Stern}},\ and\ \bibinfo {author}
  {\bibfnamefont {A.}~\bibnamefont {Forchel}},\ }\bibfield  {title} {\bibinfo
  {title} {Coupling and entangling of quantum states in quantum dot
  molecules},\ }\href@noop {} {\bibfield  {journal} {\bibinfo  {journal}
  {Science}\ }\textbf {\bibinfo {volume} {291}},\ \bibinfo {pages} {451}
  (\bibinfo {year} {2001})}\BibitemShut {NoStop}%
\bibitem [{\citenamefont {Pioro-Ladri\`ere}\ \emph {et~al.}(2003)\citenamefont
  {Pioro-Ladri\`ere}, \citenamefont {Ciorga}, \citenamefont {Lapointe},
  \citenamefont {Zawadzki}, \citenamefont {Korkusi\ifmmode~\acute{n}\else
  \'{n}\fi{}ski}, \citenamefont {Hawrylak},\ and\ \citenamefont
  {Sachrajda}}]{dqd_haw2}%
  \BibitemOpen
  \bibfield  {author} {\bibinfo {author} {\bibfnamefont {M.}~\bibnamefont
  {Pioro-Ladri\`ere}}, \bibinfo {author} {\bibfnamefont {M.}~\bibnamefont
  {Ciorga}}, \bibinfo {author} {\bibfnamefont {J.}~\bibnamefont {Lapointe}},
  \bibinfo {author} {\bibfnamefont {P.}~\bibnamefont {Zawadzki}}, \bibinfo
  {author} {\bibfnamefont {M.}~\bibnamefont {Korkusi\ifmmode~\acute{n}\else
  \'{n}\fi{}ski}}, \bibinfo {author} {\bibfnamefont {P.}~\bibnamefont
  {Hawrylak}},\ and\ \bibinfo {author} {\bibfnamefont {A.~S.}\ \bibnamefont
  {Sachrajda}},\ }\bibfield  {title} {\bibinfo {title} {Spin-blockade
  spectroscopy of a two-level artificial molecule},\ }\href
  {https://doi.org/10.1103/PhysRevLett.91.026803} {\bibfield  {journal}
  {\bibinfo  {journal} {Phys. Rev. Lett.}\ }\textbf {\bibinfo {volume} {91}},\
  \bibinfo {pages} {026803} (\bibinfo {year} {2003})}\BibitemShut {NoStop}%
\bibitem [{\citenamefont {Pioro-Ladri\`ere}\ \emph {et~al.}(2005)\citenamefont
  {Pioro-Ladri\`ere}, \citenamefont {Abolfath}, \citenamefont {Zawadzki},
  \citenamefont {Lapointe}, \citenamefont {Studenikin}, \citenamefont
  {Sachrajda},\ and\ \citenamefont {Hawrylak}}]{dqd_haw3}%
  \BibitemOpen
  \bibfield  {author} {\bibinfo {author} {\bibfnamefont {M.}~\bibnamefont
  {Pioro-Ladri\`ere}}, \bibinfo {author} {\bibfnamefont {M.~R.}\ \bibnamefont
  {Abolfath}}, \bibinfo {author} {\bibfnamefont {P.}~\bibnamefont {Zawadzki}},
  \bibinfo {author} {\bibfnamefont {J.}~\bibnamefont {Lapointe}}, \bibinfo
  {author} {\bibfnamefont {S.~A.}\ \bibnamefont {Studenikin}}, \bibinfo
  {author} {\bibfnamefont {A.~S.}\ \bibnamefont {Sachrajda}},\ and\ \bibinfo
  {author} {\bibfnamefont {P.}~\bibnamefont {Hawrylak}},\ }\bibfield  {title}
  {\bibinfo {title} {Charge sensing of an artificial ${\mathrm{h}}_{2}^{+}$
  molecule in lateral quantum dots},\ }\href
  {https://doi.org/10.1103/PhysRevB.72.125307} {\bibfield  {journal} {\bibinfo
  {journal} {Phys. Rev. B}\ }\textbf {\bibinfo {volume} {72}},\ \bibinfo
  {pages} {125307} (\bibinfo {year} {2005})}\BibitemShut {NoStop}%
\bibitem [{\citenamefont {Dybalski}\ and\ \citenamefont
  {Hawrylak}(2005)}]{dqd_haw4}%
  \BibitemOpen
  \bibfield  {author} {\bibinfo {author} {\bibfnamefont {W.}~\bibnamefont
  {Dybalski}}\ and\ \bibinfo {author} {\bibfnamefont {P.}~\bibnamefont
  {Hawrylak}},\ }\bibfield  {title} {\bibinfo {title} {Two electrons in a
  strongly coupled double quantum dot: From an artificial helium atom to a
  hydrogen molecule},\ }\href {https://doi.org/10.1103/PhysRevB.72.205432}
  {\bibfield  {journal} {\bibinfo  {journal} {Phys. Rev. B}\ }\textbf {\bibinfo
  {volume} {72}},\ \bibinfo {pages} {205432} (\bibinfo {year}
  {2005})}\BibitemShut {NoStop}%
\bibitem [{\citenamefont {Liu}\ \emph {et~al.}(2014{\natexlab{b}})\citenamefont
  {Liu}, \citenamefont {Pang}, \citenamefont {Yao},\ and\ \citenamefont
  {Yao}}]{valtr}%
  \BibitemOpen
  \bibfield  {author} {\bibinfo {author} {\bibfnamefont {G.-B.}\ \bibnamefont
  {Liu}}, \bibinfo {author} {\bibfnamefont {H.}~\bibnamefont {Pang}}, \bibinfo
  {author} {\bibfnamefont {Y.}~\bibnamefont {Yao}},\ and\ \bibinfo {author}
  {\bibfnamefont {W.}~\bibnamefont {Yao}},\ }\bibfield  {title} {\bibinfo
  {title} {Intervalley coupling by quantum dot confinement potentials in
  monolayer transition metal dichalcogenides},\ }\href
  {https://doi.org/10.1088/1367-2630/16/10/105011} {\bibfield  {journal}
  {\bibinfo  {journal} {New Journal of Physics}\ }\textbf {\bibinfo {volume}
  {16}},\ \bibinfo {pages} {105011} (\bibinfo {year}
  {2014}{\natexlab{b}})}\BibitemShut {NoStop}%
\bibitem [{\citenamefont {Pei}\ \emph {et~al.}(2017)\citenamefont {Pei},
  \citenamefont {P\'alyi}, \citenamefont {Mergenthaler}, \citenamefont {Ares},
  \citenamefont {Mavalankar}, \citenamefont {Warner}, \citenamefont {Briggs},\
  and\ \citenamefont {Laird}}]{nt3}%
  \BibitemOpen
  \bibfield  {author} {\bibinfo {author} {\bibfnamefont {T.}~\bibnamefont
  {Pei}}, \bibinfo {author} {\bibfnamefont {A.}~\bibnamefont {P\'alyi}},
  \bibinfo {author} {\bibfnamefont {M.}~\bibnamefont {Mergenthaler}}, \bibinfo
  {author} {\bibfnamefont {N.}~\bibnamefont {Ares}}, \bibinfo {author}
  {\bibfnamefont {A.}~\bibnamefont {Mavalankar}}, \bibinfo {author}
  {\bibfnamefont {J.~H.}\ \bibnamefont {Warner}}, \bibinfo {author}
  {\bibfnamefont {G.~A.~D.}\ \bibnamefont {Briggs}},\ and\ \bibinfo {author}
  {\bibfnamefont {E.~A.}\ \bibnamefont {Laird}},\ }\bibfield  {title} {\bibinfo
  {title} {Hyperfine and spin-orbit coupling effects on decay of spin-valley
  states in a carbon nanotube},\ }\href
  {https://doi.org/10.1103/PhysRevLett.118.177701} {\bibfield  {journal}
  {\bibinfo  {journal} {Phys. Rev. Lett.}\ }\textbf {\bibinfo {volume} {118}},\
  \bibinfo {pages} {177701} (\bibinfo {year} {2017})}\BibitemShut {NoStop}%
\bibitem [{\citenamefont {Laturia}\ \emph {et~al.}(2018)\citenamefont
  {Laturia}, \citenamefont {Van~de Put},\ and\ \citenamefont
  {Vandenberghe}}]{hbn}%
  \BibitemOpen
  \bibfield  {author} {\bibinfo {author} {\bibfnamefont {A.}~\bibnamefont
  {Laturia}}, \bibinfo {author} {\bibfnamefont {M.~L.}\ \bibnamefont {Van~de
  Put}},\ and\ \bibinfo {author} {\bibfnamefont {W.~G.}\ \bibnamefont
  {Vandenberghe}},\ }\bibfield  {title} {\bibinfo {title} {Dielectric
  properties of hexagonal boron nitride and transition metal dichalcogenides:
  from monolayer to bulk},\ }\href@noop {} {\bibfield  {journal} {\bibinfo
  {journal} {npj 2D Materials and Applications}\ }\textbf {\bibinfo {volume}
  {2}},\ \bibinfo {pages} {6} (\bibinfo {year} {2018})}\BibitemShut {NoStop}%
\bibitem [{\citenamefont {Davari}\ \emph
  {et~al.}(2020{\natexlab{b}})\citenamefont {Davari}, \citenamefont {Stacy},
  \citenamefont {Mercado}, \citenamefont {Tull}, \citenamefont {Basnet},
  \citenamefont {Pandey}, \citenamefont {Watanabe}, \citenamefont {Taniguchi},
  \citenamefont {Hu},\ and\ \citenamefont {Churchill}}]{Davari2020}%
  \BibitemOpen
  \bibfield  {author} {\bibinfo {author} {\bibfnamefont {S.}~\bibnamefont
  {Davari}}, \bibinfo {author} {\bibfnamefont {J.}~\bibnamefont {Stacy}},
  \bibinfo {author} {\bibfnamefont {A.}~\bibnamefont {Mercado}}, \bibinfo
  {author} {\bibfnamefont {J.}~\bibnamefont {Tull}}, \bibinfo {author}
  {\bibfnamefont {R.}~\bibnamefont {Basnet}}, \bibinfo {author} {\bibfnamefont
  {K.}~\bibnamefont {Pandey}}, \bibinfo {author} {\bibfnamefont
  {K.}~\bibnamefont {Watanabe}}, \bibinfo {author} {\bibfnamefont
  {T.}~\bibnamefont {Taniguchi}}, \bibinfo {author} {\bibfnamefont
  {J.}~\bibnamefont {Hu}},\ and\ \bibinfo {author} {\bibfnamefont
  {H.}~\bibnamefont {Churchill}},\ }\bibfield  {title} {\bibinfo {title}
  {Gate-defined accumulation-mode quantum dots in monolayer and bilayer
  ${\mathrm{w}\mathrm{se}}_{2}$},\ }\href
  {https://doi.org/10.1103/PhysRevApplied.13.054058} {\bibfield  {journal}
  {\bibinfo  {journal} {Phys. Rev. Applied}\ }\textbf {\bibinfo {volume}
  {13}},\ \bibinfo {pages} {054058} (\bibinfo {year}
  {2020}{\natexlab{b}})}\BibitemShut {NoStop}%
\bibitem [{\citenamefont {Paw\l{}owski}\ \emph {et~al.}(2016)\citenamefont
  {Paw\l{}owski}, \citenamefont {Szumniak},\ and\ \citenamefont
  {Bednarek}}]{mydrut}%
  \BibitemOpen
  \bibfield  {author} {\bibinfo {author} {\bibfnamefont {J.}~\bibnamefont
  {Paw\l{}owski}}, \bibinfo {author} {\bibfnamefont {P.}~\bibnamefont
  {Szumniak}},\ and\ \bibinfo {author} {\bibfnamefont {S.}~\bibnamefont
  {Bednarek}},\ }\bibfield  {title} {\bibinfo {title} {Electron spin rotations
  induced by oscillating rashba interaction in a quantum wire},\ }\href
  {https://doi.org/10.1103/PhysRevB.93.045309} {\bibfield  {journal} {\bibinfo
  {journal} {Phys. Rev. B}\ }\textbf {\bibinfo {volume} {93}},\ \bibinfo
  {pages} {045309} (\bibinfo {year} {2016})}\BibitemShut {NoStop}%
\bibitem [{\citenamefont {Bieniek}\ \emph {et~al.}(2018)\citenamefont
  {Bieniek}, \citenamefont {Korkusi\ifmmode~\acute{n}\else \'{n}\fi{}ski},
  \citenamefont {Szulakowska}, \citenamefont {Potasz}, \citenamefont
  {Ozfidan},\ and\ \citenamefont {Hawrylak}}]{MB1}%
  \BibitemOpen
  \bibfield  {author} {\bibinfo {author} {\bibfnamefont {M.}~\bibnamefont
  {Bieniek}}, \bibinfo {author} {\bibfnamefont {M.}~\bibnamefont
  {Korkusi\ifmmode~\acute{n}\else \'{n}\fi{}ski}}, \bibinfo {author}
  {\bibfnamefont {L.}~\bibnamefont {Szulakowska}}, \bibinfo {author}
  {\bibfnamefont {P.}~\bibnamefont {Potasz}}, \bibinfo {author} {\bibfnamefont
  {I.}~\bibnamefont {Ozfidan}},\ and\ \bibinfo {author} {\bibfnamefont
  {P.}~\bibnamefont {Hawrylak}},\ }\bibfield  {title} {\bibinfo {title} {Band
  nesting, massive dirac fermions, and valley land\'e and zeeman effects in
  transition metal dichalcogenides: A tight-binding model},\ }\href
  {https://doi.org/10.1103/PhysRevB.97.085153} {\bibfield  {journal} {\bibinfo
  {journal} {Phys. Rev. B}\ }\textbf {\bibinfo {volume} {97}},\ \bibinfo
  {pages} {085153} (\bibinfo {year} {2018})}\BibitemShut {NoStop}%
\bibitem [{\citenamefont {Rostami}\ \emph {et~al.}(2013)\citenamefont
  {Rostami}, \citenamefont {Moghaddam},\ and\ \citenamefont {Asgari}}]{azgari}%
  \BibitemOpen
  \bibfield  {author} {\bibinfo {author} {\bibfnamefont {H.}~\bibnamefont
  {Rostami}}, \bibinfo {author} {\bibfnamefont {A.~G.}\ \bibnamefont
  {Moghaddam}},\ and\ \bibinfo {author} {\bibfnamefont {R.}~\bibnamefont
  {Asgari}},\ }\bibfield  {title} {\bibinfo {title} {Effective lattice
  hamiltonian for monolayer mos${}_{2}$: Tailoring electronic structure with
  perpendicular electric and magnetic fields},\ }\href
  {https://doi.org/10.1103/PhysRevB.88.085440} {\bibfield  {journal} {\bibinfo
  {journal} {Phys. Rev. B}\ }\textbf {\bibinfo {volume} {88}},\ \bibinfo
  {pages} {085440} (\bibinfo {year} {2013})}\BibitemShut {NoStop}%
\bibitem [{\citenamefont {Ridolfi}\ \emph {et~al.}(2015)\citenamefont
  {Ridolfi}, \citenamefont {Le}, \citenamefont {Rahman}, \citenamefont
  {Mucciolo},\ and\ \citenamefont {Lewenkopf}}]{ridolfi}%
  \BibitemOpen
  \bibfield  {author} {\bibinfo {author} {\bibfnamefont {E.}~\bibnamefont
  {Ridolfi}}, \bibinfo {author} {\bibfnamefont {D.}~\bibnamefont {Le}},
  \bibinfo {author} {\bibfnamefont {T.~S.}\ \bibnamefont {Rahman}}, \bibinfo
  {author} {\bibfnamefont {E.~R.}\ \bibnamefont {Mucciolo}},\ and\ \bibinfo
  {author} {\bibfnamefont {C.~H.}\ \bibnamefont {Lewenkopf}},\ }\bibfield
  {title} {\bibinfo {title} {A tight-binding model for {MoS}2monolayers},\
  }\href {https://doi.org/10.1088/0953-8984/27/36/365501} {\bibfield  {journal}
  {\bibinfo  {journal} {Journal of Physics: Condensed Matter}\ }\textbf
  {\bibinfo {volume} {27}},\ \bibinfo {pages} {365501} (\bibinfo {year}
  {2015})}\BibitemShut {NoStop}%
\bibitem [{\citenamefont {Liu}\ \emph {et~al.}(2013)\citenamefont {Liu},
  \citenamefont {Shan}, \citenamefont {Yao}, \citenamefont {Yao},\ and\
  \citenamefont {Xiao}}]{xiao}%
  \BibitemOpen
  \bibfield  {author} {\bibinfo {author} {\bibfnamefont {G.-B.}\ \bibnamefont
  {Liu}}, \bibinfo {author} {\bibfnamefont {W.-Y.}\ \bibnamefont {Shan}},
  \bibinfo {author} {\bibfnamefont {Y.}~\bibnamefont {Yao}}, \bibinfo {author}
  {\bibfnamefont {W.}~\bibnamefont {Yao}},\ and\ \bibinfo {author}
  {\bibfnamefont {D.}~\bibnamefont {Xiao}},\ }\bibfield  {title} {\bibinfo
  {title} {Te-band tight-binding model for monolayers of group-vib transition
  metal dichalcogenides},\ }\href {https://doi.org/10.1103/PhysRevB.88.085433}
  {\bibfield  {journal} {\bibinfo  {journal} {Phys. Rev. B}\ }\textbf {\bibinfo
  {volume} {88}},\ \bibinfo {pages} {085433} (\bibinfo {year}
  {2013})}\BibitemShut {NoStop}%
\bibitem [{\citenamefont {Kadantsev}\ and\ \citenamefont
  {Hawrylak}(2012)}]{haw}%
  \BibitemOpen
  \bibfield  {author} {\bibinfo {author} {\bibfnamefont {E.~S.}\ \bibnamefont
  {Kadantsev}}\ and\ \bibinfo {author} {\bibfnamefont {P.}~\bibnamefont
  {Hawrylak}},\ }\bibfield  {title} {\bibinfo {title} {Electronic structure of
  a single {MoS}2 monolayer},\ }\href
  {https://doi.org/10.1016/j.ssc.2012.02.005} {\bibfield  {journal} {\bibinfo
  {journal} {Solid State Communications}\ }\textbf {\bibinfo {volume} {152}},\
  \bibinfo {pages} {909} (\bibinfo {year} {2012})}\BibitemShut {NoStop}%
\bibitem [{\citenamefont {\ifmmode~\dot{Z}\else \.{Z}\fi{}ebrowski}\ \emph
  {et~al.}(2017)\citenamefont {\ifmmode~\dot{Z}\else \.{Z}\fi{}ebrowski},
  \citenamefont {Peeters},\ and\ \citenamefont {Szafran}}]{ci1}%
  \BibitemOpen
  \bibfield  {author} {\bibinfo {author} {\bibfnamefont {D.~P.}\ \bibnamefont
  {\ifmmode~\dot{Z}\else \.{Z}\fi{}ebrowski}}, \bibinfo {author} {\bibfnamefont
  {F.~M.}\ \bibnamefont {Peeters}},\ and\ \bibinfo {author} {\bibfnamefont
  {B.}~\bibnamefont {Szafran}},\ }\bibfield  {title} {\bibinfo {title} {Double
  quantum dots defined in bilayer graphene},\ }\href
  {https://doi.org/10.1103/PhysRevB.96.035434} {\bibfield  {journal} {\bibinfo
  {journal} {Phys. Rev. B}\ }\textbf {\bibinfo {volume} {96}},\ \bibinfo
  {pages} {035434} (\bibinfo {year} {2017})}\BibitemShut {NoStop}%
\bibitem [{\citenamefont {Szafran}\ and\ \citenamefont {\ifmmode~\dot{Z}\else
  \.{Z}\fi{}ebrowski}(2018{\natexlab{a}})}]{ci2}%
  \BibitemOpen
  \bibfield  {author} {\bibinfo {author} {\bibfnamefont {B.}~\bibnamefont
  {Szafran}}\ and\ \bibinfo {author} {\bibfnamefont {D.}~\bibnamefont
  {\ifmmode~\dot{Z}\else \.{Z}\fi{}ebrowski}},\ }\bibfield  {title} {\bibinfo
  {title} {Spin and valley control in single and double electrostatic silicene
  quantum dots},\ }\href {https://doi.org/10.1103/PhysRevB.98.155305}
  {\bibfield  {journal} {\bibinfo  {journal} {Phys. Rev. B}\ }\textbf {\bibinfo
  {volume} {98}},\ \bibinfo {pages} {155305} (\bibinfo {year}
  {2018}{\natexlab{a}})}\BibitemShut {NoStop}%
\bibitem [{\citenamefont {Osika}\ and\ \citenamefont {Szafran}(2015)}]{ci3}%
  \BibitemOpen
  \bibfield  {author} {\bibinfo {author} {\bibfnamefont {E.~N.}\ \bibnamefont
  {Osika}}\ and\ \bibinfo {author} {\bibfnamefont {B.}~\bibnamefont
  {Szafran}},\ }\bibfield  {title} {\bibinfo {title} {Two-electron
  $n\text{\ensuremath{-}}p$ double quantum dots in carbon nanotubes},\ }\href
  {https://doi.org/10.1103/PhysRevB.91.085312} {\bibfield  {journal} {\bibinfo
  {journal} {Phys. Rev. B}\ }\textbf {\bibinfo {volume} {91}},\ \bibinfo
  {pages} {085312} (\bibinfo {year} {2015})}\BibitemShut {NoStop}%
\bibitem [{\citenamefont {Osika}\ \emph {et~al.}(2017)\citenamefont {Osika},
  \citenamefont {Chac{\'{o}}n}, \citenamefont {Lewenstein},\ and\ \citenamefont
  {Szafran}}]{ci4}%
  \BibitemOpen
  \bibfield  {author} {\bibinfo {author} {\bibfnamefont {E.~N.}\ \bibnamefont
  {Osika}}, \bibinfo {author} {\bibfnamefont {A.}~\bibnamefont {Chac{\'{o}}n}},
  \bibinfo {author} {\bibfnamefont {M.}~\bibnamefont {Lewenstein}},\ and\
  \bibinfo {author} {\bibfnamefont {B.}~\bibnamefont {Szafran}},\ }\bibfield
  {title} {\bibinfo {title} {Spin-valley dynamics of electrically driven
  ambipolar carbon-nanotube quantum dots},\ }\href
  {https://doi.org/10.1088/1361-648x/aa720e} {\bibfield  {journal} {\bibinfo
  {journal} {Journal of Physics: Condensed Matter}\ }\textbf {\bibinfo {volume}
  {29}},\ \bibinfo {pages} {285301} (\bibinfo {year} {2017})}\BibitemShut
  {NoStop}%
\bibitem [{\citenamefont {Potasz}\ \emph {et~al.}(2010)\citenamefont {Potasz},
  \citenamefont {G\"u\ifmmode~\mbox{\c{c}}\else \c{c}\fi{}l\"u},\ and\
  \citenamefont {Hawrylak}}]{ci5}%
  \BibitemOpen
  \bibfield  {author} {\bibinfo {author} {\bibfnamefont {P.}~\bibnamefont
  {Potasz}}, \bibinfo {author} {\bibfnamefont {A.~D.}\ \bibnamefont
  {G\"u\ifmmode~\mbox{\c{c}}\else \c{c}\fi{}l\"u}},\ and\ \bibinfo {author}
  {\bibfnamefont {P.}~\bibnamefont {Hawrylak}},\ }\bibfield  {title} {\bibinfo
  {title} {Spin and electronic correlations in gated graphene quantum rings},\
  }\href {https://doi.org/10.1103/PhysRevB.82.075425} {\bibfield  {journal}
  {\bibinfo  {journal} {Phys. Rev. B}\ }\textbf {\bibinfo {volume} {82}},\
  \bibinfo {pages} {075425} (\bibinfo {year} {2010})}\BibitemShut {NoStop}%
\bibitem [{\citenamefont {G\"u\ifmmode~\mbox{\c{c}}\else \c{c}\fi{}l\"u}\ \emph
  {et~al.}(2009)\citenamefont {G\"u\ifmmode~\mbox{\c{c}}\else \c{c}\fi{}l\"u},
  \citenamefont {Potasz}, \citenamefont {Voznyy}, \citenamefont {Korkusinski},\
  and\ \citenamefont {Hawrylak}}]{ci61}%
  \BibitemOpen
  \bibfield  {author} {\bibinfo {author} {\bibfnamefont {A.~D.}\ \bibnamefont
  {G\"u\ifmmode~\mbox{\c{c}}\else \c{c}\fi{}l\"u}}, \bibinfo {author}
  {\bibfnamefont {P.}~\bibnamefont {Potasz}}, \bibinfo {author} {\bibfnamefont
  {O.}~\bibnamefont {Voznyy}}, \bibinfo {author} {\bibfnamefont
  {M.}~\bibnamefont {Korkusinski}},\ and\ \bibinfo {author} {\bibfnamefont
  {P.}~\bibnamefont {Hawrylak}},\ }\bibfield  {title} {\bibinfo {title}
  {Magnetism and correlations in fractionally filled degenerate shells of
  graphene quantum dots},\ }\href
  {https://doi.org/10.1103/PhysRevLett.103.246805} {\bibfield  {journal}
  {\bibinfo  {journal} {Phys. Rev. Lett.}\ }\textbf {\bibinfo {volume} {103}},\
  \bibinfo {pages} {246805} (\bibinfo {year} {2009})}\BibitemShut {NoStop}%
\bibitem [{\citenamefont {Potasz}\ \emph {et~al.}(2012)\citenamefont {Potasz},
  \citenamefont {G\"u\ifmmode~\mbox{\c{c}}\else \c{c}\fi{}l\"u}, \citenamefont
  {W\'ojs},\ and\ \citenamefont {Hawrylak}}]{ci6}%
  \BibitemOpen
  \bibfield  {author} {\bibinfo {author} {\bibfnamefont {P.}~\bibnamefont
  {Potasz}}, \bibinfo {author} {\bibfnamefont {A.~D.}\ \bibnamefont
  {G\"u\ifmmode~\mbox{\c{c}}\else \c{c}\fi{}l\"u}}, \bibinfo {author}
  {\bibfnamefont {A.}~\bibnamefont {W\'ojs}},\ and\ \bibinfo {author}
  {\bibfnamefont {P.}~\bibnamefont {Hawrylak}},\ }\bibfield  {title} {\bibinfo
  {title} {Electronic properties of gated triangular graphene quantum dots:
  Magnetism, correlations, and geometrical effects},\ }\href
  {https://doi.org/10.1103/PhysRevB.85.075431} {\bibfield  {journal} {\bibinfo
  {journal} {Phys. Rev. B}\ }\textbf {\bibinfo {volume} {85}},\ \bibinfo
  {pages} {075431} (\bibinfo {year} {2012})}\BibitemShut {NoStop}%
\bibitem [{\citenamefont {Lepage}(1978)}]{lepage}%
  \BibitemOpen
  \bibfield  {author} {\bibinfo {author} {\bibfnamefont {G.~P.}\ \bibnamefont
  {Lepage}},\ }\bibfield  {title} {\bibinfo {title} {A new algorithm for
  adaptive multidimensional integration},\ }\href
  {https://doi.org/https://doi.org/10.1016/0021-9991(78)90004-9} {\bibfield
  {journal} {\bibinfo  {journal} {Journal of Computational Physics}\ }\textbf
  {\bibinfo {volume} {27}},\ \bibinfo {pages} {192 } (\bibinfo {year}
  {1978})}\BibitemShut {NoStop}%
\bibitem [{\citenamefont {Slater}(1930)}]{sc3}%
  \BibitemOpen
  \bibfield  {author} {\bibinfo {author} {\bibfnamefont {J.~C.}\ \bibnamefont
  {Slater}},\ }\bibfield  {title} {\bibinfo {title} {Atomic shielding
  constants},\ }\href {https://doi.org/10.1103/PhysRev.36.57} {\bibfield
  {journal} {\bibinfo  {journal} {Phys. Rev.}\ }\textbf {\bibinfo {volume}
  {36}},\ \bibinfo {pages} {57} (\bibinfo {year} {1930})}\BibitemShut {NoStop}%
\bibitem [{\citenamefont {Lesiuk}\ and\ \citenamefont {Moszynski}(2014)}]{sc2}%
  \BibitemOpen
  \bibfield  {author} {\bibinfo {author} {\bibfnamefont {M.}~\bibnamefont
  {Lesiuk}}\ and\ \bibinfo {author} {\bibfnamefont {R.}~\bibnamefont
  {Moszynski}},\ }\bibfield  {title} {\bibinfo {title} {Reexamination of the
  calculation of two-center, two-electron integrals over slater-type orbitals.
  i. coulomb and hybrid integrals},\ }\href
  {https://doi.org/10.1103/PhysRevE.90.063318} {\bibfield  {journal} {\bibinfo
  {journal} {Phys. Rev. E}\ }\textbf {\bibinfo {volume} {90}},\ \bibinfo
  {pages} {063318} (\bibinfo {year} {2014})}\BibitemShut {NoStop}%
\bibitem [{\citenamefont {Clementi}\ \emph {et~al.}(1967)\citenamefont
  {Clementi}, \citenamefont {Raimondi},\ and\ \citenamefont {Reinhardt}}]{sc4}%
  \BibitemOpen
  \bibfield  {author} {\bibinfo {author} {\bibfnamefont {E.}~\bibnamefont
  {Clementi}}, \bibinfo {author} {\bibfnamefont {D.~L.}\ \bibnamefont
  {Raimondi}},\ and\ \bibinfo {author} {\bibfnamefont {W.~P.}\ \bibnamefont
  {Reinhardt}},\ }\bibfield  {title} {\bibinfo {title} {Atomic screening
  constants from scf functions. ii. atoms with 37 to 86 electrons},\
  }\href@noop {} {\bibfield  {journal} {\bibinfo  {journal} {The Journal of
  Chemical Physics}\ }\textbf {\bibinfo {volume} {47}},\ \bibinfo {pages}
  {1300} (\bibinfo {year} {1967})}\BibitemShut {NoStop}%
\bibitem [{\citenamefont {Kresse}\ and\ \citenamefont {Joubert}(1999)}]{PAW}%
  \BibitemOpen
  \bibfield  {author} {\bibinfo {author} {\bibfnamefont {G.}~\bibnamefont
  {Kresse}}\ and\ \bibinfo {author} {\bibfnamefont {D.}~\bibnamefont
  {Joubert}},\ }\bibfield  {title} {\bibinfo {title} {From ultrasoft
  pseudopotentials to the projector augmented-wave method},\ }\href
  {https://doi.org/10.1103/PhysRevB.59.1758} {\bibfield  {journal} {\bibinfo
  {journal} {Phys. Rev. B}\ }\textbf {\bibinfo {volume} {59}},\ \bibinfo
  {pages} {1758} (\bibinfo {year} {1999})}\BibitemShut {NoStop}%
\bibitem [{\citenamefont {Perdew}\ \emph {et~al.}(1996)\citenamefont {Perdew},
  \citenamefont {Burke},\ and\ \citenamefont {Ernzerhof}}]{PBE}%
  \BibitemOpen
  \bibfield  {author} {\bibinfo {author} {\bibfnamefont {J.~P.}\ \bibnamefont
  {Perdew}}, \bibinfo {author} {\bibfnamefont {K.}~\bibnamefont {Burke}},\ and\
  \bibinfo {author} {\bibfnamefont {M.}~\bibnamefont {Ernzerhof}},\ }\bibfield
  {title} {\bibinfo {title} {Generalized gradient approximation made simple},\
  }\href {https://doi.org/10.1103/PhysRevLett.77.3865} {\bibfield  {journal}
  {\bibinfo  {journal} {Phys. Rev. Lett.}\ }\textbf {\bibinfo {volume} {77}},\
  \bibinfo {pages} {3865} (\bibinfo {year} {1996})}\BibitemShut {NoStop}%
\bibitem [{\citenamefont {Kresse}\ and\ \citenamefont
  {Furthm\"uller}(1996)}]{VASP}%
  \BibitemOpen
  \bibfield  {author} {\bibinfo {author} {\bibfnamefont {G.}~\bibnamefont
  {Kresse}}\ and\ \bibinfo {author} {\bibfnamefont {J.}~\bibnamefont
  {Furthm\"uller}},\ }\bibfield  {title} {\bibinfo {title} {Efficient iterative
  schemes for ab initio total-energy calculations using a plane-wave basis
  set},\ }\href {https://doi.org/10.1103/PhysRevB.54.11169} {\bibfield
  {journal} {\bibinfo  {journal} {Phys. Rev. B}\ }\textbf {\bibinfo {volume}
  {54}},\ \bibinfo {pages} {11169} (\bibinfo {year} {1996})}\BibitemShut
  {NoStop}%
\bibitem [{\citenamefont {Szafran}\ and\ \citenamefont {\ifmmode~\dot{Z}\else
  \.{Z}\fi{}ebrowski}(2018{\natexlab{b}})}]{szafran}%
  \BibitemOpen
  \bibfield  {author} {\bibinfo {author} {\bibfnamefont {B.}~\bibnamefont
  {Szafran}}\ and\ \bibinfo {author} {\bibfnamefont {D.}~\bibnamefont
  {\ifmmode~\dot{Z}\else \.{Z}\fi{}ebrowski}},\ }\bibfield  {title} {\bibinfo
  {title} {Spin and valley control in single and double electrostatic silicene
  quantum dots},\ }\href {https://doi.org/10.1103/PhysRevB.98.155305}
  {\bibfield  {journal} {\bibinfo  {journal} {Phys. Rev. B}\ }\textbf {\bibinfo
  {volume} {98}},\ \bibinfo {pages} {155305} (\bibinfo {year}
  {2018}{\natexlab{b}})}\BibitemShut {NoStop}%
\bibitem [{\citenamefont {Hofstadter}(1976)}]{hofst}%
  \BibitemOpen
  \bibfield  {author} {\bibinfo {author} {\bibfnamefont {D.~R.}\ \bibnamefont
  {Hofstadter}},\ }\bibfield  {title} {\bibinfo {title} {Energy levels and wave
  functions of bloch electrons in rational and irrational magnetic fields},\
  }\href {https://doi.org/10.1103/physrevb.14.2239} {\bibfield  {journal}
  {\bibinfo  {journal} {Physical Review B}\ }\textbf {\bibinfo {volume} {14}},\
  \bibinfo {pages} {2239} (\bibinfo {year} {1976})}\BibitemShut {NoStop}%
\bibitem [{\citenamefont {Barenco}\ \emph {et~al.}(1995)\citenamefont
  {Barenco}, \citenamefont {Bennett}, \citenamefont {Cleve}, \citenamefont
  {DiVincenzo}, \citenamefont {Margolus}, \citenamefont {Shor}, \citenamefont
  {Sleator}, \citenamefont {Smolin},\ and\ \citenamefont
  {Weinfurter}}]{univer}%
  \BibitemOpen
  \bibfield  {author} {\bibinfo {author} {\bibfnamefont {A.}~\bibnamefont
  {Barenco}}, \bibinfo {author} {\bibfnamefont {C.~H.}\ \bibnamefont
  {Bennett}}, \bibinfo {author} {\bibfnamefont {R.}~\bibnamefont {Cleve}},
  \bibinfo {author} {\bibfnamefont {D.~P.}\ \bibnamefont {DiVincenzo}},
  \bibinfo {author} {\bibfnamefont {N.}~\bibnamefont {Margolus}}, \bibinfo
  {author} {\bibfnamefont {P.}~\bibnamefont {Shor}}, \bibinfo {author}
  {\bibfnamefont {T.}~\bibnamefont {Sleator}}, \bibinfo {author} {\bibfnamefont
  {J.~A.}\ \bibnamefont {Smolin}},\ and\ \bibinfo {author} {\bibfnamefont
  {H.}~\bibnamefont {Weinfurter}},\ }\bibfield  {title} {\bibinfo {title}
  {Elementary gates for quantum computation},\ }\href
  {https://doi.org/10.1103/PhysRevA.52.3457} {\bibfield  {journal} {\bibinfo
  {journal} {Phys. Rev. A}\ }\textbf {\bibinfo {volume} {52}},\ \bibinfo
  {pages} {3457} (\bibinfo {year} {1995})}\BibitemShut {NoStop}%
\bibitem [{\citenamefont {Fan}\ \emph {et~al.}(2005)\citenamefont {Fan},
  \citenamefont {Roychowdhury},\ and\ \citenamefont {Szkopek}}]{swap}%
  \BibitemOpen
  \bibfield  {author} {\bibinfo {author} {\bibfnamefont {H.}~\bibnamefont
  {Fan}}, \bibinfo {author} {\bibfnamefont {V.}~\bibnamefont {Roychowdhury}},\
  and\ \bibinfo {author} {\bibfnamefont {T.}~\bibnamefont {Szkopek}},\
  }\bibfield  {title} {\bibinfo {title} {Optimal two-qubit quantum circuits
  using exchange interactions},\ }\href@noop {} {\bibfield  {journal} {\bibinfo
   {journal} {Physical Review A}\ }\textbf {\bibinfo {volume} {72}},\ \bibinfo
  {pages} {052323} (\bibinfo {year} {2005})}\BibitemShut {NoStop}%
\bibitem [{\citenamefont {Nowack}\ \emph {et~al.}(2011)\citenamefont {Nowack},
  \citenamefont {Shafiei}, \citenamefont {Laforest}, \citenamefont
  {Prawiroatmodjo}, \citenamefont {Schreiber}, \citenamefont {Reichl},
  \citenamefont {Wegscheider},\ and\ \citenamefont {Vandersypen}}]{pauli1}%
  \BibitemOpen
  \bibfield  {author} {\bibinfo {author} {\bibfnamefont {K.}~\bibnamefont
  {Nowack}}, \bibinfo {author} {\bibfnamefont {M.}~\bibnamefont {Shafiei}},
  \bibinfo {author} {\bibfnamefont {M.}~\bibnamefont {Laforest}}, \bibinfo
  {author} {\bibfnamefont {G.}~\bibnamefont {Prawiroatmodjo}}, \bibinfo
  {author} {\bibfnamefont {L.}~\bibnamefont {Schreiber}}, \bibinfo {author}
  {\bibfnamefont {C.}~\bibnamefont {Reichl}}, \bibinfo {author} {\bibfnamefont
  {W.}~\bibnamefont {Wegscheider}},\ and\ \bibinfo {author} {\bibfnamefont
  {L.}~\bibnamefont {Vandersypen}},\ }\bibfield  {title} {\bibinfo {title}
  {Single-shot correlations and two-qubit gate of solid-state spins},\
  }\href@noop {} {\bibfield  {journal} {\bibinfo  {journal} {Science}\ }\textbf
  {\bibinfo {volume} {333}},\ \bibinfo {pages} {1269} (\bibinfo {year}
  {2011})}\BibitemShut {NoStop}%
\bibitem [{\citenamefont {Hanson}\ \emph {et~al.}(2007)\citenamefont {Hanson},
  \citenamefont {Kouwenhoven}, \citenamefont {Petta}, \citenamefont {Tarucha},\
  and\ \citenamefont {Vandersypen}}]{pauli2}%
  \BibitemOpen
  \bibfield  {author} {\bibinfo {author} {\bibfnamefont {R.}~\bibnamefont
  {Hanson}}, \bibinfo {author} {\bibfnamefont {L.~P.}\ \bibnamefont
  {Kouwenhoven}}, \bibinfo {author} {\bibfnamefont {J.~R.}\ \bibnamefont
  {Petta}}, \bibinfo {author} {\bibfnamefont {S.}~\bibnamefont {Tarucha}},\
  and\ \bibinfo {author} {\bibfnamefont {L.~M.~K.}\ \bibnamefont
  {Vandersypen}},\ }\bibfield  {title} {\bibinfo {title} {Spins in few-electron
  quantum dots},\ }\href {https://doi.org/10.1103/RevModPhys.79.1217}
  {\bibfield  {journal} {\bibinfo  {journal} {Rev. Mod. Phys.}\ }\textbf
  {\bibinfo {volume} {79}},\ \bibinfo {pages} {1217} (\bibinfo {year}
  {2007})}\BibitemShut {NoStop}%
\bibitem [{\citenamefont {P\'alyi}\ and\ \citenamefont {Burkard}(2010)}]{nt1}%
  \BibitemOpen
  \bibfield  {author} {\bibinfo {author} {\bibfnamefont {A.}~\bibnamefont
  {P\'alyi}}\ and\ \bibinfo {author} {\bibfnamefont {G.}~\bibnamefont
  {Burkard}},\ }\bibfield  {title} {\bibinfo {title} {Spin-valley blockade in
  carbon nanotube double quantum dots},\ }\href
  {https://doi.org/10.1103/PhysRevB.82.155424} {\bibfield  {journal} {\bibinfo
  {journal} {Phys. Rev. B}\ }\textbf {\bibinfo {volume} {82}},\ \bibinfo
  {pages} {155424} (\bibinfo {year} {2010})}\BibitemShut {NoStop}%
\bibitem [{\citenamefont {Perron}\ \emph {et~al.}(2016)\citenamefont {Perron},
  \citenamefont {Stewart},\ and\ \citenamefont {Zimmerman}}]{pauli3}%
  \BibitemOpen
  \bibfield  {author} {\bibinfo {author} {\bibfnamefont {J.~K.}\ \bibnamefont
  {Perron}}, \bibinfo {author} {\bibfnamefont {M.~D.}\ \bibnamefont
  {Stewart}},\ and\ \bibinfo {author} {\bibfnamefont {N.~M.}\ \bibnamefont
  {Zimmerman}},\ }\bibfield  {title} {\bibinfo {title} {A new regime of
  pauli-spin blockade},\ }\href {https://doi.org/10.1063/1.4945393} {\bibfield
  {journal} {\bibinfo  {journal} {Journal of Applied Physics}\ }\textbf
  {\bibinfo {volume} {119}},\ \bibinfo {pages} {134307} (\bibinfo {year}
  {2016})}\BibitemShut {NoStop}%
\bibitem [{\citenamefont {Santos}\ and\ \citenamefont {Kaxiras}(2013)}]{eff2}%
  \BibitemOpen
  \bibfield  {author} {\bibinfo {author} {\bibfnamefont {E.~J.}\ \bibnamefont
  {Santos}}\ and\ \bibinfo {author} {\bibfnamefont {E.}~\bibnamefont
  {Kaxiras}},\ }\bibfield  {title} {\bibinfo {title} {Electric-field dependence
  of the effective dielectric constant in graphene},\ }\href@noop {} {\bibfield
   {journal} {\bibinfo  {journal} {Nano letters}\ }\textbf {\bibinfo {volume}
  {13}},\ \bibinfo {pages} {898} (\bibinfo {year} {2013})}\BibitemShut
  {NoStop}%
\bibitem [{\citenamefont {Geick}\ \emph {et~al.}(1966)\citenamefont {Geick},
  \citenamefont {Perry},\ and\ \citenamefont {Rupprecht}}]{eff1}%
  \BibitemOpen
  \bibfield  {author} {\bibinfo {author} {\bibfnamefont {R.}~\bibnamefont
  {Geick}}, \bibinfo {author} {\bibfnamefont {C.~H.}\ \bibnamefont {Perry}},\
  and\ \bibinfo {author} {\bibfnamefont {G.}~\bibnamefont {Rupprecht}},\
  }\bibfield  {title} {\bibinfo {title} {Normal modes in hexagonal boron
  nitride},\ }\href@noop {} {\bibfield  {journal} {\bibinfo  {journal} {Phys.
  Rev.}\ }\textbf {\bibinfo {volume} {146}},\ \bibinfo {pages} {543} (\bibinfo
  {year} {1966})}\BibitemShut {NoStop}%
\bibitem [{\citenamefont {Keldysh}(1979)}]{eff3}%
  \BibitemOpen
  \bibfield  {author} {\bibinfo {author} {\bibfnamefont {L.}~\bibnamefont
  {Keldysh}},\ }\bibfield  {title} {\bibinfo {title} {Coulomb interaction in
  thin semiconductor and semimetal films},\ }\href@noop {} {\bibfield
  {journal} {\bibinfo  {journal} {JETP Lett}\ }\textbf {\bibinfo {volume}
  {29}},\ \bibinfo {pages} {658} (\bibinfo {year} {1979})}\BibitemShut
  {NoStop}%
\bibitem [{\citenamefont {Schmitt-Rink}\ and\ \citenamefont
  {Ell}(1985)}]{eff4}%
  \BibitemOpen
  \bibfield  {author} {\bibinfo {author} {\bibfnamefont {S.}~\bibnamefont
  {Schmitt-Rink}}\ and\ \bibinfo {author} {\bibfnamefont {C.}~\bibnamefont
  {Ell}},\ }\bibfield  {title} {\bibinfo {title} {Excitons and electron–hole
  plasma in quasi-two-dimensional systems},\ }in\ \href
  {https://doi.org/https://doi.org/10.1016/B978-0-444-86931-9.50047-X} {\emph
  {\bibinfo {booktitle} {High Excitation and Short Pulse Phenomena}}},\
  \bibinfo {editor} {edited by\ \bibinfo {editor} {\bibfnamefont
  {M.}~\bibnamefont {Pilkuhn}}}\ (\bibinfo  {publisher} {Elsevier},\ \bibinfo
  {address} {Amsterdam},\ \bibinfo {year} {1985})\ pp.\ \bibinfo {pages} {585
  -- 596}\BibitemShut {NoStop}%
\bibitem [{\citenamefont {Van~Tuan}\ \emph {et~al.}(2018)\citenamefont
  {Van~Tuan}, \citenamefont {Yang},\ and\ \citenamefont {Dery}}]{eff5}%
  \BibitemOpen
  \bibfield  {author} {\bibinfo {author} {\bibfnamefont {D.}~\bibnamefont
  {Van~Tuan}}, \bibinfo {author} {\bibfnamefont {M.}~\bibnamefont {Yang}},\
  and\ \bibinfo {author} {\bibfnamefont {H.}~\bibnamefont {Dery}},\ }\bibfield
  {title} {\bibinfo {title} {Coulomb interaction in monolayer transition-metal
  dichalcogenides},\ }\href {https://doi.org/10.1103/PhysRevB.98.125308}
  {\bibfield  {journal} {\bibinfo  {journal} {Phys. Rev. B}\ }\textbf {\bibinfo
  {volume} {98}},\ \bibinfo {pages} {125308} (\bibinfo {year}
  {2018})}\BibitemShut {NoStop}%
\bibitem [{\citenamefont {Berkelbach}\ and\ \citenamefont
  {Reichman}(2018)}]{eff6}%
  \BibitemOpen
  \bibfield  {author} {\bibinfo {author} {\bibfnamefont {T.~C.}\ \bibnamefont
  {Berkelbach}}\ and\ \bibinfo {author} {\bibfnamefont {D.~R.}\ \bibnamefont
  {Reichman}},\ }\bibfield  {title} {\bibinfo {title} {Optical and excitonic
  properties of atomically thin transition-metal dichalcogenides},\ }\href
  {https://doi.org/10.1146/annurev-conmatphys-033117-054009} {\bibfield
  {journal} {\bibinfo  {journal} {Annual Review of Condensed Matter Physics}\
  }\textbf {\bibinfo {volume} {9}},\ \bibinfo {pages} {379} (\bibinfo {year}
  {2018})}\BibitemShut {NoStop}%
\bibitem [{\citenamefont {Cudazzo}\ \emph {et~al.}(2011)\citenamefont
  {Cudazzo}, \citenamefont {Tokatly},\ and\ \citenamefont {Rubio}}]{eff7}%
  \BibitemOpen
  \bibfield  {author} {\bibinfo {author} {\bibfnamefont {P.}~\bibnamefont
  {Cudazzo}}, \bibinfo {author} {\bibfnamefont {I.~V.}\ \bibnamefont
  {Tokatly}},\ and\ \bibinfo {author} {\bibfnamefont {A.}~\bibnamefont
  {Rubio}},\ }\bibfield  {title} {\bibinfo {title} {Dielectric screening in
  two-dimensional insulators: Implications for excitonic and impurity states in
  graphane},\ }\href@noop {} {\bibfield  {journal} {\bibinfo  {journal} {Phys.
  Rev. B}\ }\textbf {\bibinfo {volume} {84}},\ \bibinfo {pages} {085406}
  (\bibinfo {year} {2011})}\BibitemShut {NoStop}%
\end{thebibliography}%

\end{document}